\documentclass[aps,prb,times,twocolumn,longbibliography,superscriptaddress,showpacs,a4paper]{revtex4-1}
\usepackage{colortbl,marvosym}
\usepackage{amsfonts,amsmath,amssymb}
\usepackage{graphicx,wasysym}
\usepackage{tabularx,hhline}
\usepackage[dvipsnames]{xcolor}
\usepackage{multirow}% http://ctan.org/pkg/multirow
\usepackage[colorlinks,bookmarks=false,citecolor=blue,linkcolor=red,urlcolor=blue]{hyperref}
\usepackage{color}
\usepackage{ulem} % used to strike out text
\usepackage{cancel} % used to strike out text

\DeclareMathAlphabet{\mathpzc}{OT1}{pzc}{m}{it}

\makeatletter
\newcommand*{\rom}[1]{\expandafter\@slowromancap\romannumeral #1@}
\makeatother

\begin{document}
%%%%%%%%%%%%%%%%%%%%%%%%%%%%%%%%%%%%%%%%%%%%%%%%%%%%

%%%%%%%%%%%%%%%%%%%%%%%%%%%%%%%%%%%%%%%%%%%%%%%%%%%%
%\title{What color is a Graphene Quantum Dot? \\
%Exploring the shape and symmetry of GQDs through optical conductivity}
\title{Symmetry and optical selection rules in graphene quantum dots}
%%%%%%%%%%%%%%%%%%%%%%%%%%%%%%%%%%%%%%%%%%%%%%%%%%%%

%%%%%%%%%%%%%%%%%%%%%%%%%%%%%%%%%%%%%%%%%%%%%%%%%%%%
\author{Rico Pohle}
%%%%%%%%%%%%%%%%%%%%%%%%%%%%%%%%%%%%%%%%%%%%%%%%%%%%
\affiliation{
Okinawa Institute of Science and Technology Graduate University, 
Onna-son, Okinawa 904-0495, Japan}

%%%%%%%%%%%%%%%%%%%%%%%%%%%%%%%%%%%%%%%%%%%%%%%%%%%%
\author{Eleftheria G. Kavousanaki}
%%%%%%%%%%%%%%%%%%%%%%%%%%%%%%%%%%%%%%%%%%%%%%%%%%%%
\affiliation{Femtosecond Spectroscopy Unit,   
Okinawa Institute of Science and Technology Graduate University, 
Onna-son, Okinawa 904-0495, Japan}
\affiliation{Crete Center for Quantum Complexity and Nanotechnology, Department of Physics, 
University of Crete, 71003, Heraklion, Greece}

%%%%%%%%%%%%%%%%%%%%%%%%%%%%%%%%%%%%%%%%%%%%%%%%%%%%
\author{Keshav M. Dani}
%%%%%%%%%%%%%%%%%%%%%%%%%%%%%%%%%%%%%%%%%%%%%%%%%%%%
\affiliation{Femtosecond Spectroscopy Unit, 
Okinawa Institute of Science and Technology Graduate University, 
Onna-son, Okinawa 904-0495, Japan}

%%%%%%%%%%%%%%%%%%%%%%%%%%%%%%%%%%%%%%%%%%%%%%%%%%%%
\author{Nic Shannon}
%%%%%%%%%%%%%%%%%%%%%%%%%%%%%%%%%%%%%%%%%%%%%%%%%%%%
\affiliation{
Okinawa Institute of Science and Technology Graduate University, 
Onna-son, Okinawa 904-0495, Japan}

%%%%%%%%%%%%%%%%%%%%%%%%%%%%%%%%%%%%%%%%%%%%%%%%%%%%
\date{\today} 
%%%%%%%%%%%%%%%%%%%%%%%%%%%%%%%%%%%%%%%%%%%%%%%%%%%%

%%%%%%%%%%%%%%%%%%%%%%%%%%%%%%%%%%%%%%%%%%%%%%%%%%%%
\begin{abstract}
%%%%%%%%%%%%%%%%%%%%%%%%%%%%%%%%%%%%%%%%%%%%%%%%%%%%

Graphene quantum dots (GQD's) have optical properties 
which are very different from those of an extended graphene sheet.  
In this Article we explore how the size, shape and edge--structure of a 
GQD affect its optical conductivity.
Using representation theory, we derive optical selection rules for regular--shaped dots, 
starting from the symmetry properties of the current operator.
We find that, where the x-- and y--components of the current 
operator transform with the same irreducible representation (irrep) of the point 
group --- for example in triangular or hexagonal GQD's --- 
the optical conductivity is independent of the polarisation of the light.
On the other hand, where these components transform with 
different irreps --- for example in rectangular GQD's --- the optical conductivity 
depends on the polarisation of light.  
%We find that GQD's with non-commuting point--group operations --- for example dots of
%rectangular shape --- can be distinguished  from GQD's with commuting point--group 
%operations --- for example dots of triangular or hexagonal shape --- by using polarized light.
%
We carry out explicit calculations of the optical conductivity of GQD's described by a simple
tight--binding model and, for dots of intermediate size, 
%\textcolor{blue}{($10 \lesssim  L \lesssim  50\ \text{nm}$)} 
find an absorption peak in the low--frequency range of the spectrum
which allows us to distinguish between dots with zigzag and armchair edges.
We also clarify the one--dimensional nature of states at the van Hove singularity in graphene, 
providing a possible explanation for very high exciton--binding energies.
Finally we discuss the role of atomic vacancies and shape asymmetry. 

%%%%%%%%%%%%%%%%%%%%%%%%%%%%%%%%%%%%%%%%%%%%%%%%%%%%
\end{abstract}
%%%%%%%%%%%%%%%%%%%%%%%%%%%%%%%%%%%%%%%%%%%%%%%%%%%%

\pacs{xx.xx, xx.xx, xx.xx}

%%%%%%%%%%%%%%%%%%%%%%%%%%%%%%%%%%%%%%%%%%%%%%%%%%%%
\maketitle
%%%%%%%%%%%%%%%%%%%%%%%%%%%%%%%%%%%%%%%%%%%%%%%%%%%%

\let\clearpage\relax

%%%%%%%%%%%%%%%%%%%%%%%%%%%%%%%%%%%%%%%%%%%%%%%%%%
\section{Introduction}
%%%%%%%%%%%%%%%%%%%%%%%%%%%%%%%%%%%%%%%%%%%%%%%%%%

Graphene, a single sheet of carbon atoms arranged in a honeycomb lattice, became 
experimentally accessible in 2004 through perhaps the most innovative use of scotch tape 
in the 21st century \cite{Novoselov2004, Novoselov2005}.
The first of the many surprises of this ``wonder material'' was that it could be 
seen at all, %by the people who discovered it, 
using nothing more than an optical microscope \cite{Nair2008}.
And in fact, the large, universal, and approximately constant optical response of 
graphene in the visible spectrum is a signature of one of its other remarkable properties 
--- electrons with a relativistic ``Dirac'' dispersion 
\cite{Ando2002, Gusynin2007, Ziegler2007, Min2009, Stauber2008, Yuan2011b, Buividovich2012}.  

%%%%%%%%%%%%%%%%%%%%%%%%%%%%%%%%%%%%%%%%%%%%%%%%%

Graphene is also a very good conductor of DC electric current 
\cite{Novoselov2005, Geim2007, Miao2007}. 
However in this case, the conductivity measured in experiment  
is found to depend on the boundaries 
of the graphene sheet \cite{Katsnelson2006a, Tworzydlo2006, Ryu2007}, a fact 
which highlights the topological character of graphene's 
electronic states\cite{Sarma2011, Kotov2012}.
Boundary effects are even more pronounced in graphene nanostructures 
referred to as ``graphene quantum dots'' (GQD's).
GQD's have a discrete energy spectrum, and can be viewed as large, sp$^2$--bonded, 
carbon molecules, with electronic states which depend on the size, shape and 
symmetry of the dot \cite{Silva2010, GrapheneQuantumDots,Guclu2010}.  

%%%%%%%%%%%%%%%%%%%%%%%%%%%%%%%%%%%%%%%%%%%%%%%%%

The possibility of engineering the energy spectrum of a GQD, and therefore 
its optical properties, has suggested potential applications in fields ranging 
from quantum computation to solar energy 
\cite{Zhu2011, Luk2012, Son2012, Konstantatos2012, Kim2012, Jin2013, Zhang2015, Roy2015, Umrao2015, QuantumSolarCells}.
A range of different fabrication techniques are now available for GQD's 
%[say something about types of method] 
\cite{Coraux2009, Lu2011, Li2011, Mohanty2012, Olle2012, Yan2012, Mullen2014, Wang2014}.  
However, tailoring the properties of a GQD to a specific application requires 
the ability to fabricate dots with the desired shape, or to post--select for dots 
with a given shape after fabrication.
In either case, understanding the relationship between the size and shape 
of the dot, and its optical properties is paramount. 

%%%%%%%%%%%%%%%%%%%%%%%%%%%%%%%%%%%%%%%%%%%%%%%%%

In this Article, we explore how the size, shape and edge--geometry of a GQD
combine to determine its optical conductivity, paying particular attention to the 
symmetry of the dot, and the optical selection rules which follow from it.
Considering regular GQD's with a range of different shapes, we first examine 
how different point--group symmetries lead to different optical selection rules. 
We find that selection rules depend on the way in which the different 
components of the current operator transform under the symmetries 
of the dot.
Where both components of the current operator, $\hat{\mathpzc{J}}^x$ 
and $\hat{\mathpzc{J}}^y$,  transform with the same irreducible representation 
(irrep) of the point group  --- for example in triangular dots --- we find that the 
optical conductivity does not depend on the polarization of the incident light.
%
%In the case of GQD's with symmetry operations which commute 
%--- for example rectangular dots --- we find that the optical conductivity 
%depends explicitly on the polarization of the incident light. 
%
On the other hand, where the different components of the current operator
transform under different irreps  --- for example in rectangular dots ---  
the optical conductivity does depend on the polarization of the incident light.
%
%Meanwhile, GQD's whose symmetry operations do not commute 
% --- for example triangular dots ---  do not show polarization--dependence.
%
This result is illustrated through explicit, numerical calculations of the optical 
conductivity of regular GQD's within a simple tight--binding model.

%%%%%%%%%%%%%%%%%%%%%%%%%%%%%%%%%%%%%%%%%%%%%%%%%

The same numerical approach is used to explore how the optical properties of 
a GQD evolve into those of a graphene sheet, as the size of the dot is increased.
Here we find that edge--geometry plays an important role, with zigzag edges 
contributing a strong, additional feature to the optical conductivity 
within the visible spectrum, for GQD's of 
linear dimension $L > 10\ \text{nm}$, with 
spectral weight which scales as $1/L$.  
This feature is absent in GQD's with armchair edges, allowing a direct 
distinction between dots with different edge--types for dots of intermediate size.

%%%%%%%%%%%%%%%%%%%%%%%%%%%%%%%%%%%%%%%%%%%%%%%%%

We also examine how the strong peak in the optical conductivity 
of graphene in the ultraviolet,  
\mbox{at $\hbar\omega \sim 4.7~\text{eV}$ [\onlinecite{Eberlein2008, Mak2011}]},
evolves out of the spectrum of a GQD.   
Within a tight--binding model, this peak occurs at twice 
the energy of the hopping integral, and is associated with a \mbox{Van Hove} 
singularity in the single--particle density of states \cite{CastroNeto2009}. 
A very similar feature is observed in the optical conductivity of GQD's, 
where it can be traced to a highly--degenerate set of 
electronic states with \mbox{one--dimensional} character.
The \mbox{one--dimensional} nature of these states suggest a possible 
explanation for the high binding--energies of excitons in graphene 
\cite{Yang2009, Kravets2010,  Mak2011, Chae2011, Matkovic2012}.

%Our conclusions are robust since Coulomb interactions do not affect global symmetry arguments.  

%%%%%%%%%%%%%%%%%%%%%%%%%%%%%%%%%%%%%%%%%%%%%%%%%

Finally, we investigate the optical properties of GQD's with irregular shape 
and disorder, in the form of vacancies in the lattice.  
In this case, we find a polarization--dependent optical conductivity which 
depends on the details of each individual, asymmetric dot.
We also find new optical features arising from vacancies in the lattice.
Averaging over an ensemble of dots restores the polarization--independence 
of bulk graphene, but does not eliminate new features coming from vacancies.

%%%%%%%%%%%%%%%%%%%%%%%%%%%%%%%%%%%%%%%%%%%%%%%%%

While graphene is a new phenomenon, the study of the optical properties of 
two--dimensional (2D) materials has a long history. 
Theoretical studies of the optical conductivity in 2D systems date back roughly 70 years in the 
context of single graphite layers \cite{Wallace1947}, zero--gap semiconductors 
\cite{Fradkin1986} and d--wave superconductors \cite{Lee1993}. 
Nevertheless, studies explicitly in graphene experienced a sharp increase after its experimental 
realisation \cite{CastroNeto2009, Peres2010, Sarma2011, Kotov2012}.
The existence of Dirac cones in the dispersion relation classifies graphene as a semimetal with 
novel features like the presence of massless Dirac fermions \cite{Novoselov2005}, an absence 
of backscattering from electrostatic barriers known as the Klein paradox \cite{Katsnelson2006a} 
and an unconventional integer quantum Hall effect \cite{Novoselov2005, Zhang2005, Gusynin2005}, 
to name but a few.

%%%%%%%%%%%%%%%%%%%%%%%%%%%%%%%%%%%%%%%%%%%%%%%%%

The transport and optical properties of graphene have also attracted considerable interest.
An important prediction, which predates the discovery of graphene, is that its DC conductivity 
without disorder takes on the value of \cite{Shon1998, Ando2002, Noro2010}
\begin{equation}
\sigma^{DC}_{theo} = \frac{4}{\pi}  \frac{e^2}{h}  \; .
\end{equation}
%
%In the same studies it was shown that disorder highly affects this value. 
%
Early experiments reported values which were larger than this prediction 
by a factor of $\pi$, a fact which became known as the 
``mystery of the missing pi'' \cite{Novoselov2005, Geim2007}. 
Later studies explained in theory \cite{Katsnelson2006a, Tworzydlo2006, Ryu2007}, and 
confirmed in experiment  \cite{Miao2007}, that the value of $\sigma^{DC}$ strongly depends 
on the boundary conditions of the graphene sheet, highlighting the important role of topology 
in graphene's electronic states.
Disorder and interactions have been argued to also play a role \cite{Ando2002}.

%%%%%%%%%%%%%%%%%%%%%%%%%%%%%%%%%%%%%%%%%%%%%%%%%

Perhaps the most striking feature of graphene's optical conductivity is its universal value
\begin{equation}
	\sigma_0 = \frac{\pi}{2} \frac{e^2}{h}
	\label{eq:sigma0}
\end{equation}
over a wide range of frequencies which include the visible spectrum 
\cite{Ando2002, Ziegler2007, Min2009, Stauber2008, Yuan2011b, Buividovich2012}.
This universal optical conductivity is observed in experiments 
\cite{Nair2008, Mak2008, Kuzmenko2008, Mak2011, Gogoi2012}, and falls within the 
visible spectrum, making it possible to see a single layer of carbon atoms using only 
an optical microscope \cite{Nair2008}.
The optical conductivity of graphene remains nearly frequency--independent across
the visible spectrum in the presence of (weak) disorder \cite{Ostrovsky2006,Peres2006}.  
However, stronger disorder can lead to deviations from the universal value 
[Eq.~(\ref{eq:sigma0})], and contribute an additional peak at finite--energy \cite{Yuan2011b}. 

%%%%%%%%%%%%%%%%%%%%%%%%%%%%%%%%%%%%%%%%%%%%%%%%%

The other striking feature of the optical conductivity of graphene is a strong, asymmetric 
peak at energies \mbox{$\hbar\omega \sim 4.7\ \text{eV}$} \cite{Eberlein2008, Mak2011}, 
with a Fano resonance--like line--shape \cite{Chae2011}.
This peak is seen in electron loss spectroscopy \cite{Eberlein2008} and spectroscopic 
ellipsometry  \cite{Kravets2010} as well as in optical absorption  \cite{Kravets2010},  
transmission \cite{Chae2011} and reflection \cite{Mak2011, Lee2011}. 
A similar peak is seen in calculations based on a non-interacting tight--binding model, 
which can be attributed to a \mbox{van Hove} singularity in the density of states (DOS) 
\cite{Stauber2008}.
Once electron-electron interactions are taken into account, this peak is redshifted by 
\mbox{ $\Delta=400$ -- $600$ meV}, which is attributed to an excitonic state strongly coupled 
to a band continuum \cite{Yang2009, Yuan2011a, Kravets2010,  Mak2011, Chae2011, Matkovic2012}.

%%%%%%%%%%%%%%%%%%%%%%%%%%%%%%%%%%%%%%%%%%%%%%%%%

The optical properties of graphene change dramatically, once its electrons are 
spatially--confined within a nanostructure.
Recently, there has been growing interest in the properties of nanoscale flakes of 
graphene, commonly referred to as graphene quantum dots (GQD's), graphene 
nanoislands \cite{Olle2012}, or nanographene \cite{Yamamoto2006}. 
This research has also been motivated by potential applications in quantum computing 
\cite{Trauzettel2007}, bioimaging \cite{Zhu2011, Jin2013, Roy2015, Umrao2015}, 
LEDs light converters \cite{Luk2012,Son2012}, photodetectors \cite{Konstantatos2012, Zhang2015}, 
and organic solar cells \cite{QuantumSolarCells}.
GQD's can be synthesized in several ways, e.g. via fragmentation of C$_{60}$ 
molecules \cite{Lu2011}, nanoscale cutting of graphite combined with exfoliation 
\cite{Mohanty2012}, chamber pressure chemical vapor deposition (CP--CVD) 
\cite{Yan2012} and controlled decomposition of hydrocarbons \cite{Olle2012}. 
Furthermore, scanning tunneling microscopy (STM) measurements confirmed the 
confinement of electronic states in GQD's 
\cite{Ritter2009, Hamalainen2011, Subramaniam2012, Jolie2014, GrapheneQuantumDots} 
and motivated further theoretical studies.

%%%%%%%%%%%%%%%%%%%%%%%%%%%%%%%%%%%%%%%%%%%%%%%%%

Much like graphene nanoribbons \cite{CastroNeto2009}, graphene quantum dots exhibit 
metallic or insulating behavior depending on the type of their edges, namely zigzag 
or armchair\cite{Zarenia2011}.
GQD's with triangular geometry and zigzag edges show zero energy edge states 
\cite{Ezawa2007, Potasz2010}, leading to magnetic effects as edge--state 
magnetization \cite{Zhang2008, Guclu2009, Potasz2012} and spin blockade \cite{Guclu2013}. 
Studies of optical properties of GQDs have shown signatures of edge states 
\cite{Yamamoto2006}, excitonic effects on the optical absorption spectrum 
\cite{Yang2007, Guclu2010, Ozfidan2014} and edge--dependent selection rules in triangular 
dots \cite{Akola2008}.

%%%%%%%%%%%%%%%%%%%%%%%%%%%%%%%%%%%%%%%%%%%%%%%%%

Despite the huge advances made in manufacturing GQD's, there are still 
obstacles to overcome towards a complete control of size, geometry and edge type. 
It is in this context that we revisit the question of how size, shape and edge geometry
affect the optical properties of GQD's paying particular attention to the shape of the dots.   
We do not explicitly take interactions between electrons into account, but instead 
emphasize optical selection rules which are entirely determined by symmetry, 
and therefore independent of the details of model \cite{Heine, LandauLifshitz, Tinkham, Weyl, Wagner,Jones}.

%%%%%%%%%%%%%%%%%%%%%%%%%%%%%%%%%%%%%%%%%%%%%%%%%

In order to be able to compare with predictions obtained in 
the thermodynamic limit, we also calculate the optical conductivity within a simple 
tight--binding model, which is known to give a good description of 
many of the properties of bulk graphene \cite{CastroNeto2009}.  
This makes it possible to explore a large range of dot sizes, and to address issues
%, such as the emergence as the emergence of new peaks in optical conductivity 
% in the presence of disorder, 
which are not determined by symmetry alone.  
However, this broader view comes at the expense 
of neglecting correlation effects, which can be significant \cite{Kotov2012}.

%%%%%%%%%%%%%%%%%%%%%%%%%%%%%%%%%%%%%%%%%%%%%%%%%

The remainder of this Article is structured as follows: 

%%%%%%%%%%%%%%%%%%%%%%%%%%%%%%%%%%%%%%%%%%%%%%%%%

In Section~\ref{sec:TheoryModel} we briefly review the calculation 
of the optical conductivity $\sigma_\alpha(\omega)$
within linear--response theory.
We discuss the role of symmetry in the determining optical selection rules
for a general many--electron wave function.
%, in the presence of interactions.
%
We also introduce a simple, non-interacting tight--binding model, 
and describe how this can be used to make explicit predictions for the optical 
conductivity of a GQD of given size and shape.

%%%%%%%%%%%%%%%%%%%%%%%%%%%%%%%%%%%%%%%%%%%%%%%%%

In Section \ref{sec:Geometry} we use group theory to analyse the optical 
selection rules found in GQD's with triangular, hexagonal and rectangular shape. 
Each of these GQD's can be classified according to a different  
point--group, with associated irreps.
We find that, where the in--plane components of the current 
operator, $\mathpzc{J}_x$ and $\mathpzc{J}_y$, transform under the 
same irrep, the optical conductivity of the GQD, 
$\sigma_\alpha(\omega)$, does not depend on the polarization of the incident light.
On the other hand, where $\mathpzc{J}_x$ and $\mathpzc{J}_y$ 
transform under different irreps, the optical conductivity is polarisation--dependent.
%
%We find that GQD's with an abelian point--group symmetry (in this case, rectangular GQD's) 
%have an optical conductivity $\sigma_{\alpha\beta}(\omega)$ which depends on 
%the polarization of the light.  
%%
%The optical response of GQD's with a non-abelian point--group, meanwhile,
%is independent of polarization.
%
%We confirm these selection rules, by calculating the current-operator 
%(Eq.~\ref{eq:currentMomentumSpace}) in the respective symmetry-basis and resolve 
%its optical conductivity (Eq.~\ref{eq:sigma.tight.binding}) within the non-interacting 
%tight--binding model (Eq.~\ref{eq:H0}).
%
We illustrate these results for non--interacting electrons 
by making explicit comparison with the tight--binding model introduced in 
Section \ref{sec:TheoryModel}.

%%%%%%%%%%%%%%%%%%%%%%%%%%%%%%%%%%%%%%%%%

In Section \ref{sec:Edges} we use the same tight--binding model 
to explore the way in which the optical properties
of a GQD with a given shape and edge structure evolve into those of an infinite 
graphene sheet.
We find that qualitative differences persist between GQD's with 
zigzag and armchair edges, even for dots with linear--dimension 
$>10\text{nm}$.   
This suggests that optical measurements may prove a useful 
way to distinguish GQD's of different edge-types. 

%%%%%%%%%%%%%%%%%%%%%%%%%%%%%%%%%%%%%%%%%

% When computing the energy spectrum and optical conductivity of GQDs or graphene, 
% we neglect coulomb interactions by using the standard tight--binding model 
% \cite{Wallace1947, CastroNeto2009, Kotov2012}. 
% %
% This constraint does not affect the value of our results, because global symmetry 
% arguments remain unaffected by electron-electron interactions, but allows us to perform 
% calculations for clusters with system sizes in their thermodynamic limit.
%
%%%%%%%%%%%%%%%%%%%%%%%%%%%%%%%%%%%%%%%%%
%
% However coulomb interactions will have an effect on the energy spectrum, which 
% should cause a redshift in the optical conductivity as discussed before for models 
% considering electron-hole excitations in graphene \cite{Yang2009}. 
% %
% Also, effects as spin-magnetization on dot edges \cite{Guclu2013} or spin blockades 
% \cite{Guclu2013} are phenomena which will not reach the focus of our study here.   \\

%%%%%%%%%%%%%%%%%%%%%%%%%%%%%%%%%%%%%%%%%

In Section \ref{sec:1-dimWaveFunction} we identify one--dimensional wave functions  
at energies of the \mbox{Van Hove} singularity of graphene, within the tight--binding 
model introduced in Section \ref{sec:TheoryModel}
This reduced dimensionality provides a possible explanation for the unusually 
high binding--energies of excitons, seen in experiments of graphene 
\cite{Yang2009, Kravets2010,  Mak2011, Chae2011, Matkovic2012}.

%%%%%%%%%%%%%%%%%%%%%%%%%%%%%%%%%%%%%%%%%

In Section \ref{sec:Asym_Def} we discuss irregular GQD's and present how 
vacancies and asymmetry affect their optical conductivity, within the same 
tight--binding model.
While sample averaging over many randomly shaped asymmetric dots weakens 
edge-effects and recovers the optical conductivity of graphene, vacancies in the 
bulk cause additional features similar to those seen in GQD's with zigzag edges. 

%%%%%%%%%%%%%%%%%%%%%%%%%%%%%%%%%%%%%%%%%

We conclude in Section \ref{sec:conclusions} with a summary of our results, 
and a discussion of potential future avenues for research.

%%%%%%%%%%%%%%%%%%%%%%%%%%%%%%%%%%%%%%%%%%%%%%%%%%
\section{Theory of optical conductivity}    
%%%%%%%%%%%%%%%%%%%%%%%%%%%%%%%%%%%%%%%%%%%%%%%%%%
\label{sec:TheoryModel}

%%%%%%%%%%%%%%%%%%%%%%%%%%%%%%%%%%%%%%%%%%%%%%%%%%
% Fig. 1 - illustration of dots
%%%%%%%%%%%%%%%%%%%%%%%%%%%%%%%%%%%%%%%%%%%%%%%%%%

\begin{figure*} [t]
	\centering \includegraphics[width=0.98\textwidth]{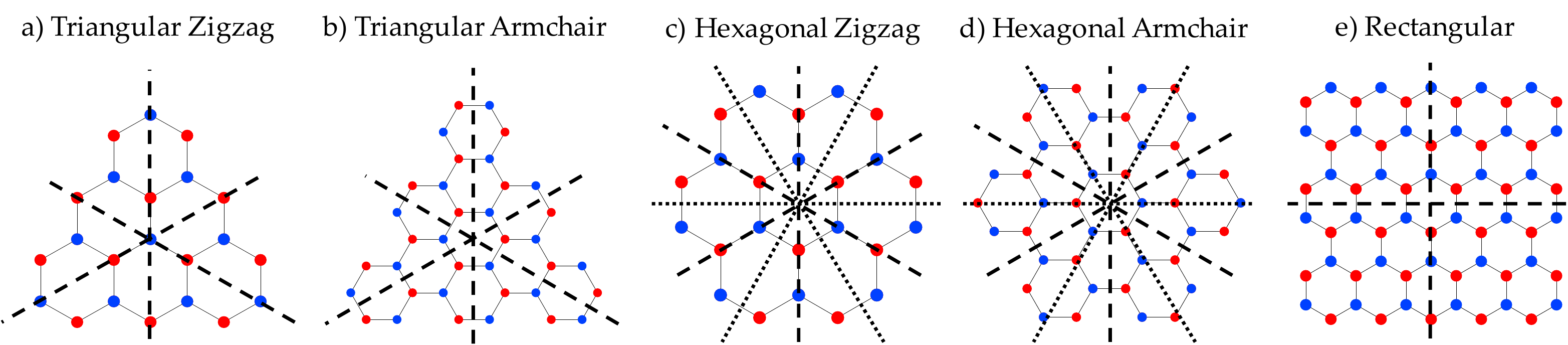}
 	\caption{
	  Regular--shaped graphene quantum dots (GQD's) considered in this Article.
          (a) Triangular zigzag,  (b) triangular armchair, (c) hexagonal zigzag, (d) 
          hexagonal armchair, and (e) rectangular. 
	  Triangular and hexagonal dots have symmetries described by the non--abelian 
	  point groups $C_{3v}$ and $C_{6v}$,  respectively, while rectangular dots have 
	  an abelian point group $C_{2v}$.
	  Dotted lines represent in--plane mirror axes, while the rotation axis (not shown) is 
	  located at the centre of each dot, perpendicular to the plane of the dot.}
\label{fig:lattice}
\end{figure*}

\subsection{Kubo formula and the role of symmetry}
%%%%%%%%%%%%%%%%%%%%%%%%%%%%%%%%%%%%%%%%%%%%%%%%%%%
\label{sec:general.theory}

In what follows we motivate the analysis of optical selection rules in terms of
the symmetries of a GQD, starting from the Kubo formula for the optical 
conductivity.
The crucial facts will be that a) optical selection rules 
are determined by the matrix elements of the current operator %$\mathpzc{J}^\alpha$ 
between the different (many--body) eigenstates of the dot, and b) this current operator 
transforms like a {\it polar vector} under the symmetries of the dot.

%%%%%%%%%%%%%%%%%%%%%%%%%%%%%%%%%%%%%%%%%%%%%%%%%%

Our chief measure of the optical properties of a GQD will be its optical conductivity.  
Formally, this is defined through
% $\sigma_{\alpha \beta} (\vec{q}, \omega)$, defined through 
 %
\begin{equation}
	\mathpzc{J}^\alpha (\vec{q}, \omega) 
	= \sum_{\beta} \sigma_{\alpha \beta} (\vec{q}, \omega) E^{\beta}(\vec{q}, \omega)   \; ,
\end{equation}
where $\sigma_{\alpha \beta} (\vec{q}, \omega)$ is the optical conductivity, and 
$\mathpzc{J}^\alpha (\vec{q}, \omega)$ is the current which flows as a result of
an applied electric field $E^{\beta}(\vec{q}, \omega)$, with frequency $\omega$, 
and direction $\alpha, \beta = x,y,z$.
The full tensor for the optical conductivity, $\sigma_{\alpha \beta} (\vec{q}, \omega)$,  
can be calculated within linear response theory, using a Kubo formula \cite{Mahan2000}.
However for the purposes of the present study, we can make a number of simplications.

%%%%%%%%%%%%%%%%%%%%%%%%%%%%%%%%%%%%%%%%%%%%%%%%%%

Firstly, we set $\vec{q} = 0$, since the wavelength of light at the 
relevant frequencies is much larger than the size of the GQD.
And, since GQD's are two--dimensional, we need only to consider the  
two--dimensional current 
\begin{eqnarray}
\vec{\mathpzc{J}} = (\mathpzc{J}^x, \mathpzc{J}^y) \; ,
\end{eqnarray}
and the in--plane components of the electric field
\begin{eqnarray}
\vec{E} = (E^x,E^y) \; .
\end{eqnarray}
We further assume that time--reversal symmetry remains unbroken, 
in which case it is sufficient to only consider the diagonal part 
of the optical conductivity
\begin{equation}
\sigma_\alpha (\omega) = \sigma_{\alpha\beta} (\omega) \delta_{\alpha\beta} \; ,
\end{equation}
such that
\begin{equation}
	\mathpzc{J}^\alpha (\omega) 
	= \sigma_\alpha (\omega) E^\alpha (\omega)   \; ,
\label{eq:linearResponse}
\end{equation}
with $\alpha = x, y$.  
And, since the temperatures of experiment are generally small
compared with the energy--scales of the dot, we further restrict 
our considerations to the real part of the optical conductivity, 
at temperature $T=0$.
Both of these conditions can be relaxed at will.

%%%%%%%%%%%%%%%%%%%%%%%%%%%%%%%%%%%%%%%%%%%%%%%%%%

With all of these restrictions in place, the Kubo formula for 
$\sigma_{\alpha} (\omega)$ reduces to the expectation value of the 
retarded, two--particle correlation function of the current operator 
$\hat{\mathpzc{J}}^\alpha(t)$, calculated within the full, many--electron 
ground state $| \psi_0 \rangle $, such that 
\begin{eqnarray}
	&&\text{Re} \big[ \sigma_\alpha (\omega) \big] \nonumber\\
	&&=   \text{Re} \left[  \frac{1}{A} \frac{1}{\hbar \omega} \int_0^{\infty} dt \ e^{i \omega t}  \ 
	\langle   \psi_0 | \big[ \hat{\mathpzc{J}}^{\dagger\alpha} (t) , 
	\hat{ \mathpzc{J} }^\alpha (0) \big] |   \psi_0  \rangle   \right]    \; ,	\nonumber\\
\label{eq:kubo.formula}
\end{eqnarray}
where $A$ is the area of the dot.

%%%%%%%%%%%%%%%%%%%%%%%%%%%%%%%%%%%%%%%%%%%%%%%%%%

Up to this point we have not placed any restriction 
on the model used to describe the GQD.
Quite generally, we can consider this to be described by any 
Hamiltonian $\hat{\mathpzc{H}}$, which respects the symmetries 
of a given dot.
The many--electron eigenstates of this Hamiltonian 
\begin{eqnarray}
	  \hat{\mathpzc{H}}  | \Psi_n \rangle 
	  &= E_n  | \Psi_n \rangle   \; , 
	  \label{eq:eigenstates.H} 
\end{eqnarray}
will also, automatically, respect the symmetries of the dot, 
since all symmetry operators must commute with $\hat{\mathpzc{H}}$.   
And we can use the completeness of these eigenstates
to express Eq.~(\ref{eq:kubo.formula}) as 
\begin{eqnarray}
	\text{Re} \big[ \sigma_{\alpha} (\omega) \big] = \lim_{\gamma \to 0} \  
	\frac{2 \gamma \hbar}{A} \frac{1}{ \hbar \omega} 
	\sum_{n} 
	 | \mathpzc{J}^\alpha_{n0} |^2 \delta(\hbar \omega - E_{n0}) 
\nonumber\\
\label{eq:general.result.sigma}
\end{eqnarray}
where
\begin{eqnarray}
\mathpzc{J}^\alpha_{nm} = \langle \Psi_n   \mid  \hat{\mathpzc{J}}^\alpha \mid  \Psi_m \rangle \; ,
\label{eq:many.electron.matrix.element}
\end{eqnarray}
are the matrix elements of the 
current operator $ \hat{\mathpzc{J}}$ from one many--body 
state to another, and 
\begin{eqnarray}
E_{nm} =  E_n - E_m 
\end{eqnarray}
is the energy of the associated optical transition.

%We note that 
%%
%\begin{eqnarray}
%\lim_{\gamma \to 0} \frac{\gamma}{\pi} \frac{1}{(\hbar\omega - E)^2 + \gamma^2}
%   = \delta(\hbar\omega -E)
%\end{eqnarray}
%%
%is a representation of a delta function.

%%%%%%%%%%%%%%%%%%%%%%%%%%%%%%%%%%%%%%%%%%%%%%%%%%

It follows from Eq.~(\ref{eq:general.result.sigma}) that the details of the 
optical response of a given GQD will depend on the precise form of the 
current operator $\hat{\mathpzc{J}}^\alpha$, and on the detailed form 
of the many--electron eigenstates $| \Psi_n \rangle $, which can 
generally only be found approximately.
None the less, it is possible to make some statements 
about $\sigma_{\alpha} (\omega)$ 
without seeking a full solution for $| \Psi_n \rangle $.
In particular, the optical selection rules which govern the polarisation--dependence
of $\sigma_{\alpha} (\omega)$ depend only on the way in which the current operator 
 $\hat{\mathpzc{J}}^\alpha$ , and the many--electron 
eigenstates $| \Psi_n \rangle$, transform under the symmetries of the dot 
[\onlinecite{Weyl, Heine, LandauLifshitz, Tinkham, Wagner,Jones}].
Simply put, the matrix element 
\begin{eqnarray}
\mathpzc{J}^\alpha_{nm} = \langle  \Psi \mid \Psi^\prime \rangle   \; ,
\end{eqnarray}
[cf. Eq.~(\ref{eq:many.electron.matrix.element})] must vanish if the ket 
\begin{eqnarray}
\mid \Psi^\prime \rangle = \hat{\mathpzc{J}}^\alpha \mid  \Psi_m \rangle \; , 
\end{eqnarray}
has a different symmetry from the bra 
\begin{eqnarray}
\langle  \Psi \mid = \langle  \Psi_n \mid \; .
\end{eqnarray}
It follows that identifying allowed optical transitions reduces 
to an exercise in symmetry analysis for the state 
$\mid \Psi^\prime \rangle$, created by the action 
of the polar vector $\hat{\mathpzc{J}}^\alpha$, 
on an eigenstate $ \mid  \Psi_m \rangle$, within a GQD of given shape.
This is pursued in Section~\ref{sec:Geometry}, for the different 
GQD's illustrated in Fig.~\ref{fig:lattice}.

%%%%%%%%%%%%%%%%%%%%%%%%%%%%%%%%%%%%%%%%%%%%%%%%%%

%\textcolor{blue}{In exploring how this works}, it will prove convenient to diagonalise $\hat{\mathpzc{H}}$
%[cf. Eq.~(\ref{eq:eigenstates.H})], in a basis of eigenstates $| \Psi_n  \rangle$ which 
%are simultaneous eigenstates of a given symmetry operation $\hat{\mathpzc{M}}_\lambda$, i.e.  
%%
%\begin{align}
%	  \hat{\mathpzc{M}}_\lambda | \Psi_n  \rangle &= \mu^\lambda_n | \Psi_n  \rangle  \; ,
%\end{align}
%%
%where $\mu^\lambda_n$ is the corresponding eigenvalue.  
%%
%This is always possible, since for any valid model of the GQD, whether interacting
%or non--interacting,
%%
%\begin{align}
%	 [ \hat{\mathpzc{H}}, \hat{\mathpzc{M}_\lambda} ] = 0
%\end{align}
%%
%We return to this point in Section~\ref{sec:Geometry}.
%%, where we explicitly 
%%determine the optical selection rules which follow for a given symmetry of dot.

%%%%%%%%%%%%%%%%%%%%%%%%%%%%%%%%%%%%%%%%%
\subsection{Optical Conductivity of a GQD within the tight--binding model}
%%%%%%%%%%%%%%%%%%%%%%%%%%%%%%%%%%%%%%%%%
\label{sec:tight-binding.model}

Many-body interactions can have significant effects in 
the electronic properties of graphene \cite{Kotov2012}, 
and will also have impact on the optical spectra of GQD's.
None the less, as discussed in Section~\ref{sec:general.theory}, above, 
optical selection rules are a special case, determined by 
symmetry alone. 
For this reason it is instructive to calculate the optical conductivity 
$\sigma_\alpha (\omega)$ explicitly for a non--interacting GQD, 
where optical transitions can be discussed in terms of the 
energy levels for individual electrons.  
Doing so also gives us a handle on the evolution of optical 
conductivity with the size of the GQD, and enables us to 
approach questions which are not determined by symmetry alone.
The framework needed to do this is introduced below.

%%%%%%%%%%%%%%%%%%%%%%%%%%%%%%%%%%%%%%%%%
%\subsection{\textcolor{red}{Tight-binding model}}
%%%%%%%%%%%%%%%%%%%%%%%%%%%%%%%%%%%%%%%%%

A good starting point to understand many of the electronic properties 
of graphene is the simple tight--binding model 
\cite{Wallace1947, CastroNeto2009, Kotov2012}: 
\begin{equation}
	\hat{\mathpzc{H}}_0 = -t \sum_{\langle i j \rangle, s} 
	\big( \hat{\mathpzc{a}}^{\dagger}_{i,s} \hat{\mathpzc{b}}_{j,s} 
	+  \hat{\mathpzc{b}}^{\dagger}_{j,s} \hat{\mathpzc{a}}_{i,s}  \big)  \; ,
\label{eq:H0} 	
\end{equation}
where $t$ is a hopping parameter and the sum on $\langle i j \rangle$
runs over all nearest--neighbour bonds on the honeycomb lattice.
The operators $\hat{\mathpzc{a}}^{\dagger}_{i, s}$ ($\hat{\mathpzc{b}}^{\dagger}_{j, s}$) 
and $\hat{\mathpzc{a}}_{i, s}$ ($\hat{\mathpzc{b}}_{j, s}$) respectively create and 
annihilate electrons with spin $s = \uparrow, \downarrow$ at site $i$ ($j$) 
of sub--lattice $\mathpzc{a}$ ($\mathpzc{b}$).
To describe a finite--sized system such as a GQD (Fig.~\ref{fig:lattice}), 
we assume zero hopping beyond the edges of the dot (open boundaries), 
which in real systems may be realized by passivating dangling bonds 
with hydrogen atoms.

%%%%%%%%%%%%%%%%%%%%%%%%%%%%%%%%%%%%%%%%%

% Here we restrict our selves to a non-interacting tight-binding model, which 
% explicitly ignores coulomb-interactions.  
% %
% This approach allows us to study clusters of graphene with open boundary 
% conditions in order to simulate Graphene Quantum Dots (GQDs) with passivated 
% dangling bonds of sizes large enough to be comparable to a system in its 
% thermodynamic limit.  
% Symmetry properties of GQDs will not be affected by the choice of the model.  
% %
% Nevertheless we would expect an overall shift in energies, causing a redshifted 
% peak in the optical absorption spectrum after including electron-hole interactions, 
% as already shown for graphene \cite{Yang2009}.

%%%%%%%%%%%%%%%%%%%%%%%%%%%%%%%%%%%%%%%%%

In the case of the tight--binding model  $\hat{\mathpzc{H}}_0$ 
[Eq.~(\ref{eq:H0})], it is sufficient to consider single--electron 
eigenstates $ | \psi_{n, s} \rangle $, with energy $\epsilon_n$, 
satisfying
\begin{align}
	  \hat{\mathpzc{H}}_0  | \psi_{n, s} \rangle 
	  &= \epsilon_n  | \psi_{n, s} \rangle   \; . 
	  %\\
%	  \hat{\mathpzc{M}} | \psi_{n, s}  \rangle &= \mu_n | \psi_{n, s}  \rangle \; .
\label{eq:eigenstates.H0} 
\end{align}
These single--electron eigenstates can in turn be written as
\begin{eqnarray}
	  | \psi_{n, s} \rangle &=& \sum_i c_{i, n, s} | \phi_{i, s} \rangle  \; , \label{eq:EFHam}
	  \label{eq:single.electron.eigenstates}
\end{eqnarray}
where the coefficients
\begin{equation}
	c_{i, n, s}  = \langle \psi_{n, s}  |  \phi_{i, s} \rangle   	  \label{eq:coefficients}
\end{equation}
can be expressed in terms of the local atomic states through 
\begin{equation}
	 \langle{\vec{r}} \ | \ \phi_{i, s } \rangle = \phi_{i, s}(\vec{r})  = w(\vec{r} - \vec{R}_i) \chi_{s}     \; ,
\label{eq:Wannier}
\end{equation}
a Wannier orbital $w(\vec{r} - \vec{R}_i)$ with  
\begin{equation}
	\chi_{\uparrow} = \left( \begin{array}{c} 1 \\ 0 \end{array} \right) \; ,  \; 
	\chi_{\downarrow} = \left( \begin{array}{c} 0 \\ 1 \end{array} \right) 
\end{equation}
spinors representing the electron's spin degree of freedom.

%%%%%%%%%%%%%%%%%%%%%%%%%%%%%%%%%%%%%%%%%

The most commonly--quoted measure of the single--electron properties 
is the density of states (DOS), given here by 
\begin{equation}
	g(\epsilon)
	= \lim_{\gamma \to 0} \frac{\gamma}{\pi} \sum_{n=1}^N \frac{1}{(\epsilon - \epsilon_n )^2 + \gamma^2}
	= \sum_{n=1}^N \delta(\epsilon - \epsilon_n)  
	\; ,
\label{eq:DOS}
\end{equation}
where the sum $\sum_{n=1}^N$ runs over all possible single--electron
eigenstates of a dot with $N$ sites.
For purposes of plotting the DOS, it is convenient to 
work with a finite value of $\gamma$. 
This is equivalent to convoluting the DOS with a normalised Lorentzian, 
of full--width at half--maximum $2\gamma$.

%%%%%%%%%%%%%%%%%%%%%%%%%%%%%%%%%%%%%%%%%%
%\subsection{Optical Conductivity of a GQD within tight--binding model}
%%%%%%%%%%%%%%%%%%%%%%%%%%%%%%%%%%%%%%%%%%

It is also possible to calculate the optical conductivity within the tight--binding 
model $\hat{\mathpzc{H}}_0$ [Eq.~(\ref{eq:H0})], starting from the Kubo 
formula, Eq.~(\ref{eq:kubo.formula}).
An essential ingredient for this analysis is the correct form 
of the current operator $\hat{\mathpzc{J}}^\alpha$.
Since electrical charge is a conserved quantity, 
the form of the current operator is determined
by the equation of continuity for tight--binding electrons on a lattice.
And the equation of continuity, in turn, is 
determined by the structure of $\hat{\mathpzc{H}}_0$.
It follows that the correct form of the current operator is given 
by~\cite{Cabib1973, Kuzemsky2011} 
\begin{equation}	
	\hat{\mathpzc{J}}^\alpha = -i t \Big( \frac{e}{\hbar} \Big) \sum_{ij, s} 
	\big(R^\alpha_i - R^\alpha_j \big) 
	\big( \hat{\mathpzc{a}}^{\dagger}_{i, s} \hat{\mathpzc{b}}_{j, s}  
	- \hat{\mathpzc{b}}^{\dagger}_{j, s} \hat{\mathpzc{a}}_{i, s}\big )  \; ,	
\label{eq:currentRealSpace}	
\end{equation}
where $R^\alpha_i$ is the $\alpha$--component of the 
position vector to the lattice site~$i$.
We note that making an inconsistent choice of the current operator 
can lead to incorrect values of the optical conductivity, and in particular a deviation 
from the universal value in Eq.~(\ref{eq:sigma0}).

%%%%%%%%%%%%%%%%%%%%%%%%%%%%%%%%%%%%%%%%%

For purposes of calculation, we will use Eq.~(\ref{eq:coefficients}) to express 
the current operator $\hat{\mathpzc{J}}^\alpha$ in terms of the 
basis $|  \phi_{i, s} \rangle $ 
\begin{equation}
	 \mathpzc{J}^\alpha_{nm} = - i t \Big( \frac{e}{\hbar} \Big) 
	\sum_{ij} c^*_{ni} c_{mj}  (R^\alpha_i - R^\alpha_j) \; ,
\label{eq:single.electron.matrix.element}
\end{equation}
where, for brevity, we drop spin indices and write
\begin{equation}
c_{i n}  = c_{i n, s} \; .
\end{equation}
It follows from the general result, Eq.~(\ref{eq:general.result.sigma}), 
that the real part of the optical conductivity is given by
\begin{equation}
	\text{Re} \big[ \sigma_{\alpha} (\omega) \big] = \lim_{\gamma \to 0} \  
	\frac{2 \gamma}{\omega A}  \sum_{nm} | \mathpzc{J}^\alpha_{mn}|^2 
	\frac{ f(\epsilon_m) - f(\epsilon_n) }{(\hbar \omega - \epsilon_{nm}) ^ 2 + \gamma^2}   \; ,
\label{eq:sigma.tight.binding}
\end{equation}
where 
\begin{equation}
\epsilon_{nm} =  \epsilon_n - \epsilon_m  \; ,
\end{equation}
%
%is the difference in energy of states of an individual electron, 
and the Fermi function $f(\epsilon_n)$ should be evaluated at $T=0$.
We note that the numerical results for $\sigma_{\alpha}(\omega)$ presented 
in this Article have been calculated for a small but finite value of $\gamma$, 
mimicking the finite lifetime of electronic states.  
\\

%%%%%%%%%%%%%%%%%%%%%%%%%%%%%%%%%%%%%%%%%

%So far as optical selection rules are concerned, all of the discussion of 
%Section~\ref{sec:general.theory} holds, with the considerable simplification 
%that we now need only consider single--electron matrix element of the current 
%operator in Eq.~(\ref{eq:general.result.sigma}).
%%%
%And, since any valid tight--binding model {\it must} commute with all symmetries 
%of dot, we can always chose the single--electron basis $ | \psi_{n, s}  \rangle$ 
%in Eq.~(\ref{eq:single.electron.eigenstates}), such that it also respects a given 
%symmetry operation, i.e.
%%
%\begin{align}
%	  \hat{\mathpzc{M}} | \psi_{n, s}  \rangle &= \mu_n | \psi_{n, s}  \rangle \; .
%\end{align}
%%
%For a given dot, and symmetry operation $\hat{\mathpzc{M}}$, the 
%energy eigenvalues $\epsilon_n$, and the corresponding coefficients $c_{i, n, s}$, 
%can be found numerically.
%%
%This makes it possible both to calculate the optical conductivity $\sigma_{\alpha}$, 
%and to classify the matrix--elements involved according to the symmetries 
%of relevant single--electron states.
%%
%We return to this point in Section~\ref{sec:SymTri}, below.

%%%%%%%%%%%%%%%%%%%%%%%%%%%%%%%%%%%%%%%%%
\subsection{Optical conductivity of an infinite graphene sheet 
                    within the tight--binding model}
%%%%%%%%%%%%%%%%%%%%%%%%%%%%%%%%%%%%%%%%%

For many purposes, it is also interesting to compare the optical response of a 
GQD with that of an infinite graphene sheet.
Within the tight--binding model, we can evaluate this by imposing periodic boundary 
conditions on the lattice, and consider its thermodynamic limit with $N \to \infty$.   
The result follows from Eq.~(\ref{eq:sigma.tight.binding})~: for $T =  0$, 
in the limit $q \to 0$, transitions are only allowed between states with the 
same crystal momentum and we find, consistent with the 
literature\cite{Buividovich2012}, 
\begin{widetext}
\begin{equation}
	\text{Re} \big[ \sigma_{\alpha} (\omega) \big] 
	= \lim_{\gamma \to 0}  \ \frac{1}{3 \sqrt{3}a^2} \frac{t^2 e^2}{\hbar^2 }  \frac{\gamma}{\omega} \frac{1}{N}   
	\sum^{N}_{k} \frac{| \big( \vec\mu_k \big)_{\alpha} \nu_k^* 
	+ \big( \vec\mu_k^* \big)_{\alpha} \nu_k |^2}{|\nu_k|^2}    \\
	 \Bigg[ \frac{1}{(\hbar \omega - 2t |\nu_k|)^2 
	 + \gamma^2} - \frac{1}{(\hbar \omega 
	 + 2t |\nu_k|)^2 + \gamma^2} \Bigg]   \ ,
\label{eq:conductivityFinal_energySpace}
\end{equation}
\end{widetext}
where the $\sum_k$ runs over all $k$--values in the Brillouin zone, 
\begin{align}
	\nu_k = \sum\limits_{j=1}^{3}e^{i \vec{k}\vec{\delta}_j}    \quad , \quad
	\vec{\mu}_k = \sum\limits_{j=1}^{3}  \vec{\delta}_{j} e^{i \vec{k} \vec{\delta}_j}  \ ,
\end{align}
and the lattice vectors $\vec{\delta}_j$ are given by
\begin{equation}
\vec{\delta}_1 =  \frac{a}{2} \{-\sqrt{3}, -1 \} \; , \;
\vec{\delta}_2 =  \frac{a}{2} \{\sqrt{3}, -1 \} \; , \;
\vec{\delta}_3 =  a \big\{0, 1 \}  \; , \;  \nonumber
\end{equation}
where $a$ is the 
%\textcolor{red}{honeycomb} 
lattice constant. 
We return to this result in Fig.~\ref{fig:SizeEvo} of Sec.~\ref{sec:Edges}. 

%\begin{align}
%\vec{\delta}_1 &=  \frac{a}{2} \{-\sqrt{3}, -1 \} \; ,\nonumber\\
%\vec{\delta}_2 &=  \frac{a}{2} \{\sqrt{3}, -1 \} \; ,\\ 
%\vec{\delta}_3 &=  a \big\{0, 1 \} \; .\nonumber
%\end{align}

%%%%%%%%%%%%%%%%%%%%%%%%%%%%%%%%%%%%%%%%%
\section{The role of  symmetry }   
%%%%%%%%%%%%%%%%%%%%%%%%%%%%%%%%%%%%%%%%%
\label{sec:Geometry}

Graphene quantum dots (GQD's) with regular shapes can be 
considered as macroscopic molecules, classified by their point--group symmetry. 
And, as with conventional molecules, different point groups lead to different optical 
selection rules, and therefore to different optical properties 
[cf. Section~\ref{sec:general.theory}].
In the following Section we use representation theory to derive the optical selection 
rules associated with GQD's of different symmetry.
This is a standard application of group theory, and our derivation 
closely parallels text--book treatments of optical selection rules in 
atoms~\cite{Weyl, Heine, LandauLifshitz, Wagner, Jones} and molecules ~\cite{Tinkham}.
We consider GQD's of the type shown in Fig.~\ref{fig:lattice}, because of their
experimental availability and prominence in the existing literature 
\cite{Hamalainen2011, Lu2011, Olle2012, Yan2012, Mohanty2012}. 

%%%%%%%%%%%%%%%%%%%%%%%%%%%%%%%%%%%%%

Our main finding relates to the polarisation--dependence of optical spectra 
for dots of different geometry, and can be summarised as follows. 
The GQD's shown in Fig.~\ref{fig:lattice} can be divided into two groups.       
Dots (a)--(d) are described by the point groups $C_{3v}$ (triangular dots) 
and $C_{6v}$ (hexagonal dots).
In both of these cases, the in--plane components of the current operator,  
$\hat{\mathpzc{J}}^x$ and $\hat{\mathpzc{J}}^y$, transform with the 
same irrep of the point group \cite{Weyl,Heine,LandauLifshitz,Tinkham,Wagner,Jones}.
It follows that the optical conductivity is independent of polarization of the incident light. 
%
%In this case, we find that the optical conductivity is 
%{\it independent} of the polarization of the incident light.  
%
Meanwhile, the rectangular dot, (e), is described by the point group 
$C_{2v}$.
In this case, $\hat{\mathpzc{J}}^x$ and $\hat{\mathpzc{J}}^y$ transform with
different irreps, and the optical conductivity depends on the polarisation
of the applied light.
In what follows, we show explicitly how this result can be obtained 
for GQD's with triangular and rectangular symmetry.
The case of hexagonal dots is discussed in Appendix~\ref{appendix:SymHex}.

%%%%%%%%%%%%%%%%%%%%%%%%%%%%%%%%%%%%%

In order to make these results concrete, we also calculate the optical 
conductivity explicitly within the minimal tight--binding 
model introduced in Sec.~\ref{sec:tight-binding.model}.
In this case, because the electrons are non--interacting, 
it is possible to relate selection rules explicitly to transitions 
between different single--electron states, of a given symmetry.
Optical spectra calculated within a tight--binding model 
must be approached with some caution, as they cannot hope  
to capture every feature of a GQD with interacting electrons.  
None the less, the tight--binding model remains a valid point 
of comparison for optical selection rules, since these are  
determined by symmetry alone [Section~\ref{sec:general.theory}], 
and interactions are not expected to lead to spontaneous 
changes of symmetry in a finite--size dot \cite{plischke-book,huang-book}.

%%%%%%%%%%%%%%%%%%%%%%%%%%%%%%%%%%%%
\subsection{Optical selection rules for triangular graphene quantum dots}  		 
%%%%%%%%%%%%%%%%%%%%%%%%%%%%%%%%%%%%%%%%%%
\label{sec:SymTri}

%%%%%%%%%%%%%%%%%%%%%%%%%%%%%%%%%%%%
\subsubsection{Group--Theory Analysis}  		 
%%%%%%%%%%%%%%%%%%%%%%%%%%%%%%%%%%%%%%%%%%
\label{sec:group.theory.triangular.GQD}

As discussed in Section~\ref{sec:general.theory}, the optical selection rules for 
a given GQD follow from the matrix elements of the current operator 
\begin{eqnarray}
\langle \Psi_n   \mid  \hat{\mathpzc{J}}^\alpha \mid  \Psi_m \rangle \; , \nonumber
\end{eqnarray}
[cf. Eq.~(\ref{eq:many.electron.matrix.element})].  
We therefore need to understand how both, the many--body eigenstates 
$\mid  \Psi_m \rangle$ and $\mid  \Psi_n \rangle$, and the current operator 
$\hat{\mathpzc{J}}^\alpha$, transform under the symmetries of the dot.

%%%%%%%%%%%%%%%%%%%%%%%%%%%%%%%%%%%%%%%%%%
%  Table I - character table of C_{3v}
%%%%%%%%%%%%%%%%%%%%%%%%%%%%%%%%%%%%%%%%%%

\setlength{\tabcolsep}{0.5em} % for the horizontal padding
{\renewcommand{\arraystretch}{1.2}% for the vertical padding
\begin{table} [t]
\centering
    \begin{tabular}{ | c | c | c | c | c  | p{1cm} | }     
    \hline
    	$C_{3v}$ 	& $E$ 	& $2C_3$ & $3{\sigma}_v$ 	&	\text{polar vectors}	\\ \hline
    	$A_1$ 	& 1 		& 1 		& 1 				&	z 				\\ \hline
    	$A_2$ 	& 1 		& 1 		& -1 				&					\\ \hline
	$E$ 		& 2		& -1  	& 0 				&	(x, y)				\\ \hline
    \end{tabular}
    \caption{
    Character Table of the point--group $C_{3v}$ 
    describing the symmetry of triangular graphene quantum dots (GQD's)
    of the type shown in Fig.~\ref{fig:lattice}(a) and Fig.~\ref{fig:lattice}(b).
    Eigenstates of a triangular GQD can be classified according to the 
    irreducible representations (irreps) $A_1$, $A_2$ and $E$, 
     while the $x$ and $y$ components of the current 
     operator $\hat{\mathpzc{J}}$ (a polar vector) transform with $E$  
    [cf. \onlinecite{Weyl,Heine,LandauLifshitz,Tinkham,Wagner,Jones}].
    The corresponding symmetry operations are the identity ($E$), $2 \times \frac{2 \pi}{3}$ 
    rotations in the principle axis ($2C_3$) and 3 reflections on symmetry axes ($3{\sigma}_v$),
    as shown in Fig.~\ref{fig:lattice}(a) and (b).
    }	
    \label{tab:CharacterTableTriangle}
\end{table}

%%%%%%%%%%%%%%%%%%%%%%%%%%%%%%%%%%%%%%%%%%

GQD's with triangular geometry have reflection symmetry about three 
different axes in the plane, as well as a three--fold rotation symmetry 
about the center of the dot, as illustrated in Fig.~\ref{fig:lattice}(a) 
and Fig.~\ref{fig:lattice}(b).   
The corresponding symmetry operations can be labelled 
$E$ (identity), $C_3$ (rotation) and ${\sigma}_v$ (reflection). 
These symmetry operations form the group $C_{3v}$
\cite{Weyl,Heine,LandauLifshitz,Tinkham,Wagner,Jones}, with the 
caveat that reflection and rotation operators do not commute.   
This group has two one--dimensional irreducible representations (irreps), 
$A_1$ and $A_2$, and one two--dimensional irrep, $E$, 
as listed in Table~\ref{tab:CharacterTableTriangle}.

%%%%%%%%%%%%%%%%%%%%%%%%%%%%%%%%%%%%%%%%%%

It follows from fundamental theorems of quantum mechanics 
\cite{Weyl,Heine,LandauLifshitz,Tinkham,Wagner,Jones}, 
that all possible many--body eigenstates of a triangular GQD,  
$\mid  \Psi_n \rangle$, as well as the current operator 
$\hat{\mathpzc{J}}$, can be classified in terms 
of the irreps of $C_{3v}$, and must transform like these irreps   
under the symmetry operations of the dot.  
Moreover, eigenstates transforming with different irreps must
be orthogonal, i.e. for eigenstates associated with the irreps 
$\lambda, \lambda^\prime$  
\begin{eqnarray}
\langle \Psi_{n \in \lambda} \mid  \Psi_{n \in \lambda^\prime} \rangle = 0 
   \; , \; \text{if} \;
   \lambda \ne \lambda^\prime \; .
\label{eq:orthogonality.of.irreps}
\end{eqnarray}
In the case of $\mid  \Psi_n \rangle$, the irrep depends on which 
eigenstate is considered.
However the symmetry properties of $\hat{\mathpzc{J}}$ are 
fully determined by the fact that it is a polar vector, confined to
the $(x,y)$ plane.
And within $C_{3v}$, all polar vectors transform with the $E$ irrep 
[\onlinecite{Weyl,Heine,LandauLifshitz,Tinkham,Wagner,Jones}] 
--- cf. Table~\ref{tab:CharacterTableTriangle}.

%%%%%%%%%%%%%%%%%%%%%%%%%%%%%%%%%%%%%%%%%%

From this starting point, the determination of optical selection rules 
is a standard application of the theory of represenations 
\cite{Weyl,Heine,LandauLifshitz,Tinkham,Wagner,Jones}.  
The matrix element  
$\langle \Psi_n   \mid  \hat{\mathpzc{J}}^\alpha \mid  \Psi_m \rangle$ 
will be zero unless the final state, $\langle \Psi_n \mid$, has the same 
symmetry as (some component of) the state 
$\hat{\mathpzc{J}} \mid  \Psi_m \rangle$.   
This in turn will transform as a direct product of the irrep $E$ 
associated with the current operator $\hat{\mathpzc{J}}$, 
and the irrep associated with the initial state $\mid  \Psi_m \rangle$.
Such direct products are in general reducible (in the
group--theoretical sense), and can be decomposed in terms of
the irreps of $C_{3v}$, using the information in 
Table~\ref{tab:CharacterTableTriangle}, and the fact
that the character of a product representation is given 
by the product of the characters of its components 
\cite{Weyl,Heine,LandauLifshitz,Tinkham,Wagner,Jones}.

%%%%%%%%%%%%%%%%%%%%%%%%%%%%%%%%%%%%%%%%%%

We consider as an example an initial state $\mid  \Psi_m \rangle$ 
transforming with the irrep $A_1$. 
In this case the direct product associated with 
the matrix element $\mathpzc{J}^\alpha_{nm}$ 
can be broken down as follows
\begin{eqnarray}
	\mathpzc{J}^\alpha_{nm} 	
	=  \langle \Psi_n \mid \underbrace{\hat{\mathpzc{J}}^\alpha \mid  \Psi_{m \in A_1} \rangle}_{E \times A_1 = E}
	= \langle \Psi_n \underbrace{\mid \Psi^\prime \rangle}_{E} 
\end{eqnarray}
It follows from Eq.~(\ref{eq:orthogonality.of.irreps}) that 
contributions to the optical conductivity will vanish 
unless the final state $\langle \Psi_n \mid$ transforms with the irrep $E$.
This implies that the allowed transitions $m \to n$ must 
satisfy the selection rule 
\begin{equation}
	A_1 	\rightarrow E \; . 	
\end{equation}
In the same spirit, we can evaluate the direct product for eignstates
transforming with the other irreps of $C_{3v}$.
For an initial state with symmetry $A_2$, we have
\begin{eqnarray}
	\mathpzc{J}^\alpha_{nm} 	 
	&=&  \langle \Psi_n \mid \underbrace{\hat{\mathpzc{J}}^\alpha \mid  \Psi_{m \in A_2} \rangle}_{E \times A_2 = E} 
	=  \langle \Psi_n \underbrace{\mid \Psi^\prime \rangle}_{E} \; ,
\end{eqnarray}	
with associated selection rule 
\begin{eqnarray}
   A_2 \rightarrow E \; .
\end{eqnarray}
Meanwhile, for an initial state with symmetry $E$, we have
\begin{eqnarray}
	 \mathpzc{J}^\alpha_{nm} 		 
	&=&  \langle \Psi_n \mid \underbrace{\hat{\mathpzc{J}}^\alpha \mid  \Psi_{m \in E} \rangle}_{E \times E = A_1 + A_2 + E} 	
	=  \langle \Psi_n \underbrace{\mid \Psi^\prime \rangle}_{A_1 + A_2 + E} \; 
\end{eqnarray}
with associated selection rule 
\begin{eqnarray}
	E \rightarrow \{A_1, A_2, E \}  \; .
\end{eqnarray}
It follows that the complete list of allowed optical transitions, taking into 
account all possible starting states, reads 
\begin{equation}
	A_1 	\longleftrightarrow E \; ;	\quad \quad 	
	A_2 	\longleftrightarrow E \; ;	\quad \quad 
	E	\longleftrightarrow E \;,
\label{eq:SelRulesTri} 
\end{equation}
regardless of the details of the Hamiltonian describing the dot.

%%%%%%%%%%%%%%%%%%%%%%%%%%%%%%%%%%%%%%%%%%

In the analysis above, we did not need to specify which component 
of the current operator $\hat{\mathpzc{J}}^\alpha$ we had in mind, 
since for a GQD with point--group symmetry $C_{3v}$, both 
transform with $E$ [cf. Table~\ref{tab:CharacterTableTriangle}].  
And our main result, about the polarisation--independence of the 
optical conductivity, follows directly from this fact.
We note that the optical selection rules which we have derived here
for a GQD apply equally to any other system with point--group $C_{3v}$, 
regardless of the underlying model.
This means that they will also hold for other two--dimensional systems with 
the same symmetry, such as PbSe nanocrystal quantum dots \cite{Goupalov2009} 
and defect--based ZnO \cite{Jungwirth2016} --- 
even in the presence of electron--electron interactions.

%%%%%%%%%%%%%%%%%%%%%%%%%%%%%%%%%%%%%%%%%%%%%%%%%%
%   Fig. 2 & Fig. 3 - triangular GQD
%%%%%%%%%%%%%%%%%%%%%%%%%%%%%%%%%%%%%%%%%%%%%%%%%%

\begin{figure*}    
	\centering 
	\begin{minipage}[t]{.47\textwidth}  
		\centering 
		\includegraphics[width=\textwidth]{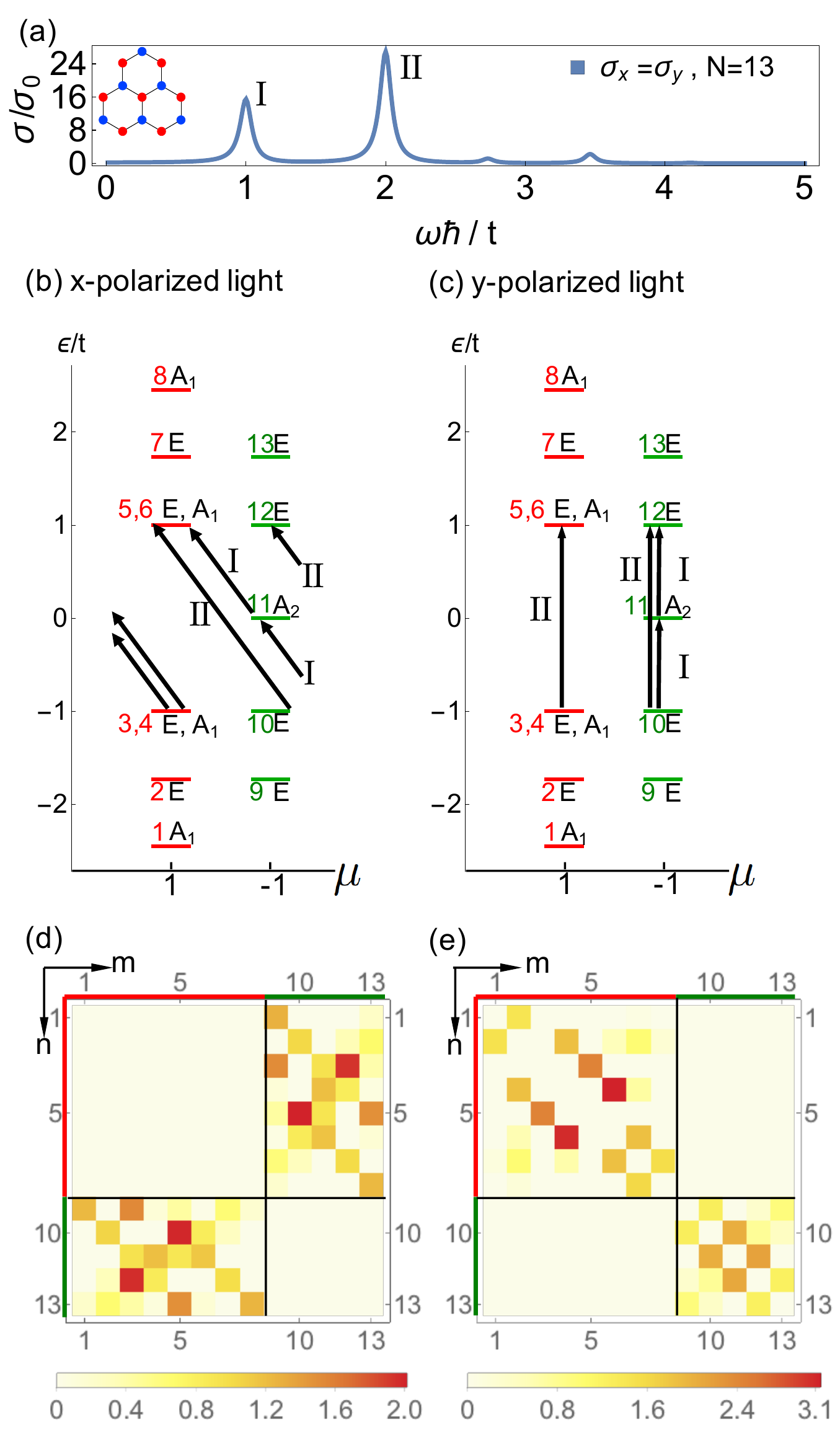}
		\caption{
		Optical selection rules for the smallest possible triangular graphene quantum dot (GQD) 
		with zigzag edges ($N=13$), in linearly--polarized light.
      	  	(a) Optical conductivity $\sigma_{\alpha}(\omega)$, showing the equivalence of results for 
       	 	$x$-- and $y$--polarized light.    
		(b) and (c) Spectrum of the corresponding tight--binding model [Eq.~(\ref{eq:H0})], 
		in the mirror basis Eq.~(\ref{eq:mirror}), showing the different 
		allowed transitions for $x$-- and $y$--polarized light.  
         	(d) and (e) Matrix elements of the corresponding current operators 
      		$|\hat{\mathpzc{J}}_{nm}^{x}|^2$ and  $|\hat{\mathpzc{J}}_{nm}^{y}|^2$ 
		[cf.~Eq.~(\ref{eq:single.electron.matrix.element})], 
        		in units of $(e t / \hbar)^2$.  
          	Results for $\sigma_{\alpha}(\omega)$ were calculated from Eq.~(\ref{eq:sigma.tight.binding}), 
         	with a Lorentzian of FWHM $2\gamma = 0.1\ \text{t}$.
          	Eigenstates are labelled according to their quantum number $n=1\ldots 13$, 
         	eigenvalue $\mu_n = \pm 1$ [Eq.~(\ref{eq:mirror})] and corresponding irrep 
        	  	[cf. Table~\ref{tab:CharacterTableTriangle}].  
		 }
		\label{fig:SymTri_Mir}
	\end{minipage}
	\quad \quad
	\begin{minipage}[t]{.47\textwidth} 
		\centering 
		\includegraphics[width=\textwidth]{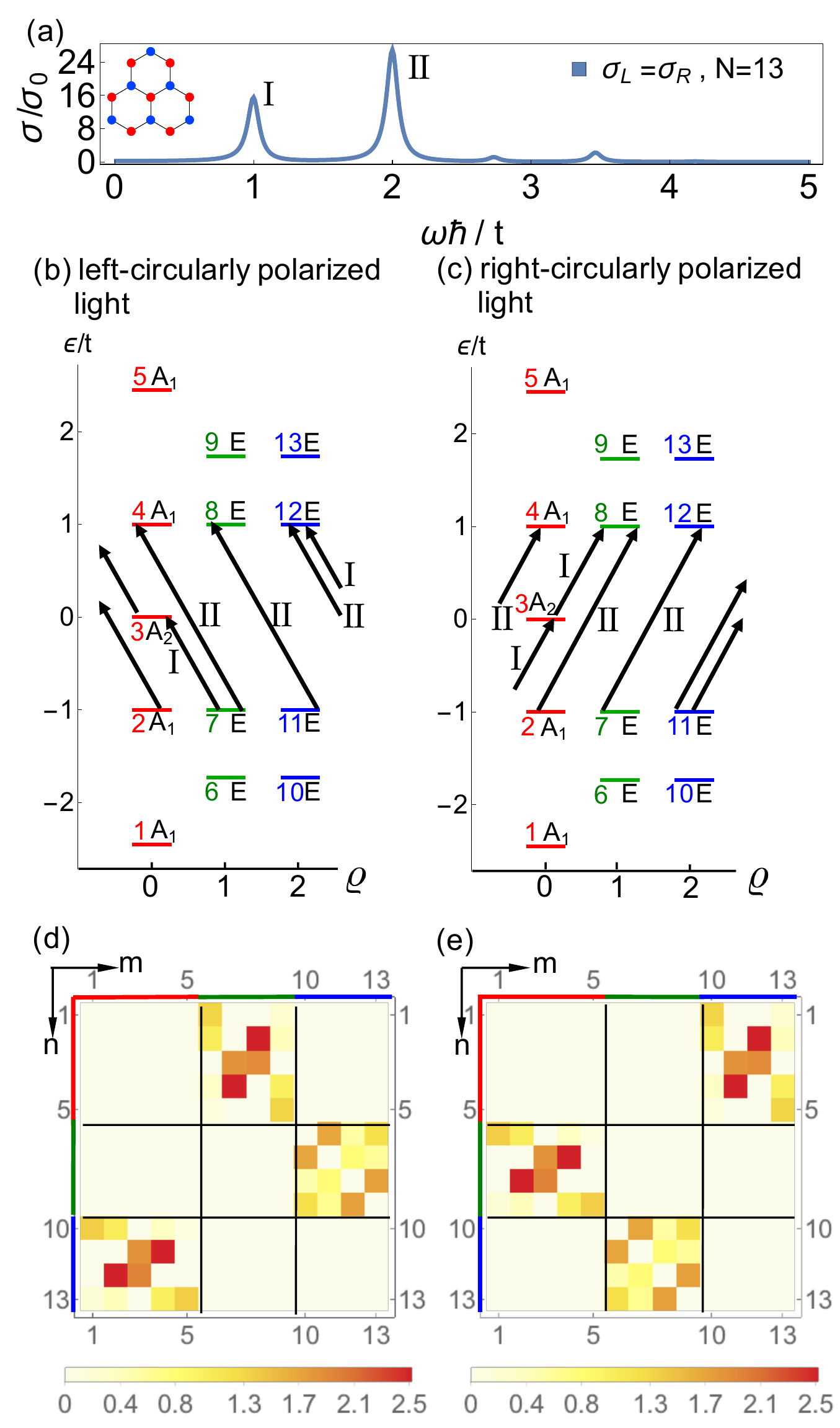}
		\caption{
		Optical selection rules for the smallest possible triangular graphene quantum dot (GQD) 
		with zigzag edges ($N=13$), in circularly--polarized light.
		(a) Optical conductivity $\sigma_{\alpha}(\omega)$, showing equivalence of results for 
		left-- and right--circularly polarized light.
		(b) and (c) Spectrum of the corresponding tight--binding model Eq.~(\ref{eq:H0}),
		in the rotation basis Eq.~(\ref{eq:RotationOperator}), showing the different allowed transitions
		for left-- and right--circularly polarized light. 
		(d) and (e) Matrix elements of the corresponding current operators 
		$|\hat{\mathpzc{J}}_{nm}^{L}|^2$ and $|\hat{\mathpzc{J}}_{nm}^{R}|^2$ 
		[cf.~Eq.~(\ref{eq:single.electron.matrix.element})], in units of $(e t / \hbar)^2$.
		Results for $\sigma_{\alpha}(\omega)$ were calculated from Eq.~(\ref{eq:sigma.tight.binding}), 
		with a Lorentzian of FWHM $2\gamma = 0.1\ \text{t}$.
		Eigenstates are labelled according to their quantum number $n=1\ldots 13$, 
     	    	eigenvalue $\mu_n  = e^{i\frac{2\pi}{3} \rho_n} \; , \; \rho_n = 0,1,2 $ 
		[Eq.~(\ref{eq:EV_RotationOperator})] and corresponding irrep 
		[cf. Table~\ref{tab:CharacterTableTriangle}]. 
      	   }
	\label{fig:SymTri_Rot}
	\end{minipage}		
\end{figure*}

%%%%%%%%%%%%%%%%%%%%%%%%%%%%%%%%%%%%%%%%%%
\subsubsection{Illustration of optical selection rules for linearly--polarised light}      
%%%%%%%%%%%%%%%%%%%%%%%%%%%%%%%%%%%%%%%%%%
\label{sec:mir}

We can make these results more tangible, and somewhat 
easier to compare with experiments, by examining how optical transitions occur in 
a specific microscopic model for a GQD with triangular symmetry.  
The simplest model, which we can consider is the tight--binding 
model $\hat{\mathpzc{H}}_0$ [Eq.~\ref{eq:H0}], as introduced in 
Section~\ref{sec:tight-binding.model}.
And in this case, since the system is non--interacting, we can 
analyze all optical transitions at the level of a single electron.

%%%%%%%%%%%%%%%%%%%%%%%%%%%%%%%%%%%%%%%%%

We first consider results for linearly--polarised light incident 
on a triangular GQD of the type shown in Fig.~\ref{fig:lattice}(a).   
The optical conductivity $\sigma_\alpha(\omega)$  
can be calculated using Eq.~(\ref{eq:sigma.tight.binding}).
Results for a triangular GQD containing 13 sites are 
shown in Fig.~\ref{fig:SymTri_Mir}(a); for purposes 
of visualisation, $\sigma_\alpha(\omega)$ has been 
convoluted with a Lorentzian 
of FWHM $2\gamma = 0.1\ \text{t}$.
For this (very) small GQD, the optical response is 
dominated by peaks at (I) \mbox{$\hbar \omega = t$} 
and (II) \mbox{$ \hbar \omega  = 2t$}. 
As already noted in Section~\ref{sec:group.theory.triangular.GQD}, 
these results are independent of whether the light is polarised along 
the x-- or y--axis.

%%%%%%%%%%%%%%%%%%%%%%%%%%%%%%%%%%%%%%%%%

In Fig.~\ref{fig:SymTri_Mir}(b) and Fig.~\ref{fig:SymTri_Mir}(c) 
we show the optical transitions associated with each of 
these peaks, for the two possible polarisations of light.   
Eigenstates %of $\hat{\mathpzc{H}}_0$ [Eq.~\ref{eq:H0}] 
have been labelled according to their irrep ($A_1$, $E$), 
and further classified according to their eigenvalue under 
the mirror--symmetry operation 
\begin{equation}   
	\hat{\mathpzc{M}_y} | \psi_n \rangle  
	= \mu_n | \psi_n \rangle, \quad \mu_n = 1, -1 	\; ,
	 \label{eq:mirror}	
\end{equation}
where $\hat{\mathpzc{M}_y}$ corresponds to a reflection 
along the vertical y--axis, as shown in Fig.~\ref{fig:lattice} (a).
Each of the peaks in Fig.~\ref{fig:SymTri_Mir}(a) correspond 
to transitions of a single electron, and it is immediately 
apparent that all of the associated transitions satisfy the 
selection rules in Eq.~(\ref{eq:SelRulesTri}).
In addition, the fact that $\hat{\mathpzc{J}}^x$ 
($\hat{\mathpzc{J}}^y$) is odd (even) under reflection 
$\mathpzc{M}_y$ imposes the additional condition 
$\mu_n \mu_m = -1$ ($\mu_n \mu_m = 1$)
on the transitions for $x$--polarised ($y$--polarised) light.
However, this additional condition cannot have any effect 
on the final result for $\sigma_\alpha(\omega)$
[Fig.~\ref{fig:SymTri_Mir}(a)], since both 
components of $\hat{\mathpzc{J}}^\alpha$ 
transform with the same irrep %of $C_{3v}$ 
[cf. Table~\ref{tab:CharacterTableTriangle}, Section~\ref{sec:group.theory.triangular.GQD}].

%%%%%%%%%%%%%%%%%%%%%%%%%%%%%%%%%%%%%%%%%

It is also informative to view these transitions in terms 
of the matrix elements of the current operator 
$\mathpzc{J}^\alpha_{nm}$, expressed now in terms 
of the single--election eigenstates of 
$\hat{\mathpzc{H}}_0$ [Eq.~\ref{eq:H0}]
--- cf. Eq.~(\ref{eq:single.electron.matrix.element}).
The norm of these matrix elements, 
$|\hat{\mathpzc{J}}_{nm}^\alpha|^2$ is plotted in 
Fig.~\ref{fig:SymTri_Mir}(d) and Fig.~\ref{fig:SymTri_Mir}(e).
Once again, states have been labelled according
to their eigenvalue $\mu_n$ under the reflection $\mathpzc{M}_y$.
The block off--diagonal form of Fig.~\ref{fig:SymTri_Mir}(d)
[block--diagonal form of Fig.~\ref{fig:SymTri_Mir}(e)] therefore 
reflects the condition $\mu_n \mu_m = -1$ ($\mu_n \mu_m = 1$).  
And, while the structure of these matrices looks very different, 
the final result for $\sigma_\alpha(\omega)$ is independent 
of polarisation, as it must be --- cf. Fig.~\ref{fig:SymTri_Mir}(a).

%%%%%%%%%%%%%%%%%%%%%%%%%%%%%%%%%%%%%%%%%

%\textcolor{red}{[\bf Rico:  would it make sense to replace 
%Fig.~\ref{fig:SymTri_Mir}(d) and Fig.~\ref{fig:SymTri_Mir}(e) 
%with plots where the eigenstates have also been classified according 
%to irreps of $C_{3v}$ ?]}

%%%%%%%%%%%%%%%%%%%%%%%%%%%%%%%%%%%%%%%%%%
\subsubsection{Illustration of optical selection rules for circularly--polarised light}      
%%%%%%%%%%%%%%%%%%%%%%%%%%%%%%%%%%%%%%%%%%
\label{sec:rot}

We can also use the example of a non--interacting tight--binding model 
[Section~\ref{sec:tight-binding.model}], to illustrate how the optical 
selection rules of a triangular GQD function for circularly--polarised light.  
The main result is easy to anticipate:
in the absence of magnetic field, circularly--polarised light can always 
be decomposed into linear components.
And since the optical response of a triangular GQD is independent of the 
linear polarisation of light, it will also be independent of the circular 
polarisation of light.

%%%%%%%%%%%%%%%%%%%%%%%%%%%%%%%%%%%%%%%%%%

In Fig.~\ref{fig:SymTri_Rot}(a) we show results for the optical conductivity 
of a triangular GQD with 13 sites, calculated using 
Eq.~(\ref{eq:sigma.tight.binding}), for circularly--polarised light.
As expected, the result is independent of the (circular) polarisation 
of the light, and is identical to that found for linearly--polarised light,  
Fig.~\ref{fig:SymTri_Mir}(a).  
The optical transitions associated with each of the features in 
$\sigma_\alpha(\omega)$ are set out in 
Fig.~\ref{fig:SymTri_Rot}(b) (left--polarised light) 
and Fig.~\ref{fig:SymTri_Rot}(c) (right--polarised light), 
where eigenstates have been labelled according to 
their irrep, and classified according to their eigenvalue 
under the rotation operator 
\begin{equation}			
	\hat{\mathpzc{R}}_{2\pi/3} | \psi_n \rangle 
	=   \mu_n  | \psi_n \rangle \; ,
 \label{eq:RotationOperator}
\end{equation}
where 
\begin{equation}			
	\mu_n  = e^{i\frac{2\pi}{3} \rho_n} \; , \; \rho_n = 0,1,2 \; .
 	\label{eq:EV_RotationOperator}
\end{equation}
Once again, all transitions satisfy the selection rules derived
from symmetry, Eq.~(\ref{eq:SelRulesTri}). 
The transitions for left--polarised light also satisfy the anti--cyclic condition 
\mbox{$\rho= \{ 0\rightarrow 2 \rightarrow 1 \rightarrow 0 \}$}, 
while those for right--polarised light satisfy 
the cyclic condition 
\mbox{$\rho=\{ 0\rightarrow 1 \rightarrow 2 \rightarrow 0 \}$}.
Once again, these conditions reflect the way in which the 
current operator transforms under the rotation 
$\hat{\mathpzc{R}}_{2\pi/3}$.   
The same anti--cyclic and cyclic structure is visible in plots of 
matrix elements, Fig.~\ref{fig:SymTri_Rot}(d) and Fig.~\ref{fig:SymTri_Rot}(e).   

%%%%%%%%%%%%%%%%%%%%%%%%%%%%%%%%%%%%%%%%%

Given the role played by rotation symmetry, it is natural to think of the 
selection rules for circularly--polarised light, in terms of the exchange 
of angular momentum between a photon and the electrons in a GQD.
However this intuition must be approached with a little caution, as 
$\hat{\mathpzc{R}}_{2\pi/3}$ generates discrete, and not continuous
rotations of the dot~\footnote{Conservation of angular momentum follows 
from a continuous rotation symmetry, through Noether's theorem.}. 
None the less, working with eigenstates of $\hat{\mathpzc{R}}_{2\pi/3}$, 
we find that states with $\rho = 1,2$ have a net circulation of current 
on their bonds, which is to say a net magnetic moment. 
All of these states transform with the $E$--irrep, and are Kramers doublets, 
degenerate pairs of states whose current--flow is related 
by time--reversal symmetry.
An example of the current flow within such a Kramers doublet, 
calculated within a non--interacting tight--binding model is shown 
in Fig.~\ref{fig:KramersDoubletTri} in  Appendix~\ref{appendix:KramersD}.
Applying a magnetic field lifts the degeneracy of these Kramers
doublets --- a subject explored in more detail in Ref.~\onlinecite{Kavousanaki2015}.

%%%%%%%%%%%%%%%%%%%%%%%%%%%%%%%%%%%%%%%%%%%%%%%%
\subsection{Optical selection rules for rectangular graphene quantum dots}  
%%%%%%%%%%%%%%%%%%%%%%%%%%%%%%%%%%%%%%%%%%%%%%%%%%
\label{sec:SymRect}

In Section~\ref{sec:SymTri}, we explored the optical selection rules which 
arise in the simplest example of a GQD where the planar ($x, y$) components 
of the current operator transform under the same irrep.
%
% --- the triangular GQD shown in Fig.~\ref{fig:lattice}(a), with point group  $C_{3v}$.}
%
We now consider what happens in the simplest example where the planar 
components of the current operator transform under different irreps 
---  the rectangular GQD shown in Fig.~\ref{fig:lattice}(e), with point group, $C_{2v}$.
In this case, we will find that the optical selection rules, 
and associated optical conductivity $\sigma_{\alpha}(\omega)$, {\it does}
depend on the polarisation of the incident light.   

%%%%%%%%%%%%%%%%%%%%%%%%%%%%%%%%%%%%%%%%%%%%%%%%
\subsubsection{Group--theory Analysis}  
%%%%%%%%%%%%%%%%%%%%%%%%%%%%%%%%%%%%%%%%%%%%%%%%%%
\label{sec:group.theory.rectangular.GQD}

The symmetry analysis for a rectangular GQD exactly follows 
the template of Section~\ref{sec:group.theory.triangular.GQD}, 
with one vital difference --- we must now work with the 
representations appropriate to the symmetries of a rectangle, $C_{2v}$, 
rather than those of a triangle, $C_{3v}$.
The representations of the point--group $C_{2v}$ are 
listed in Table~\ref{tab:CharacterTableRectangle}. 
The group $C_{2v}$ comprises the identity ($E$), $\pi$--rotations 
about the two principal axes in the plane ($C_2$), and reflections 
($\sigma_v$ and $\sigma_{v'}$) about the same two axes, 
as shown in Fig.~\ref{fig:lattice}(e). 
All of these symmetry operations commute, and as a result
the group has only one--dimensional irreps : 
$A_1, A_2, B_1$ and $B_2$.
And, crucially, the way in which a polar vector 
(such as the current operator $\hat{\mathpzc{J}}_\alpha$) 
transforms within $C_{2v}$ is also different from the way 
in which it transforms in $C_{3v}$.
As shown in Table~\ref{tab:CharacterTableRectangle}, each of
the different components of a polar vector transform with different 
irreps of $C_{2v}$.   
And, since different polarisations of light couple to different 
components of  $\hat{\mathpzc{J}}_\alpha$, this opens the 
door to a polarisation--dependent optical response.

%%%%%%%%%%%%%%%%%%%%%%%%%%%%%%%%%%%%%%%%%%%%%%%%%%
%  Table II - character table of C_{2v}
%%%%%%%%%%%%%%%%%%%%%%%%%%%%%%%%%%%%%%%%%%%%%%%%%%

\setlength{\tabcolsep}{0.5em} % for the horizontal padding
\renewcommand{\arraystretch}{1.2}% for the vertical padding
\begin{table} [t]
\centering
    \begin{tabular}{ | c | c | c | c | c | c | p{1cm} | }     
    \hline
    	$C_{2v}$	& $E$ 	& $C_2(z)$	& $\sigma_v(xz)$ 	& $\sigma_{v'}(yz)$ 	& 	polar vectors	\\ \hline
    	$A_1$	& 1		& 1 			& 1 				& 1 				& 	z			\\ \hline
    	$A_2$ 	& 1 		& 1 			& -1 				& -1 				&				\\ \hline
	$B_1$ 	& 1 		& -1  		& 1 				& -1 				&	x			\\ \hline
	$B_2$ 	& 1 		& -1  		& -1 				& 1 				&	y			\\ \hline
    \end{tabular}
    \caption{
    Character Table of the point--group $C_{2v}$, describing the 
    symmetry of rectangular graphene quantum dots (GQD's) of the type shown in 
    Fig.~\ref{fig:lattice}(e).  
    Eigenstates of a triangular GQD transform with irreducible representations 
    (irreps) $A_1$, $A_2$, $B_1$ and $B_2$, 
    while the $x$ and $y$ components of the current operator $\hat{\mathpzc{J}}$ 
    (a polar vector) transform with $B_1$ and $B_2$ 
    [cf. \onlinecite{Weyl,Heine,LandauLifshitz,Tinkham,Wagner,Jones}].
    The corresponding symmetry operations are identity ($E$), $\pi$--rotation about the principal 
    axes ($C_2$) and two reflections on symmetry axes ($\sigma_v$ and $\sigma_{v'}$), as 
    shown in Fig.~\ref{fig:lattice}(e).
    }	
    \label{tab:CharacterTableRectangle}
\end{table}

%%%%%%%%%%%%%%%%%%%%%%%%%%%%%%%%%%%%%%%%%%%%%%%%%%

Once again, we can proceed from the character Table to optical selection rules 
by applying the general rules for the products of representations \cite{Weyl,Heine,LandauLifshitz,Tinkham,Wagner,Jones}.  
The two components of the current operator, $\hat{\mathpzc{J}}^x$
and $\hat{\mathpzc{J}}^y$, transform with $B_1$ and $B_2$, respectively.
And, following the steps of Section~\ref{sec:group.theory.triangular.GQD}, 
the products entering into optical matrix elements can be decomposed as 
\begin{center}
	\begin{tabular}{c c c c c c c c c l}
	$B_1$ 	& $\times$	& $A_1$	&	$ \rightarrow B_1$	  \; ,
	& \quad \quad 	&	$B_2$	& $\times$	& $A_1$	&	$ \rightarrow B_2$  \; ,  
	\\
	$B_1$ 	& $\times$	& $A_2$	&	$ \rightarrow B_2$	  \; ,
	& \quad \quad 	&	$B_2$	& $\times$	& $A_2$	&	$ \rightarrow B_1$  \; ,  
	\\
	$B_1$ 	& $\times$	& $B_1$	&	$ \rightarrow A_1$	 \; ,
	& \quad \quad 	&	$B_2$	& $\times$	& $B_1$	&	$ \rightarrow A_2$  \; ,  
	\\	
	$B_1$ 	& $\times$	& $B_2$	&	$ \rightarrow A_2$	 \; ,
	& \quad \quad 	&	$B_2$	& $\times$	& $B_2$	&	$ \rightarrow A_1$  \; .  
	\\
	\end{tabular}
\end{center}
These results lead to the polarisation--dependent selection rules
\begin{eqnarray}
   \sigma_x (\omega) \; : \;  A_1 \longleftrightarrow B_1 \; , \;  A_2 \longleftrightarrow B_2 \; ,   
    \label{eq:rectangular.dot.sigma.x} 
\end{eqnarray}
and 
\begin{eqnarray}
   \sigma_y (\omega) \; : \;  A_1 \longleftrightarrow B_2 \; , \;  A_2 \longleftrightarrow B_1 \; . 
    \label{eq:rectangular.dot.sigma.y} 
\end{eqnarray}
As a direct consequence, in general, for a rectangular GQD, 
\begin{eqnarray}
\sigma_x(\omega) \ne \sigma_y(\omega) \; .
\end{eqnarray}

%%%%%%%%%%%%%%%%%%%%%%%%%%%%%%%%%%%%%%%%%%%%%%%%
\subsubsection{Illustration of optical selection rules for linearly--polarised light}  
%%%%%%%%%%%%%%%%%%%%%%%%%%%%%%%%%%%%%%%%%%%%%%%%%%

%%%%%%%%%%%%%%%%%%%%%%%%%%%%%%%%%%%%%%%%%%%%%%%%%%
% Fig. 4 - selection rules for a rectangular GQD
%%%%%%%%%%%%%%%%%%%%%%%%%%%%%%%%%%%%%%%%%%%%%%%%%%

\begin{figure} 
\centering
	\includegraphics[width=0.48\textwidth]{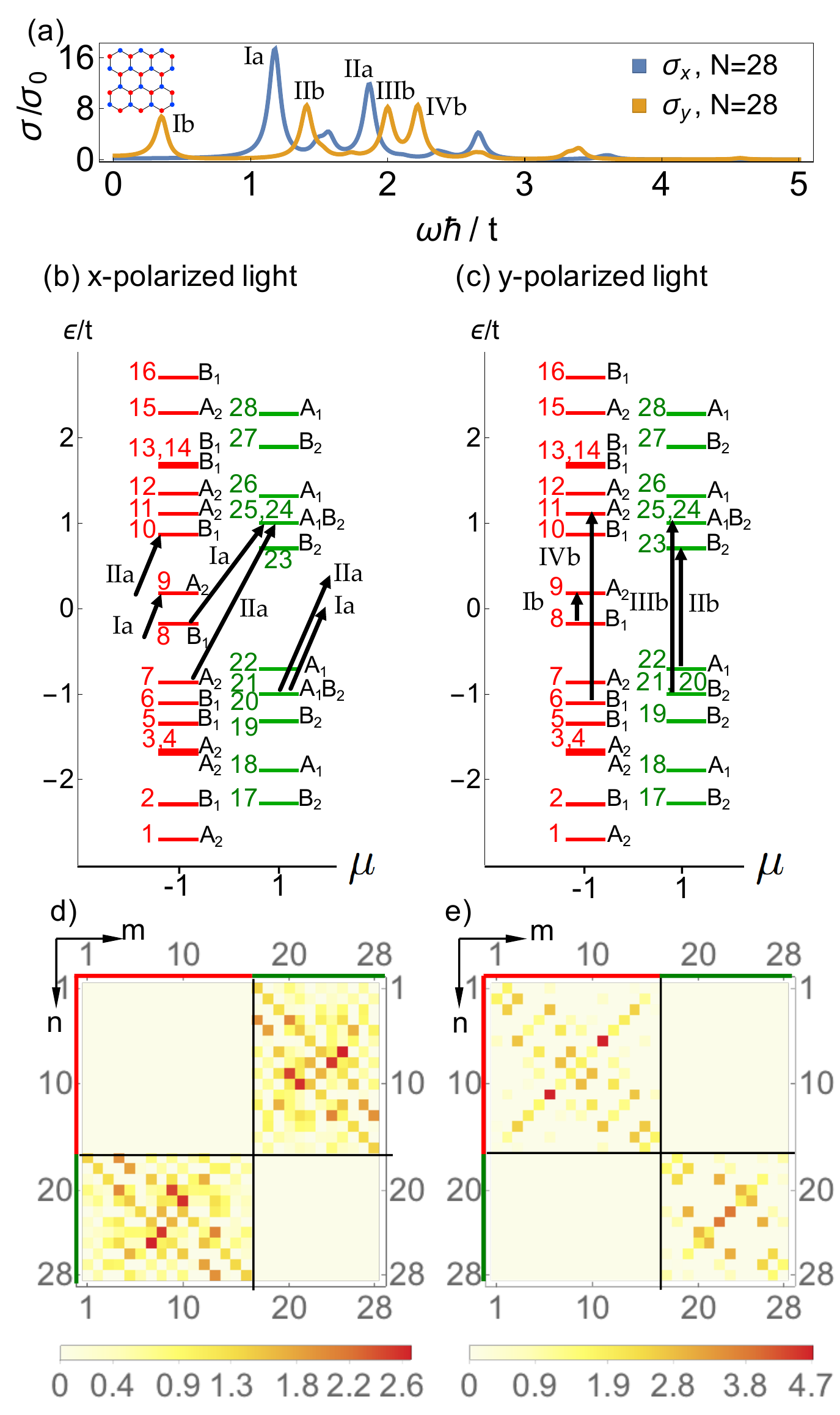}
 	 \caption{
	 Optical selection rules for a rectangular graphene quantum dot (GQD) of size 
	 $N = 28$ in linearly--polarized light.
         (a) Optical conductivity $\sigma_{\alpha}(\omega)$, showing different results for 
         $x$-- and $y$--polarized light.    
	 (b) and (c) Spectrum of the corresponding tight--binding model Eq.~(\ref{eq:H0}), 
	 in the mirror basis Eq.~(\ref{eq:mirror}), showing the different 
	 allowed transitions for $x$-- and $y$--polarized light.  
         (d) and (e) Matrix elements of the corresponding current operators 
          $|\mathpzc{J}_{nm}^{x}|^2$ and $|\mathpzc{J}_{nm}^{y}|^2$
          [cf.~Eq.~(\ref{eq:single.electron.matrix.element})]
          in units of $(e t / \hbar)^2$.
         Results for $\sigma_{\alpha}(\omega)$ were calculated from Eq.~(\ref{eq:sigma.tight.binding}), 
         with a Lorentzian of FWHM $2\gamma = 0.1\ \text{t}$.
         Eigenstates are labelled according to their quantum number $n=1\ldots 28$, 
         eigenvalue $\mu = \pm 1$ [Eq.~(\ref{eq:mirror})] and corresponding irrep 
         [cf. Table~\ref{tab:CharacterTableRectangle}].
	}
  	\label{fig:SymRec}
\end{figure}

%%%%%%%%%%%%%%%%%%%%%%%%%%%%%%%%%%%%%%%%%%%%%%%%%%

Once again, it is instructive to illustrate the general, group--theoretical 
result for the specific example of a rectangular GQD described 
by the tight--binding model introduced in Section~\ref{sec:tight-binding.model}.
The corresponding results for the optical conductivity $\sigma_\alpha(\omega)$ of
a rectangular GQD of size $N = 28$ are shown in Fig.~\ref{fig:SymRec}(a).
As expected, results for $x$--polarized light (blue line) are dramatically 
different from those for $y$--polarized light (orange line), 
with peaks at different values of $\omega$, labelled I--IV, signalling the 
different allowed optical transitions.  
These results are in contrast with the polarisation--independent
optical conductivity obtained for a triangular GQD, seen in Fig.~\ref{fig:SymTri_Mir}.

%%%%%%%%%%%%%%%%%%%%%%%%%%%%%%%%%%%%%%%%%%%%%%%%%%

The optical transitions associated with each of these features in 
$\sigma_\alpha(\omega)$ are shown in Fig.~\ref{fig:SymRec}(b) 
and Fig.~\ref{fig:SymRec}(c).  
Once again, states have been labelled according to their irreps, 
and further classified according their eigenvalues under the reflection 
operator $\hat{\mathpzc{M}_y}$ [Eq.~\ref{eq:mirror}].
The transitions associated with specific peaks in $\sigma_x(\omega)$
and $\sigma_y(\omega)$ are labelled I--III.  
All transitions contributing to $\sigma_x(\omega)$ satisfy
the optical selection rules Eq.~\ref{eq:rectangular.dot.sigma.x}, 
while transitions contributing to $\sigma_y(\omega)$ satisfy
the optical selection rules Eq.~\ref{eq:rectangular.dot.sigma.y}.
The corresponding matrix elements are illustrated in 
Fig.~\ref{fig:SymTri_Rot}(d) and Fig.~\ref{fig:SymTri_Rot}(e).   

%%%%%%%%%%%%%%%%%%%%%%%%%%%%%%%%%%%%%%%%%%%%%%%%
\subsubsection{Effect of sublattice symmetry--breaking}
%%%%%%%%%%%%%%%%%%%%%%%%%%%%%%%%%%%%%%%%%%%%%%%%%%

One of the possibilities discussed in bulk graphene is that interactions 
could drive a many--electron instability which would spontaneously break 
the symmetry between the two sites in the unit cell of the honeycomb 
lattice \cite{Kotov2012}.
On general grounds, such symmetry breaking is not 
expected to occur spontaneously in a finite--size system 
\cite{plischke-book,huang-book}.
None the less, it is instructive to consider what effect this broken 
symmetry would have, if it was induced by an external field or 
perturbation of the graphene structure.

%%%%%%%%%%%%%%%%%%%%%%%%%%%%%%%%%%%%%%%%%%%%%%%%%%

In the case of the rectangular GQD's considered above, breaking the 
symmetry between the A-- and B--sublattices of sites would reduce 
the point--group symmetry from C$_{2v}$ to C$_2$.
Within the group C$_{2v}$, the x--component of the current 
operator $\mathpzc{J}^\alpha$ transforms with the irrep B$_1$, 
while the y--component transforms with B$_2$ 
[cf. Table~\ref{tab:CharacterTableRectangle}].
As a consequence, the optical conductivity $\sigma_\alpha(\omega)$
depends on the polarization $\alpha$ of the incident light 
[cf. Sec.~\ref{sec:group.theory.rectangular.GQD}].

%%%%%%%%%%%%%%%%%%%%%%%%%%%%%%%%%%%%%%%%%%%%%%%%%%

In contrast, the group C$_2$, has only two irreps, A and B, 
and both the x-- and y--components 
of $\mathpzc{J}^\alpha$ transform with the same irrep, B.
It follows that, once the symmetry between the A-- and B--sublattices 
is broken, the optical conductivity of rectangular dot is  
{\it independent} of the polarization of the incident light.
This is a profound change, and emphasizes the power of optical 
selection rules in probing the symmetry of a GQD.
\\

%%%%%%%%%%%%%%%%%%%%%%%%%%%%%%%%%%%%%%%%%%%%%%%%%%
\subsection{Optical selection rules for GQD's of other symmetry}
%%%%%%%%%%%%%%%%%%%%%%%%%%%%%%%%%%%%%%%%%%%%%%%%%%

In Section~\ref{sec:SymTri} and Section~\ref{sec:SymRect} we have seen how 
optical selection rules work for a GQD with triangular and rectangular 
point--group symmetry.
%
%In Section~\ref{sec:SymTri} and Section~\ref{sec:SymRect} we have seen how 
%optical selection rules work for  simplest example of a GQD with a non--abelian point group, 
%and the simplest example of a GQD with an abelian point group.

%%%%%%%%%%%%%%%%%%%%%%%%%%%%%%%%%%%%%%%%%%%%%%%%%%

In the case of the triangular GQD, with non--abelian point group $C_{3v}$,
the optical conductivity $\sigma_\alpha(\omega)$ was found to be 
polarisation--independent, with \mbox{$\sigma_x(\omega) = \sigma_y(\omega)$}. 
This result followed from the fact that both components of the current operator 
$\hat{\mathpzc{J}}^\alpha$ transform with the same two--dimensional 
irrep, $E$, and therefore transform in the same way under the 
symmetries of the dot.

%%%%%%%%%%%%%%%%%%%%%%%%%%%%%%%%%%%%%%%%%%%%%%%%%%

In the case of the rectangular GQD, with abelian point group $C_{2v}$,
the optical conductivity $\sigma_\alpha(\omega)$ was found to be 
polarisation--dependent, with \mbox{$\sigma_x(\omega) \ne \sigma_y(\omega)$}.   
This result followed from the fact that $C_{2v}$ supports only 
one--dimensional irreps, and the different components of the 
current operator $\hat{\mathpzc{J}}^\alpha$ must therefore 
transform with different irreps, and so in different ways under 
the symmetries of the dot.

%%%%%%%%%%%%%%%%%%%%%%%%%%%%%%%%%%%%%%%%%%%%%%%%%%

It might seem counter intuitive that the simpler symmetry  
structure, $C_{2v}$, should lead to a more complex optical 
response.
However it is precisely the simplicity of the group, with 
only one--dimensional irreps, which allows the different components 
of the current operator to transform in different ways.
It is the larger and more complex symmetry group $C_{3v}$ which 
permits both components of the current operator 
to transform in the same way.
And, by the same token, these results generalise straight--forwardly
to dots of different 
symmetry : 
GQD's with a point--group such that the x-- and y--components 
of the current operator transform under the 
same irrep, will show an optical conductivity which is independent of  
polarisation.
Meanwhile, dots with a point--group for which the x-- and y--components 
of the current operator transform under different 
irreps, will have a polarisation--dependent response.
%
%And, by the same token, these results generalise straight--forwardly
%to dots of different shape : dots with an abelian point--group symmetry 
%will have an optical conductivity which is independent of the polarisation, 
%while dots with abelian point--group symmetry will have a 
%polarisation--dependent response.
%
We do not develop this theme further here, but illustrate 
results for another example, the hexagonal GQD in Appendix~\ref{appendix:SymHex}.

%%%%%%%%%%%%%%%%%%%%%%%%%%%%%%%%%%%%%%%%%%%%%%%%%
\section{The role of edge types}  \label{sec:Edges}
%%%%%%%%%%%%%%%%%%%%%%%%%%%%%%%%%%%%%%%%%%%%%%%%%%

%%%%%%%%%%%%%%%%%%%%%%%%%%%%%%%%%%%%%%%%%%%%%%%%%%
% Table III - finite size--scaling of degeneracies 
%%%%%%%%%%%%%%%%%%%%%%%%%%%%%%%%%%%%%%%%%%%%%%%%%%
\begin{table*} 
%%%%%%%%%%%%%%%%%%%%%%%%%%%%%%%%%%%%%%%%%%%%%%%%%%
\begin{center}
    \begin{tabular}{|l|l|c|c|c|c|c|}     
    \hline
       		&			& Triangular Zigzag 			& Triangular Armchair 			& Hexagonal Zigzag 			& Hexagonal Armchair 		& Rectangular  				\\ \hline 
    $L$	&			& $\sqrt{3} (N_h + 1)a$		& $3 N_h a$		 			& $\sqrt{3} (N_h -1/3) a$		& $(3 N_h - 2) a$		 	& $ L_x = \sqrt{3} (2N_h -1) a $ \\ 
    		&			&						&							&						&						& $ L_y = 2.5 N_h a$	\\ \hline 
    $N $ 	&			& $(N_{h} + 1)^2 + 2 N_{h}$	& $\frac{3}{4} (N_{h} + 2) N_{h}$ 	& $6 N_{h}^2$ 				& $9 N_{h} ((N_{h}/2) - 1) + 6$ 	& $2(4N_h^2 - N_h)$   		\\ \hline
    $N_0$ 	&			& $N_{h} - 1$ 				& $0$ 						& $0$ 					& $0$  					& $0$ 					\\ \hline
    $N_t$ 	&  $N_h$ even 	& $3$ 					& $0$ 						& $3$ 					& $3 N_{h}$ 	 			& $N_h$ 					\\ \cline{2-7}
          	&  $N_h$ odd 	& $3$ 					& $\frac{1}{2} (3N_{h} + 1)$	 	& $N_{h} + (-1)^{(N_{h}+1)/2}$ 	& $4N_h -3$  				& $N_h + 1$ 				\\ \hline
    \end{tabular}
\end{center}
    \caption{    
    Finite--size scaling of degeneracies for different types of regularly--shaped GQD's, within 
    the tight--binding model, Eq.~(\ref{eq:H0}).   
    $L$ is the length of the edge of the GQD, $a$ is the %\textcolor{red}{honeycomb} 
    lattice constant, $N_h$ the number of hexagons per side, 
    $N$ is the total number of sites, $N_0$ the number of zero--energy states, and $N_t$ the number 
    of states at energy $\epsilon=t$. 
   For rectangular GQD's $N_h$ refers to the number of hexagons on the side with the armchair edges.}
    \label{tab:scalingbehavior}
%%%%%%%%%%%%%%%%%%%%%%%%%%%%%%%%%%%%%%%%%%%%%%%%%%
\end{table*}
%%%%%%%%%%%%%%%%%%%%%%%%%%%%%%%%%%%%%%%%%%%%%%%%%%

% section intro

In Section \ref{sec:Geometry}, we explored the role of symmetry on the optical 
conductivity, and have shown that the polarization of light can be used to distinguish between 
dots of different shape.  

%In Section \ref{sec:Geometry}, we have explored the role of symmetry on the optical 
%conductivity, and shown that the polarization of light can be used to distinguish between 
%dots of abelian and non--abelian point-group symmetry.  

%%%%%%%%%%%%%%%%%%%%%%%%%%%%%%%%%%%%%%%%%%%%%%%%%%

In this section, we explore the role of edge types in GQD's by studying the optical 
conductivity $\sigma_{\alpha}(\omega)$ for triangular, hexagonal and rectangular GQD's 
with zigzag and armchair edges and various sizes. 
Using exact diagonalization of the tight--binding model [Eq.~(\ref{eq:H0})], 
we show the size evolution of $\sigma_{\alpha}(\omega)$ for dots with a total number of sites 
up to $N \approx 25,000$ ($\sim 40$~nm) and find features which allow us to separate 
between dots of zigzag and armchair edges. 

%%%%%%%%%%%%%%%%%%%%%%%%%%%%%%%%%%%%%%%%%%%%%%%%%%

% 0 energy states in DOS

Fig.~\ref{fig:DosCondALL} shows an overview of the density of states 
and optical conductivity for the five different regular--shaped GQD's 
considered in this Article (Fig.~\ref{fig:lattice}), for sizes up to $N\approx 10,000$ sites.
Triangular GQD's with zigzag edges [Fig.~\ref{fig:DosCondALL}(a)] show a peak 
in the DOS at zero energy for all dot sizes, due to a large number of states localized 
on the edge that have exactly $\epsilon=0$ within the tight--binding model \cite{Ezawa2007, Zhou2012}.
The number $N_0$  of these ``zero--energy'' states increases linearly with dot size: 
$N_0=N_h-1$\cite{Zarenia2011} [see Table~\ref{tab:scalingbehavior}], where $N_h$ 
is the number of hexagons per side of the GQD.
In the thermodynamic limit the number of edge-states is negligible compared to 
the total number of states ($N_0/N\rightarrow 0$), thus recovering graphene's zero 
DOS at $\epsilon=0$.

%%%%%%%%%%%%%%%%%%%%%%%%%%%%%%%%%%%%%%%%%%%%%%%%%%

As seen in Fig.~\ref{fig:DosCondALL}(c) and (e), the DOS shows also a feature 
at zero--energy in hexagonal zigzag and rectangular GQD's for sizes $N \gtrsim 300$.
In fact, for these dots, there are no states with the exact value $\epsilon = 0$ 
(Table~\ref{tab:scalingbehavior}). 
However a finite number of states in the vicinity of zero energy approach zero for 
large dot sizes, resulting in a distinct peak in the DOS.
On the other hand, for dots without any zigzag edges [Fig.~\ref{fig:DosCondALL}(b) and (d)], 
the density of states at the chemical potential, $g(\epsilon=0)$ is zero for all sizes.

%%%%%%%%%%%%%%%%%%%%%%%%%%%%%%%%%%%%%%%%%%%%%%%%%%

% t states in DOS

Graphene's DOS exhibits two characteristic peaks at $\epsilon=\pm t$ (\mbox{Van Hove} 
singularities)\cite{CastroNeto2009}.  
Fig.~\ref{fig:DosCondALL} clearly shows that these peaks are recovered for large dots 
in all shapes and edge types.
They are due to the presence of states with energy exactly $\epsilon = \pm t$.
For most dots, their number $N_t$ scales linearly with dot size 
(see Table~\ref{tab:scalingbehavior}).
However, unlike the case for zero--energy states, their contribution to the DOS at $\pm t$ 
does not disappear in the thermodynamic limit.
States with energies close to $\pm t$ converge to \mbox{$\epsilon=\pm t$} in the 
thermodynamic limit, causing the \mbox{Van Hove} singularities known in graphene.
We have found analytical expressions for the wave functions of states with exactly 
$\epsilon = \pm t$ and will discuss them more detailed in Section~\ref{sec:1-dimWaveFunction}.

%%%%%%%%%%%%%%%%%%%%%%%%%%%%%%%%%%%%%%%%%%%%%%%%%%

% 2t peak in optical conductivity

Transitions between states of \mbox{$\epsilon = \pm t$} cause a characteristic 
peak in the optical conductivity for \mbox{$\hbar\omega=2t$}. 
The size--evolution shows that the intensity of this peak is shape and edge-type
dependent, though particularly pronounced in dots with armchair edges. 
However, for large enough sizes all GQD's show an asymmetric Fano resonance--like line-shape, 
similar to the one in graphene \cite{Fano1961, Mak2011, Chae2011}.

%%%%%%%%%%%%%%%%%%%%%%%%%%%%%%%%%%%%%%%%%%%%%%%%%%

% t peak in optical conductivity

Interestingly, all GQD's with at least some zigzag edges, show an additional peak 
at \mbox{$\hbar\omega=t$} at least for sizes larger than \mbox{$N>1000$}, because 
of transitions between zero--energy states and states at $\epsilon=\pm t$. 
This peak does not appear in the case of dots with purely armchair edges, thus 
providing a way to differentiate between the two edge types.

%%%%%%%%%%%%%%%%%%%%%%%%%%%%%%%%%%%%%%%%%%%%%%%%%%

Rectangular dots show a different behaviour of the optical conductivity for $x$-- 
and $y$--polarized light [Fig.~\ref{fig:DosCondALL}(e)--(f)], as explained in 
Sec.~\ref{sec:SymRect}.
When the electric field is in parallel with the zigzag edges ($x$--axis for these dots), 
the additional peak at \mbox{$\hbar\omega=t$} shows up.
This is due to the fact that the zero--energy states are localised on the zigzag edges, 
causing a non-zero current.
On the other hand, for $y$--polarized light the optical conductivity does not exhibit this feature.

%%%%%%%%%%%%%%%%%%%%%%%%%%%%%%%%%%%%%%%%%%%%%%%%%%
% Fig. 5 - evolution of \sigma(\omega) with size
%%%%%%%%%%%%%%%%%%%%%%%%%%%%%%%%%%%%%%%%%%%%%%%%%%

\begin{figure*}
  \centering
  \includegraphics[width=0.99\textwidth]{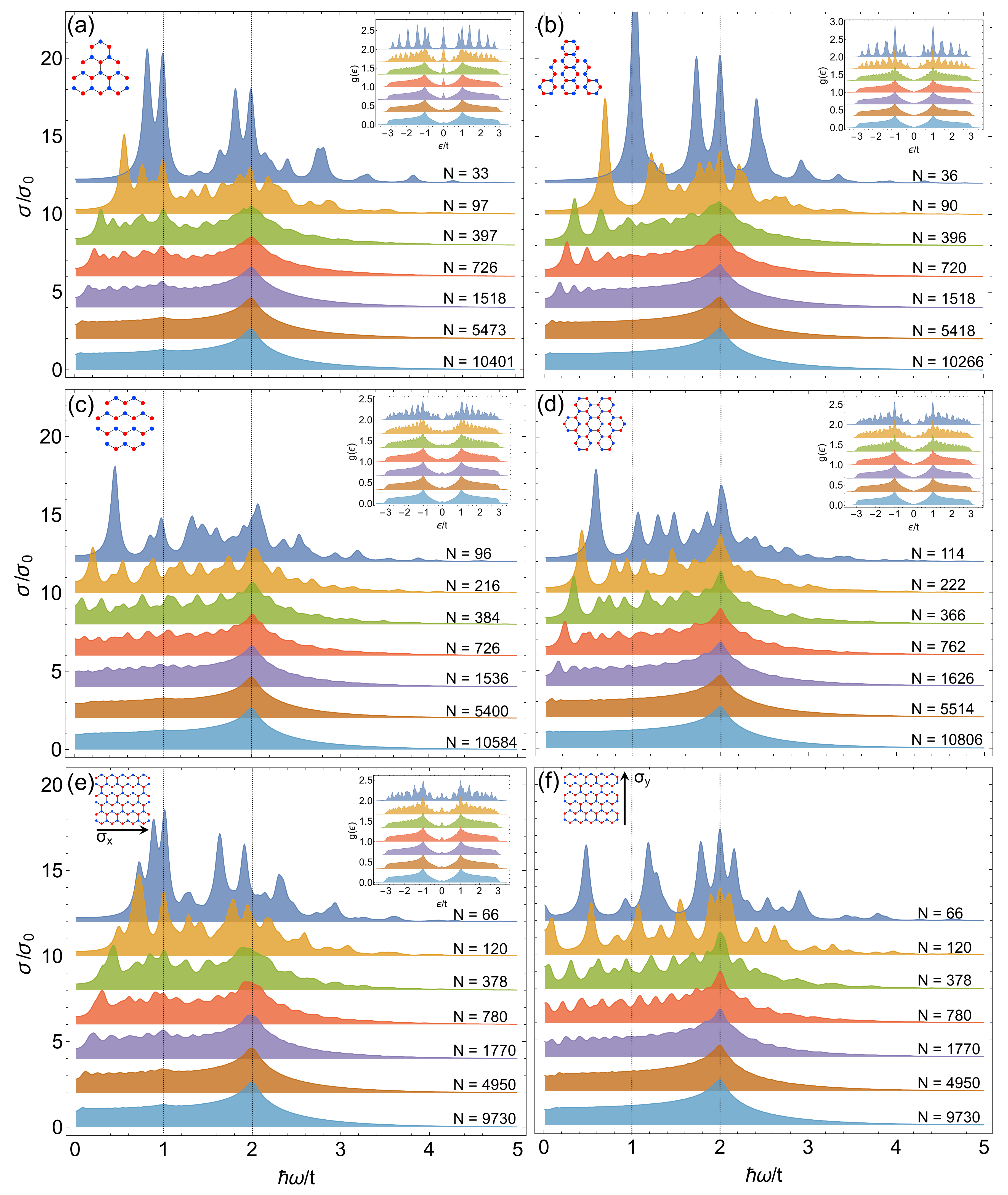}
  \caption{
    Evolution of the optical conductivity $\sigma_{\alpha}(\omega)$ and density of states $g(\epsilon)$ 
    of  the regular--shaped graphene quantum dots (GQD's) shown in Fig.~\ref{fig:lattice}, as a 
    function of the size of the dot.
    (a) Triangular zigzag, (b) triangular armchair, (c) hexagonal zigzag, (d) hexagonal armchair, 
    and (e)--(f) rectangular GQD for both $x$-- and $y$--polarized light.
    We find that states with $\epsilon = \pm t$ exist in all dots, but are generally more pronounced in 
    dots with armchair edges, causing dominant peaks at $g(\epsilon = \pm t)$. 
    For all dots with zigzag edges, there are states in the vicinity of $\epsilon = 0$, causing an additional 
    absorption peak at $\hbar \omega / t = 1$.  
    All results were calculated within the tight--binding model Eq.~(\ref{eq:H0}), and have been convoluted 
    with a Lorentzian of FWHM $2\gamma = 0.1\ \text{t}$
    [cf. Eq.~(\ref{eq:DOS}), Eq.~(\ref{eq:sigma.tight.binding})].
    For better visualisation each plot has been shifted along the vertical axis. 
  }
  \label{fig:DosCondALL}
\end{figure*}

%%%%%%%%%%%%%%%%%%%%%%%%%%%%%%%%%%%%%%%%%%%%%%%%%%
% Fig. 6 - \sigma(\omega) at low frequencies
%%%%%%%%%%%%%%%%%%%%%%%%%%%%%%%%%%%%%%%%%%%%%%%%%%

\begin{figure*}
  \centering
  \includegraphics[width=0.95\textwidth]{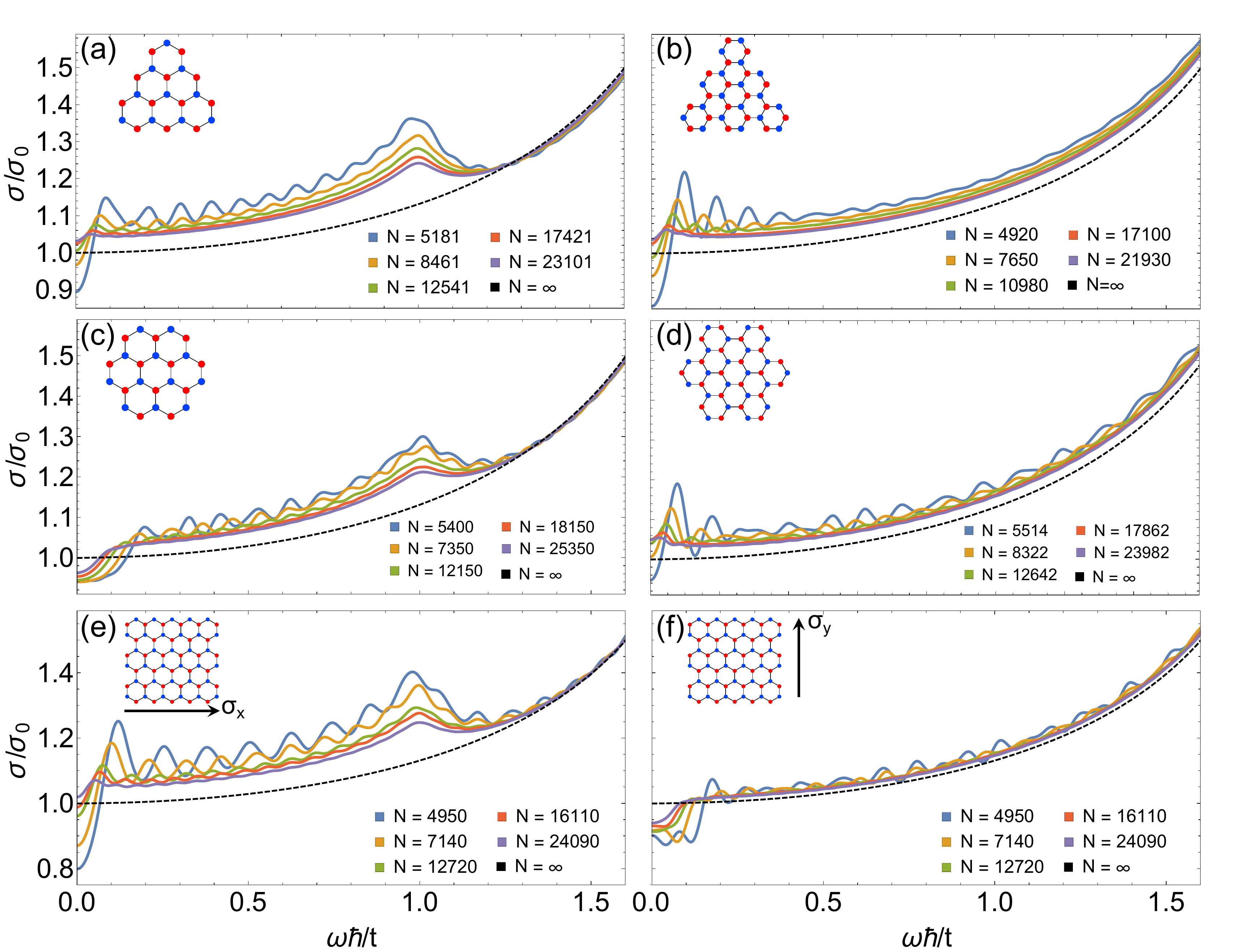}
  \caption{
    Low frequency part of the optical conductivity $\sigma_{\alpha}(\omega)$ as shown in 
    Fig.~\ref{fig:DosCondALL}. 
    (a) Triangular zigzag, (b) triangular armchair, (c) hexagonal zigzag, (d) hexagonal armchair, 
    (e)--(f) rectangular graphene quantum dot for both $x$-- and $y$--polarized light, 
    for sizes up to $N \approx 25, 000$.
    The peak at $\hbar \omega / t = 1$ in $\sigma_{\alpha}(\omega)$ for dots with zigzag edges 
    (polarization parallel to the zigzag edge) is distinguishable from the featureless conductivity 
    of dots with armchair edges.
    We compare the results to the analytical solution for a graphene sheet in its thermodynamic limit 
    [see Eq.~(\ref{eq:conductivityFinal_energySpace}) --- dashed line].
     All results were calculated within the tight--binding model Eq.~(\ref{eq:H0}), and have been
     convoluted with a Lorentzian of FWHM $2\gamma = 0.1\ \text{t}$
     [cf. Eq.~(\ref{eq:sigma.tight.binding})].
  }
  \label{fig:SizeEvo}
\end{figure*}

%%%%%%%%%%%%%%%%%%%%%%%%%%%%%%%%%%%%%%%%%%%%%%%%%%
%  Figure 7 - finite--size scaling of dominant features in \sigma(\omega)
%%%%%%%%%%%%%%%%%%%%%%%%%%%%%%%%%%%%%%%%%%%%%%%%%%

\begin{figure}  
  	\includegraphics[width=0.49\textwidth]{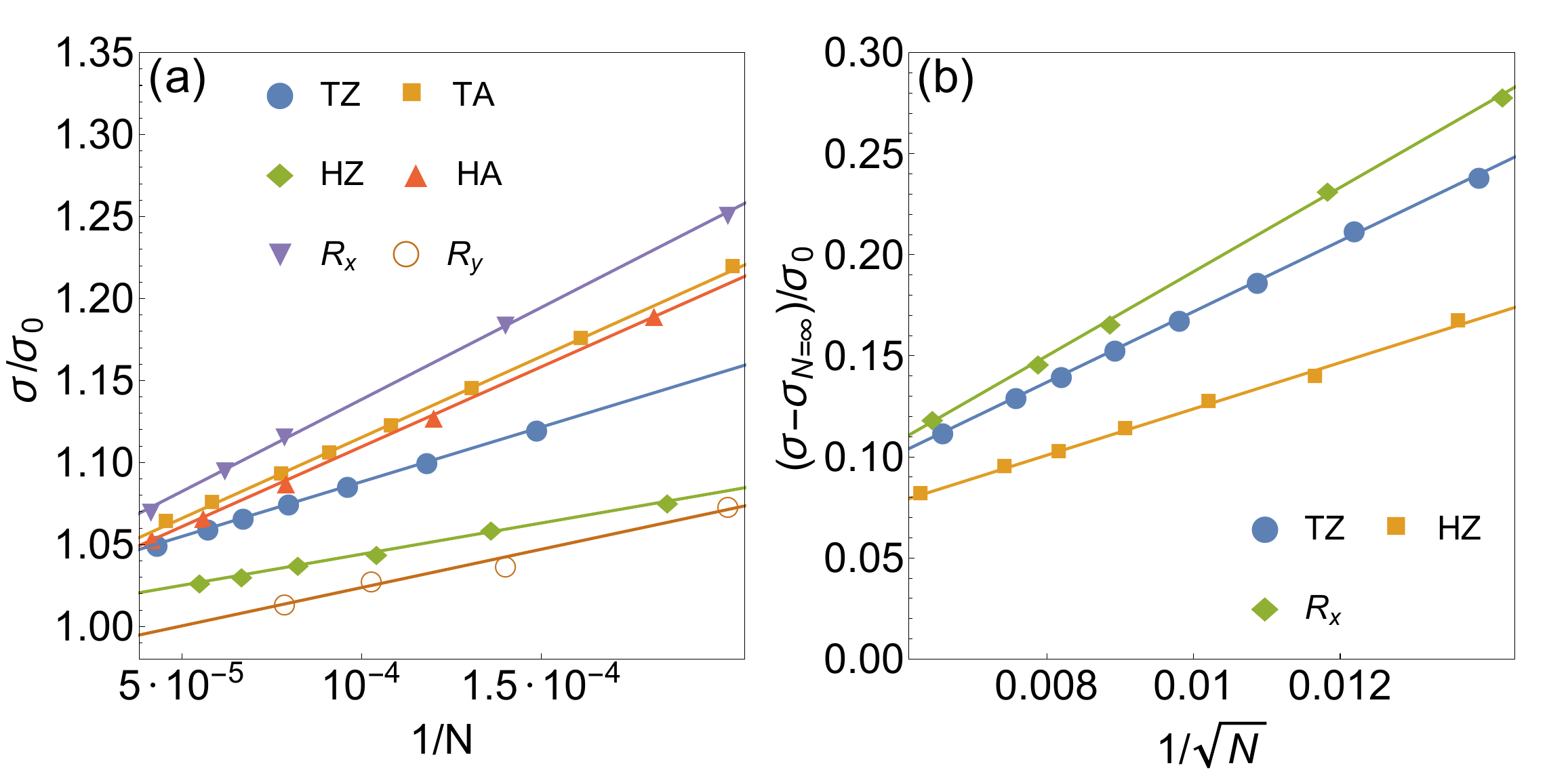}
	\caption{
	Finite--size scaling of the dominant features in the optical conductivity  $\sigma_{\alpha}(\omega)$ 
	from Fig.~\ref{fig:SizeEvo}.
	(a) Linear scaling of the peak maximum for $\hbar \omega / t < 0.2$ with the inverse 
	system size $1/N$, showing the trend for all graphene quantum dots (GQD's) to reach the universal 
	optical conductivity $\sigma_0$ [Eq.~(\ref{eq:sigma0})] in the thermodynamic limit at low frequencies.
	(b)  Scaling of the enhancement in optical conductivity at $\hbar \omega/ t  = 1$, with the linear 
	dimension of the dot \mbox{$1/\sqrt{N}$}, for a dot with zig--zag edges.   
	For \mbox{$N \to \infty$} we recover the result for bulk graphene, 
	Eq.~(\ref{eq:conductivityFinal_energySpace}).  
	}
 	\label{fig:FiniteSizeScaling}
\end{figure}

%%%%%%%%%%%%%%%%%%%%%%%%%%%%%%%%%%%%%%%%%%%%%%%%%%

To get further insight in the size evolution of the peak at $\hbar\omega=t$, we plot the optical 
conductivity for dot sizes of $N\approx 5,000-25,000$ ($\approx 20-40$~nm) in Fig.~\ref{fig:SizeEvo} 
and compare to results of the analytical solution for graphene in its thermodynamic limit 
[see Eq.~(\ref{eq:conductivityFinal_energySpace})]. 

%%%%%%%%%%%%%%%%%%%%%%%%%%%%%%%%%%%%%%%%%%%%%%%%%%

The right panels of Fig.~\ref{fig:SizeEvo} show, that the optical conductivity for GQD's with 
armchair edges converges slowly to the graphene limit, uniformly for all energies.
On the other hand, \mbox{$\sigma_{\alpha}(\omega)$} for dots with zigzag edges has a more complicated 
behaviour [left panels of Fig.~\ref{fig:SizeEvo}].
The peak at \mbox{$\hbar\omega=t$} is well distinguishable from the background for sizes 
up to $N\approx 25,000$ ($\sim 40$ nm).
For values lower in frequency than this peak, the difference to the graphene limit is larger than 
in the case of armchair edges.
This can be explained by transitions  occurring between states of $\epsilon < t$ and states of
 zero--energy, which are absent in armchair dots. 
For \mbox{$\hbar\omega>t$}, $\sigma_{\alpha}(\omega)$ coincides with the graphene limit.

%%%%%%%%%%%%%%%%%%%%%%%%%%%%%%%%%%%%%%%%%%%%%%%%%%

% Finite Size Scaling
As seen in Fig.~\ref{fig:SizeEvo}, the plateau of the optical conductivity exhibits strong oscillations. 
However, in the thermodynamic limit, it approaches the universal conductivity of graphene, 
$\sigma_0$ [Eq.~(\ref{eq:sigma0})].
This is seen in Fig.~\ref{fig:FiniteSizeScaling}(a) where the finite--size scaling of the peak at 
$\hbar\omega / t < 0.2$ is shown for all dots under consideration. 
For all GQD's, this peak approaches the universal conductivity $\sigma_0$ within 
$2 \%$.

%%%%%%%%%%%%%%%%%%%%%%%%%%%%%%%%%%%%%%%%%%%%%%%%%%

%By fitting with a linear function we find values of the optical conductivity $\sigma/\sigma_0$ for infinite large dots as follows: triangular zigzag $1.021(1)$, triangular armchair $1.016(1)$, hexagonal zigzag $1.006(1)$, hexagonal armchair $1.011(1)$, rectangular dots for x-polarization $1.024(1)$ and rectangular dots for y-polarization $0.977(1)$.

%%%%%%%%%%%%%%%%%%%%%%%%%%%%%%%%%%%%%%%%%%%%%%%%%%

In Fig.~\ref{fig:FiniteSizeScaling}(b) we show the finite--size scaling of the peak at 
$\hbar \omega / t = 1$ for GQD's with zigzag edges. 
%
%In subtract $\sigma_{pbc}(\omega)$ of graphene with PBC from GQD's with zigzag edges $\sigma_{tz}(\omega)$ and scale the peak over the inverse system size.
%
We find that this peak approaches the value of graphene linearly due to its 
inverse system length \mbox{$1/L \sim 1/\sqrt{N}$}.
%Fitting the peak with a quadratic function shows a convergence of $\lim_{N \to \infty} \big(\sigma_{tz}(\omega) - \sigma_{pbc}(\omega) \big)$ for triangular zigzag to $0.059(1)$, hexagonal zigzag to $0.051(1)$ and rectangular GQD's polarized in x-direction to $0.57(1)$.

%%%%%%%%%%%%%%%%%%%%%%%%%%%%%%%%%%%%%%%%%%%%%%%%%%

In summary, we discussed the optical conductivity of GQD's for a variety of shapes, edge types and 
sizes up to the thermodynamic limit. 
We showed that polarization along a zigzag edge causes a distinct peak in 
$\sigma_{\alpha}(\omega)$ at \mbox{$\hbar \omega =t$}.
This effect may be used as a way to distinguish between GQD's with zigzag and armchair edges.
 
%%%%%%%%%%%%%%%%%%%%%%%%%%%%%%%%%%%%%%%%%%%%%%%%%
\section{Localized states with energy $\epsilon/t = \pm 1$}  
%%%%%%%%%%%%%%%%%%%%%%%%%%%%%%%%%%%%%%%%%%%%%%%%%
\label{sec:1-dimWaveFunction}

%%%%%%%%%%%%%%%%%%%%%%%%%%%%%%%%%%%%%%%%%%%%%%%%%
% Fig. 8 - example of 1D t-state in small trianglular dot
%%%%%%%%%%%%%%%%%%%%%%%%%%%%%%%%%%%%%%%%%%%%%%%%%

\begin{figure}  [t]
\centering
	\includegraphics[width=0.48\textwidth]{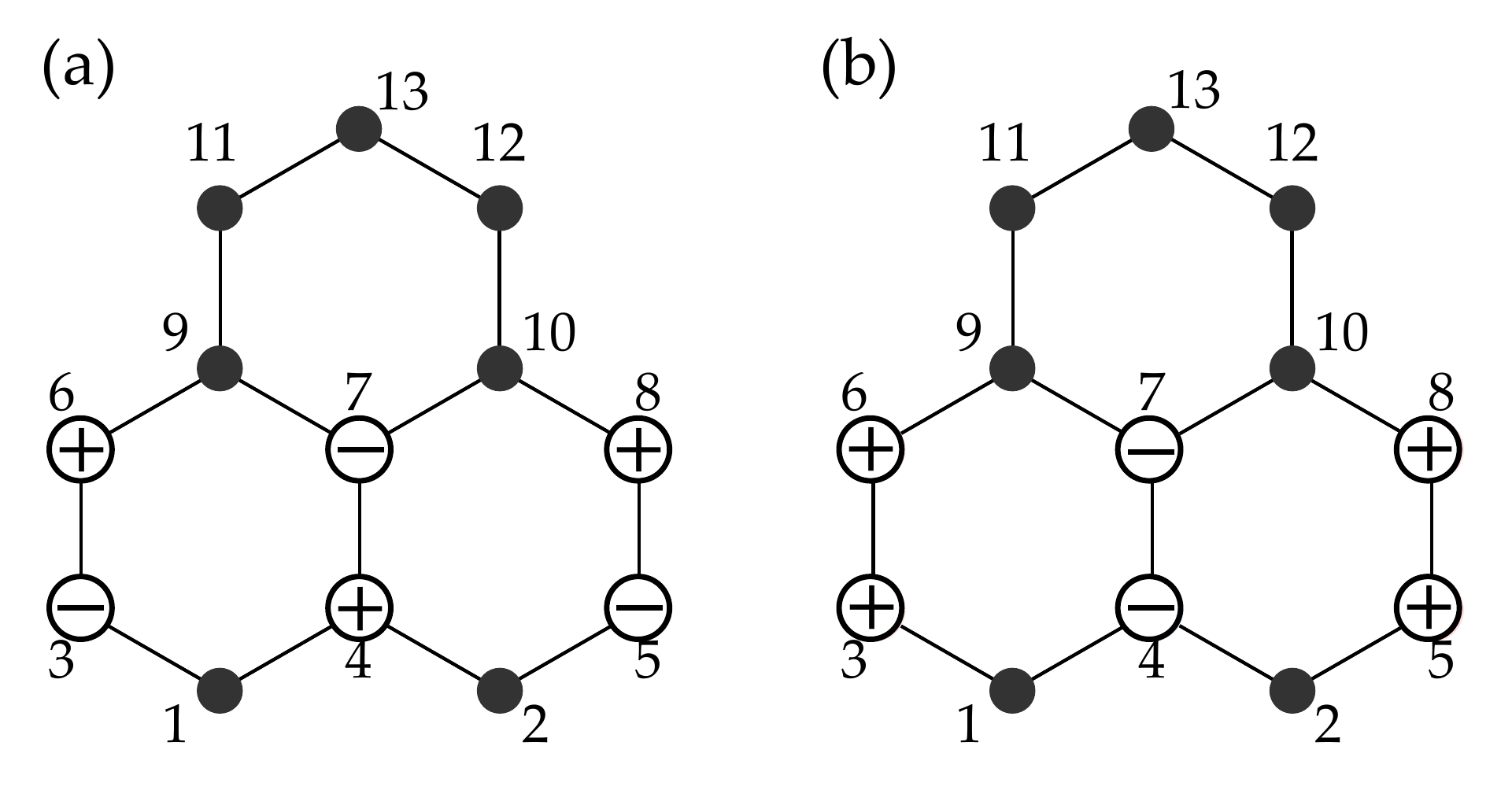}
  	\caption{
	An example of a one--dimensional wave function at energy $\epsilon=\pm t$ in 
	the smallest triangular graphene quantum dot with zigzag edges.
	(a) Anti-binding ($\epsilon = t$) wave function [see Eq. (\ref{eq:1Dwf})], 
	(b) binding ($\epsilon = -t$) wave function. 
	Plus (minus) denotes the probability amplitude of \mbox{$c_{\pm t, i} = +1$}
	\mbox{($c_{\pm t, i} = -1$)} of the wave function on site $i$ [as defined in 
	Eq.~(\ref{eq:1DwfDef})]. 
	For all other sites we find  $c_{\pm t, i} = 0$. 
	Such configurations ensure that hopping outside the highlighted sites vanishes.
	}
\label{fig:1DWF}
\end{figure}

%%%%%%%%%%%%%%%%%%%%%%%%%%%%%%%%%%%%%%%%%%%%%%%%%%

As discussed in Sec.~\ref{sec:Edges}, within the tight--binding model ${\mathcal H}_0$ 
[Eq.~\ref{eq:H0}], all GQD's exhibit states with an exact energy $\epsilon=\pm t$. 
The number of such states depends on the shape and edge type of the GQD and, in most cases, 
scales linearly with the dot size (Table~\ref{tab:scalingbehavior}).
In this section we shed light onto the microscopic nature of these states.
We present an analytical solution for their wave functions and show that these states exhibit an 
one--dimensional (1D) character.
%providing a possible explanation for very high exciton binding energies in graphene. 

%%%%%%%%%%%%%%%%%%%%%%%%%%%%%%%%%%%%%%%%%%%%%%%%%%

The fact that the energy of these states, $\epsilon=\pm t$, coincides with the hopping energy between 
adjacent sites in the tight--binding Hamiltonian [Eq.~(\ref{eq:H0})], motivates us to seek for 
wave functions for which electron hopping takes place on single bonds in the honeycomb lattice.
Fig.~\ref{fig:1DWF} shows an example of such a wave function for a triangular zigzag GQD with 
energy $\epsilon=t$ (anti--binding) and $\epsilon=-t$ (binding).
The wave function $|\psi_t\rangle$ is a linear combination of Wannier functions [Eq.~(\ref{eq:EFHam})],
\begin{equation}
|\psi_t\rangle = \sum_i c_{t,i}|\phi_i\rangle\ ,
\label{eq:1DwfDef}
\end{equation}
where $c_{t,i}$ are appropriately chosen to be $\pm 1$ at the ``$+$'' and ``$-$'' sites respectively.
In the case of Fig.~\ref{fig:1DWF}(a), the wave function is explicitly written as
\begin{equation}
	|\psi_t \rangle \propto  - |\phi_{3} \rangle + |\phi_{6} \rangle + |\phi_{4} \rangle - |\phi_{7} 
	\rangle - |\phi_{5} \rangle + |\phi_{8} \rangle  \, ,
\label{eq:1Dwf}
\end{equation}
which can be directly shown to satisfy $\hat{\mathpzc{H}} |\psi_t \rangle = t ~ |\psi_t \rangle$.
Eq.~(\ref{eq:1Dwf}) has the energy $\epsilon=t$ because electron hopping is non--zero only
between pairs of atoms, highlighted in Fig.~\ref{fig:1DWF}(a). 
Hopping to, e.g. site 1 is suppressed because of the opposite coefficients on 
sites 3 and 4.
In a similar way, we can construct a binding wave function with $\epsilon = -t$ as seen in 
Fig.~\ref{fig:1DWF}(b).	
We note that these wave functions are 1D--like, since they are non--zero only on pairs of sites 
along an 1D ladder.
This is consistent with their 1D scaling behaviour shown in Table~\ref{tab:scalingbehavior}. 

%%%%%%%%%%%%%%%%%%%%%%%%%%%%%%%%%%%%%%%%%%%%%%%%%
% Fig. 9 - example of t-states in dots with different edges 
%%%%%%%%%%%%%%%%%%%%%%%%%%%%%%%%%%%%%%%%%%%%%%%%%

\begin{figure} 
\centering
	\includegraphics[width=0.48 \textwidth]{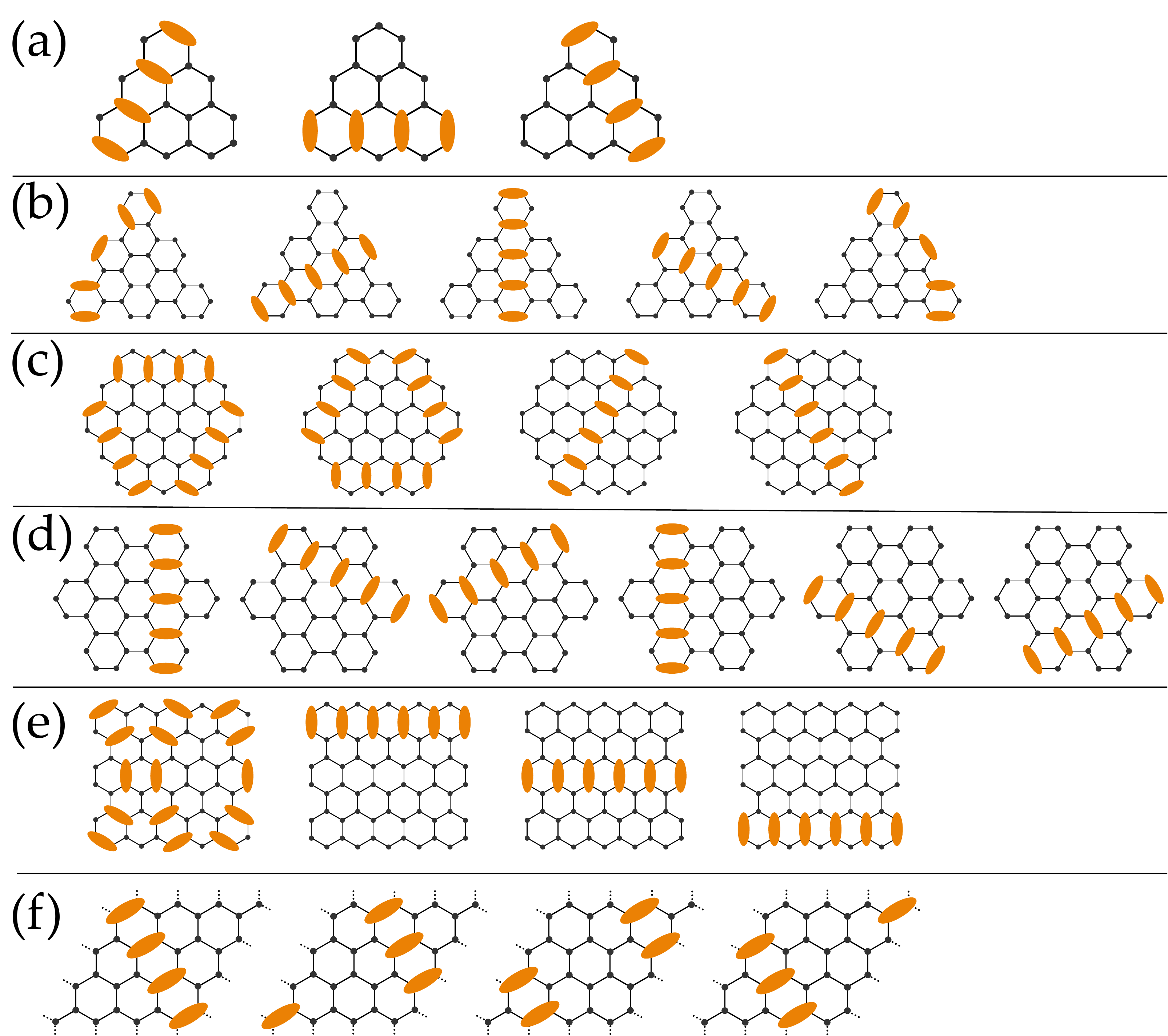}    
  	\caption{	
	Examples for one--dimensional wave functions in graphene quantum dots (GQD's) 
	and bulk graphene.
	(a) Triangular zigzag, (b) triangular armchair, (c) hexagonal zigzag, (d) hexagonal armchair, 
	(e) rectangular and (f) cluster with periodic boundary conditions (graphene).
	Boundary conditions set the number and shape of allowed one--dimensional states [see scaling
	behaviour in Tab.~\ref{tab:scalingbehavior}], which do not necessarily have to proceed in
	a straight line, as seen in the first example of (e).
        }
	\label{fig:1DAll}
\end{figure}

%%%%%%%%%%%%%%%%%%%%%%%%%%%%%%%%%%%%%%%%%%%%%%%%%%%%%%%%

The key for the existence of such states is the presence of a pair of atoms on the edge 
of the GQD for which each atom has one bond less.
This allows the creation of these 1D--ladder states that live on a line of pairs of 
carbon atoms.
Fig.~\ref{fig:1DAll} shows 1D wave functions for all GQD's considered in 
Sec.~\ref{sec:Edges} and for graphene clusters with periodic boundary conditions. 
It is clear that the number of 1D wave functions depends strongly on the 
boundary conditions. 
%
%If the edges do not show a pair of sites connected to just one neighboring 
%site, we do not find exactly 1-dimensional states with energy $\epsilon = \pm t$. 
%
Therefore hexagonal GQD's with armchair edges [Fig.~\ref{fig:1DAll}(d)] favour the 
presence of these states, which exist for all system sizes [see Table~\ref{tab:scalingbehavior}].

%%%%%%%%%%%%%%%%%%%%%%%%%%%%%%%%%%%%%%%%%%%%%%%%%%%%%%%%

We should note that these 1D wave functions do not necessarily exist on a 
straight line, as seen in Fig.~\ref{fig:1DAll}(b), (c) and (e).
In these cases, boundaries allow a bent 1D string of site pairs within the 
dot, leading to the rather complicated scaling functions of these states 
in Table~\ref{tab:scalingbehavior}.
Such 1D wave functions also exist in graphene clusters with periodic 
boundary conditions, as shown in Fig.~\ref{fig:1DAll}(f).

%%%%%%%%%%%%%%%%%%%%%%%%%%%%%%%%%%%%%%%%%%%%%%%%%%%%%%%%

The one--dimensional character of states at \mbox{$\epsilon=\pm t$} provides a 
plausible explanation for the large exciton binding energies of \mbox{ $\sim 500$~meV}, 
found in graphene \cite{Yang2009, Kravets2010, Mak2011, Chae2011, Matkovic2012}, 
since confinement in 1D greatly enhances binding.   
Furthermore, optical resonances in carbon nanotubes are predicted to arise from 
strongly--correlated 1D excitons \cite{Ando1997}.  
Their binding energies have been predicted with \mbox{$\sim 400$~meV} \cite{Wang2005}, 
comparably close to the results for graphene.
A very crude estimate of the local on--site Hubbard $U$ 
has been done by describing these 1D states with a two--site Hubbard model.
Details of this estimate are given in Appendix~\ref{appendix:HubbardU}.

%%%%%%%%%%%%%%%%%%%%%%%%%%%%%%%%%%%%%%%%%%%%%%%%%%%%%%%%

In summary, we have shown that eigenstates with energy \mbox{$\epsilon=\pm t$} 
show a 1D character and can exist in both, GQD's and graphene.
Their 1D nature provides a possible explanation for the unusually high exciton--binding 
energies found at the M--point in graphene \cite{Yang2009, Kravets2010, Mak2011, Chae2011, Matkovic2012}.

%%%%%%%%%%%%%%%%%%%%%%%%%%%%%%%%%%%%%%%%%%%%%%%%%
\section{Disorder in a graphene quantum dot : The role of vacancies and asymmetry}  
%%%%%%%%%%%%%%%%%%%%%%%%%%%%%%%%%%%%%%%%%%%%%%%%%%
\label{sec:Asym_Def}

%%%%%%%%%%%%%%%%%%%%%%%%%%%%%%%%%%%%%%%%%%%%%%%%%%
%   Fig. 10
%%%%%%%%%%%%%%%%%%%%%%%%%%%%%%%%%%%%%%%%%%%%%%%%%%

\begin{figure*}
	\centering
	\includegraphics[width=0.9\textwidth]{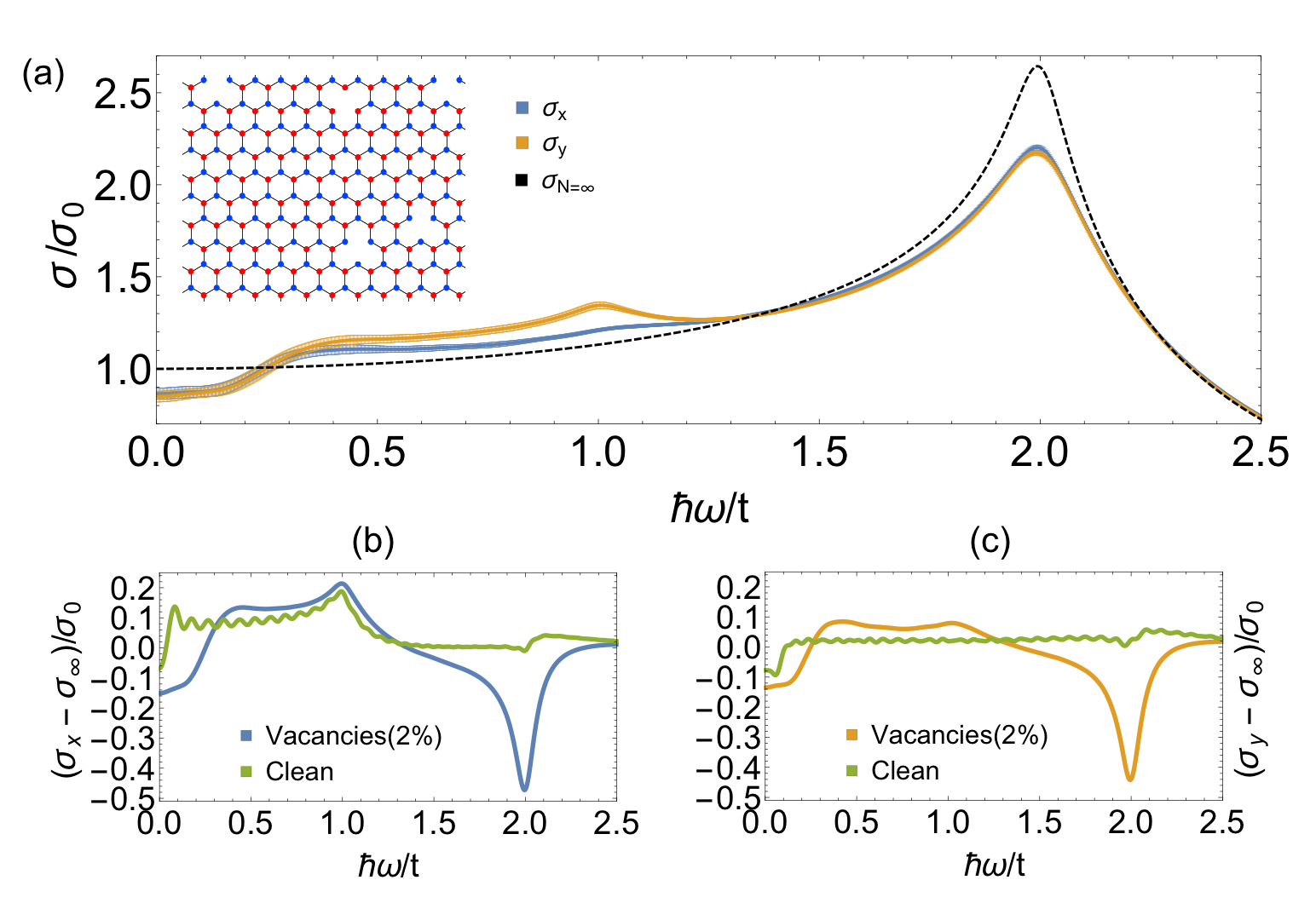}
	\caption{
	  Effect of random site vacancies on the optical conductivity 
	  of a rectangular graphene quantum dot (GQD).
	  (a) Optical conductivity of an ensemble of disordered GQD's, 
	  for $x$--polarized (blue curve) and $y$--polarized (yellow curve) light. 
	  Error bars show the standard deviation of results with the ensemble, while   
	  the dashed line shows equivalent results for an infinite graphene sheet, without vacancies.
	  (b) Difference between optical conductivity of clean and disordered GQD's 
	  and an infinite graphene sheet, for $x$--polarized light.   
          (c) Equivalent results for $y$--polarized light.   
          The disordered GQD shows additional features in the visible spectrum, 
	  while the dominant absorption peak at $\hbar \omega/t = 2$ is reduced.
	  Results were calculated from Eq.~(\ref{eq:sigma.tight.binding}), within a 
	  tight--binding model for a GQD with $N ~\approx10,000$ sites, setting 
	  a Lorentzian of FWHM $2\gamma = 0.1\ \text{t}$.  
	  Disorder--averages were calculated for $2 \%$ of randomly introduced 
	  vacancies, and averaged over $\approx100$ realisations.
	  }
\label{fig:Defects}
\end{figure*}

%%%%%%%%%%%%%%%%%%%%%%%%%%%%%%%%%%%%%%%%%%%%%%%%%%
%   Fig. 11
%%%%%%%%%%%%%%%%%%%%%%%%%%%%%%%%%%%%%%%%%%%%%%%%%%

\begin{figure*}
	\centering
	\includegraphics[width=0.9\textwidth]{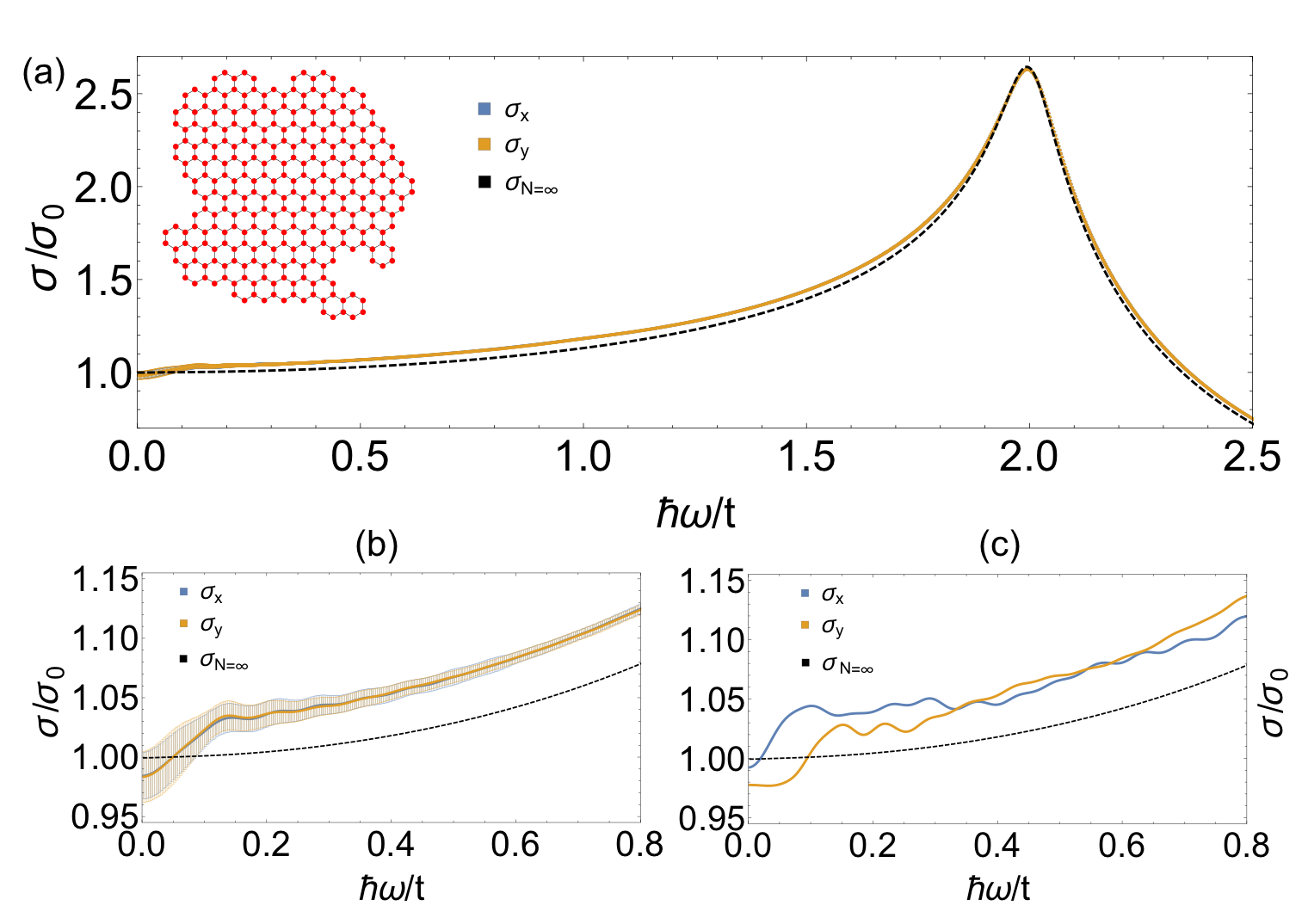}
	\caption{
	  Effect of edge--disorder on the optical conductivity of a graphene quantum dot (GQD).
	  (a) Optical conductivity of an ensemble of disordered GQD's, 
	  for $x$--polarized (blue curve) and $y$--polarized (yellow curve) light. 
	  Both curves lay on top of each other.
	  The dashed line shows equivalent results for an infinite graphene sheet, without disorder.
	  (b) Detail of the optical conductivity of disordered GQD's at low frequencies.   
	  Error bars show the standard deviation within the ensemble.	  
          (c) Optical conductivity of an individual disordered GQD at low frequencies, 
          showing the difference between $x$-- and $y$--polarized light. 
	  Results were calculated from Eq.~(\ref{eq:sigma.tight.binding}), within a 
	  tight--binding model for a GQD with \mbox{$N~\approx10,800$} sites, setting 
	  a Lorentzian of FWHM $2\gamma = 0.1\ \text{t}$.  
	  Disorder--averages were calculated for an ensemble of $\approx100$ realisations.
         }
\label{fig:Asym}
\end{figure*}

%%%%%%%%%%%%%%%%%%%%%%%%%%%%%%%%%%%%%%%%%%%%%%%%%%

Even though GQD's can be produced in predefined regular shapes \cite{Lu2011, Li2011, Mohanty2012, Olle2012, Yan2012}, 
techniques like chemical vapour deposition (CVD) or temperature programmed growth (TPG) result in dots with irregular 
shapes\cite{Coraux2009}.
Furthermore, surface contamination with adatoms during the fabrication process can introduce scattering potentials, 
which change the electronic properties of the GQD's.
In the previous sections, we explained how the geometry of GQD's impacts their optical properties and showed
that the presence of zigzag edges results in an additional peak in the optical conductivity at low frequencies. 
It is therefore useful to examine to which extent these features can survive in the case of irregular shaped dots.

Here we discuss two of such cases: (a) rectangular GQD's with single atom vacancies, and
(b) GQD's with asymmetric shape.
We concentrate on the minimal tight--binding model introduced in 
Section~\ref{sec:tight-binding.model}.

%%%%%%%%%%%%%%%%%%%%%%%%%%%%%%%%%%%%%%%%%%%%%%%%%%

\subsection{The role of vacancies}

%%%%%%%%%%%%%%%%%%%%%%%%%%%%%%%%%%%%%%%%%%%%%%%%%%
% Notes from literature
%%%%%%%%%%%%%%%%%%%%%%%%%%%%%%%%%%%%%%%%%%%%%%%%%%

% In experiments, graphene has always all sorts of disorder or impurities such as holes, ripples, 
% adatoms, etc. \cite{Yuan2010}. 
%
% For the case of graphene, vacancies are prototype examples of the resonant scatterers.21,24 
% \cite{Yuan2010}
% 
% Numerous adatoms  and  admolecules including  the  important  case  of hydrogen atoms 
% covalently bonded with carbon atoms provide other examples 25-27 \cite{Yuan2010}.
%
% A vacancy can be regarded as an atom lattice point with and on-site energy v->inf or with its 
% hopping parameters to  other  sites  being  zero.  
% % 
% In  the  numerical  simulation,  the simplest way to implement a vacancy it to remove the 
% atom at the vacancy site. Introducing vacancies in a graphene sheet will  create  a  zero  energy  
% modes midgap  state.19,31,32 The exact analytical wave function associated with the zero mode 
% induced  by  a  single  vacancy  in  a  graphene  sheet  was  obtained  in  Ref.33,  showing  a  
% quasilocalized  character  with the amplitude of the wave-function decaying as inverse 
% distance to the vacancy.  
% %
% Graphene with a finite concentration of vacancies was studied numerically in Ref.31. 
% % 
% The number of the midgap states increases with the concentration of the vacancies 
% \cite{Yuan2010}

%%%%%%%%%%%%%%%%%%%%%%%%%%%%%%%%%%%%%%%%%%%%%%%%%%

The role of vacancies on the electronic properties of graphene and its nanostructures 
has been extensively studied \cite{Pereira2006, Peres2006, Ostrovsky2014, Sanyal2016}.
Moreover, it has been shown that metallic adatoms on the graphene's surface --- a common 
source of contamination during fabrication --- can be treated theoretically as an atomic vacancy
on the graphene lattice in the case of strong local scattering potentials. \cite{Yuan2010}

%%%%%%%%%%%%%%%%%%%%%%%%%%%%%%%%%%%%%%%%%%%%%%%%%%

Here, we discuss the effect of vacancies on the optical conductivity of rectangular GQD's.
As explained in Sec.~\ref{sec:Edges}, for clean, vacancy--free dots, 
optical excitation, polarized along the zigzag edges captures signatures of zero--energy, edge modes
in the form of a peak at \mbox{$\hbar \omega = t$}.
This peak is absent for polarization along the armchair edges, allowing a distinction between 
different edge types. 
We now calculate the optical conductivity for dots with a number of randomly introduced vacancies 
on the lattice.
In our model, we set electron hopping to/from the vacancy sites to zero, which is equivalent to an
infinite on--site energy at the vacancies.

%%%%%%%%%%%%%%%%%%%%%%%%%%%%%%%%%%%%%%%%%%%%%%%%%%

Fig.~\ref{fig:Defects}(a) shows the optical conductivities \mbox{$\sigma_x(\omega)$} and
\mbox{$\sigma_y(\omega)$} for rectangular GQD's of size \mbox{$N ~\approx10,000$} 
and \mbox{$2 \%$} vacancies.
In comparison with the graphene limit, \mbox{$\sigma_{N=\infty}(\omega)$}
[Eq.~(\ref{eq:conductivityFinal_energySpace})], both polarization directions show 
an enhanced shoulder in the visible region of the spectrum and a decrease of intensity 
at \mbox{$\hbar \omega/t = 2$}.

%%%%%%%%%%%%%%%%%%%%%%%%%%%%%%%%%%%%%%%%%%%%%%%%%%%%%%%%

This is more clearly seen in Fig.~\ref{fig:Defects}(b) and (c) where the difference 
$\sigma(\omega) - \sigma_{N=\infty}(\omega)$ for both clean GQD's and dots with $2\%$ 
vacancies is plotted.
The dominant peak at \mbox{$\hbar \omega/t = 2$} is significantly reduced by the presence 
of vacancies.
Similarly, for \mbox{$\hbar\omega/t \approx 1$}, the vacancies result in an enhanced 
shoulder for $\sigma_y$, which is absent in the clean GQD.
For $\sigma_x$, this shoulder is only slightly larger than the one already present for 
vacancy--free dots. 

%%%%%%%%%%%%%%%%%%%%%%%%%%%%%%%%%%%%%%%%%%%%%%%%%%%%%%%%

These features can be explained by the fact that every single vacancy on the graphene lattice 
creates a zigzag edge around it, which can be seen as an inverse zigzag triangular dot.
Therefore additional zero--energy states can be formed along these edges, which allow scattering to
the highly degenerate states in the vicinity of the \mbox{Van Hove} singularity and result in the
enhanced shoulder in the visible region of the spectrum.

%%%%%%%%%%%%%%%%%%%%%%%%%%%%%%%%%%%%%%%%%%%%%%%%%%%%%%%%

On the other hand, the presence of vacancies destroys some of 1D--wave functions with energy 
$\epsilon=\pm t$ [see Sec.~\ref{sec:1-dimWaveFunction}].
Consequently, the dominant absorption peak at $\hbar \omega = 2t$, which is created by 
transitions between such states, is reduced.

%%%%%%%%%%%%%%%%%%%%%%%%%%%%%%%%%%%%%%%%%%%%%%%%%%%%%%%%

Our findings are consistent with previous results showing that vacancies in graphene result in the formation 
of localised states \cite{Pereira2006}.
Also, calculations in disordered graphene show that it exhibits mid--gap states in the density of states 
\cite{Ugeda2011, Ugeda2010} and an additional peak in the optical conductivity \cite{Yuan2011b}.

%%%%%%%%%%%%%%%%%%%%%%%%%%%%%%%%%%%%%%%%%%%%%%%%%%%%%%%%

\subsection{The role of asymmetry }
Asymmetry is a certain issue in the fabrication of GQD's.
Techniques like chemical vapour deposition (CVD) and temperature programmed growth (TPG) 
produce GQD's of various sizes and shapes \cite{Coraux2009}.
Even techniques that can create GQD's with a predefined shape are not free of errors 
\cite{Lu2011, Li2011, Mohanty2012, Olle2012, Yan2012}.

%%%%%%%%%%%%%%%%%%%%%%%%%%%%%%%%%%%%%%%%%%%%%%%%%%%%%%%%

Here we address this case, by calculating the optical conductivity $\sigma_{\alpha}(\omega)$ for asymmetric 
dots, showing a random mixture of armchair and zigzag edges. 

%%%%%%%%%%%%%%%%%%%%%%%%%%%%%%%%%%%%%%%%%%%%%%%%%%%%%%%%

Fig.~\ref{fig:Asym}(a) shows the optical conductivity $\sigma_{\alpha}(\omega)$ for $x$-- and $y$--polarized  
light for asymmetric GQD's of size $N \approx 10,800$, averaged over $\approx100$ dots.
The mean value of the optical conductivity is very close to the infinite graphene limit, since edge--effects 
are averaged out. 
The small offset from the graphene limit is due to finite--size effects.

%%%%%%%%%%%%%%%%%%%%%%%%%%%%%%%%%%%%%%%%%%%%%%%%%%%%%%%%

The effect of averaging is shown more clearly in Figs.~\ref{fig:Asym}(b) and (c).
Even though on average, $\sigma_x$ and $\sigma_y$ coincide [Fig.~\ref{fig:Asym}(b)], for each 
individual dot they do not [Fig.~\ref{fig:Asym}(c)], due to the absence of symmetry.
%

%%%%%%%%%%%%%%%%%%%%%%%%%%%%%%%%%%%%%%%%%%%%%%%%%
\section{Conclusions}    
%%%%%%%%%%%%%%%%%%%%%%%%%%%%%%%%%%%%%%%
\label{sec:conclusions}

The discovery of Graphene, more than 10 years ago 
\cite{Novoselov2004, Novoselov2005}, has sparked a rennaisance 
in the study of two--dimensional materials, and their potential 
technological applications \cite{Avouris2012, Luo2013}.
Graphene quantum dots (GQD's), offer yet another new opportunity, 
to control the properties of a graphene sheet by restricting its size 
and shape \cite{Silva2010, GrapheneQuantumDots}.  
In particular, the ability to control the optical properties of GQD's has 
potential applications in fields ranging from quantum computation to 
solar energy\cite{Zhu2011, Jin2013, Roy2015, Umrao2015, Luk2012,
Son2012, Konstantatos2012, Zhang2015, QuantumSolarCells}.
However, tailoring the properties of a GQD to a specific application 
requires the ability to fabricate dots with the desired shape, or to 
post--select for dots with a given shape after fabrication.
In either case, understanding the relationship between the size and shape of 
the dot and its physical properties is paramount.

%%%%%%%%%%%%%%%%%%%%%%%%%%%%%%%%%%%%%%%

In this Article, we have explored the role that size, shape, edge--type 
and atomic vacancies play in the optical conductivity of GQD's.
Using group theory, we determined the optical selection
rules, which follow from the symmetry of a regular--shaped GQD.
We find that the optical response is {\it independent} of the polarization
of the incident light in GQD's of symmetry, where the 
in--plane ($x, y$) components of the current operator transform under 
the same irrep.
This has been shown on the example of triangular and hexagonal GQD's
[cf. Fig.~\ref{fig:SymTri_Mir}, Fig.~\ref{fig:SymTri_Rot}].
Meanwhile the optical conductivity is {\it polarization--dependent} 
in GQD's of symmetry, where the in--plane ($x, y$) components of 
the current operator transform under different irreps, as shown 
on the example of rectangular GQD's [cf. Fig.~\ref{fig:SymRec}].

%We find that the optical response is {\it independent} of the polarization
%of the incident light in triangular or hexagonal GQD's, which have a 
%non--abelian point--group symmetry 
%[cf. Fig.~\ref{fig:SymTri_Mir}, Fig.~\ref{fig:SymTri_Rot}]. 
%%
%Meanwhile the optical conductivity is {\it polarization--dependent} 
%in rectangular GQD's, which have an abelian point--group 
%symmetry [cf. Fig.~\ref{fig:SymRec}].

%%%%%%%%%%%%%%%%%%%%%%%%%%%%%%%%%%%%%%%

We have also explored the optical conductivity $\sigma_{\alpha}(\omega)$ of GQD's 
within a simple tight--binding model [Eq.~(\ref{eq:H0})], 
known to give a good account of the properties of bulk graphene 
[\onlinecite{CastroNeto2009}].
For small GQD's, $\sigma_{\alpha}(\omega)$ depends strongly on
the type of the considered dot, and has many non--universal features
[cf. Fig.~\ref{fig:DosCondALL}].   
These features evolve with size, and the optical response of GQD's 
of intermediate size ($N \gtrsim 5,000$ sites, $L \gtrsim 20$nm) 
has much in common with the response of bulk graphene.
In particular this shows a strong peak for UV light, as observed 
in graphene.
None the less, for dots of this size, $\sigma_{\alpha}(\omega)$ still retains 
tell--tale features which provide important information about 
edge--geometry of the dot.
In particular, we find an additional peak in the optical conductivity 
at the UV end of the visible spectrum in GQD's with zigzag edges 
[cf. Fig.~\ref{fig:SizeEvo}]. 

%%%%%%%%%%%%%%%%%%%%%%%%%%%%%%%%%%%%%%%

Within a tight--binding model, both of these peaks in the optical 
conductivity are intimately connected with the existence of states with 
energy \mbox{$\epsilon = \pm t$}.
We have explored the nature of the wave functions of these states
in different--shaped GQD's, and find that they have a one--dimensional 
character [cf. Fig.~\ref{fig:1DWF} and Fig.~\ref{fig:1DAll} ].  
Equivalent one--dimensional states also exist for clusters with 
periodic boundary conditions, where they occur at the M--point 
in the Brillouin zone [i.e. ${\bf k_1} = (\frac{2 \pi}{3 \sqrt{3}} , \frac{2\pi}{3})$ and 
${\bf k_2} = (\frac{4 \pi}{3 \sqrt{3}} , 0)$], 
and are associated with \mbox{Van Hove} singularities 
in the single--particle density of states.  
The one--dimensionality of these wave functions provides
a very natural explanation for large binding energies of excitons
formed of particles and holes near the M--point 
\cite{Yang2009, Kravets2010, Mak2011, Chae2011, Matkovic2012}.

%%%%%%%%%%%%%%%%%%%%%%%%%%%%%%%%%%%%%%%

Finally, we discussed the effect of atomic vacancies and  shape--asymmetry
in the optical response of GQD's. 
We showed that atomic vacancies in the lattice enhance the peak in 
the optical conductivity arising from zigzag edges [cf. Fig.~\ref{fig:Defects}].  
This is a signature of additional localization around the vacancy sites.
In the case of completely asymmetric GQD's, the optical conductivity is 
polarization--dependent, although those effects may not be measurable for large 
distributions of randomly shaped dots [cf. Fig.~\ref{fig:Asym}].

%%%%%%%%%%%%%%%%%%%%%%%%%%%%%%%%%%%%%%%

An important open question for future studies of the optical properties
of GQD's is the effect of electron--electron interactions.
There is already a substantial literature on the effect of interaction 
in bulk graphene, where the fact that electrons are restricted to 
two dimensions, and have a Dirac--like dispersion, leads to many 
departures from conventional Fermi--liquid behaviour~\cite{Kotov2012}. 
%especially in the response to impurities and disorder~\cite{Kotov2012}.
%
Given this, it is reasonable to ask how the optical properties of 
GQD's might change, if interactions were included ?

%%%%%%%%%%%%%%%%%%%%%%%%%%%%%%%%%%%%%%%

The optical selection rules derived in Section~\ref{sec:Geometry} 
follow from symmetry alone [cf. Section~\ref{sec:general.theory}].   
For this reason they apply equally to any GQD 
with a given symmetry, regardless of whether it is described 
by a simple tight--binding model [Section~\ref{sec:tight-binding.model}], 
or a more general model of interacting electrons which respects
the symmetries of the dot.
And, while it is possible that interactions could drive changes in 
the symmetry of an infinite graphene sheet \cite{Kotov2012}, 
such spontaneous symmetry--breaking is not expected in a finite--size 
GQD \cite{plischke-book,huang-book}.

%%%%%%%%%%%%%%%%%%%%%%%%%%%%%%%%%%%%%%%

None the less, interactions are known to have a profound effect 
on electrons on the edges of a graphene sheet \cite{Kotov2012}.
And, since the optical response of a GQD comprises a discrete set 
of peaks, even small changes in individual energy levels coming from 
interactions will be directly visible in $\sigma_\alpha (\omega)$.
Moreover, the precursors of any bulk symmetry--breaking 
may also manifest themselves in the spectrum of a finite--size GQD, 
in much the same way as they do in the finite--size spectra of interacting 
quantum spin models \cite{Bernu1994}.
For all of these reasons, we anticipate that interactions will have a
significant effect on many of the optical properties of GQD's.

%%%%%%%%%%%%%%%%%%%%%%%%%%%%%%%%%%%%%%%

In small GQD's, magnetic effects are likely to be important.
In this case, interactions can generate local moments at the edges of 
of a dot \cite{Fujita1996, fernandez-rossier07, Feldner2010, schmidt10, luitz11, Sasaki2008, Kotov2012, Guclu2013}.   
In addition, interactions will split many of the individual peaks found 
in non-interacting calculations, where the electrons' spin plays no role 
[Section~\ref{sec:tight-binding.model}].  
The resulting optical conductivity $\sigma_\alpha(\omega)$ could, in 
principle, be calculated from Eq.~(\ref{eq:general.result.sigma}), by 
writing the interacting Hamiltonian $\hat{\mathpzc{H}}$ as a matrix 
and diagonalising this numerically.
In this way, it is possible to determine both the many--electron eigenstates 
$ | \Psi_n \rangle $ [cf. Eq.~(\ref{eq:eigenstates.H})], 
and optical matrix elements $\mathpzc{J}^\alpha_{nm}$ 
[cf. Eq.~(\ref{eq:many.electron.matrix.element})].
Such exact--diagonalisation approaches have already been used to 
study edge--magnetism in GQD's and nano--ribbons 
\cite{Feldner2010,luitz11,Guclu2013}.
However the exponential growth in the size of the Hilbert space, 
and the need to determine eigenstates and matrix elements spanning 
a wide range of energies, will limit exact--diagonalisation studies 
of $\sigma_\alpha(\omega)$ to dots with a very small number of electrons.
Moreover, care must always be taken that the current operator 
$\hat{\mathpzc{J}}^\alpha$ [Eq.~(\ref{eq:many.electron.matrix.element})]
is defined in a way which is consistent with the 
Hamiltonian used \cite{Cabib1973, Kuzemsky2011}.

%%%%%%%%%%%%%%%%%%%%%%%%%%%%%%%%%%%%%%%

In larger GQD's, where the optical response is a smooth function of 
frequency, interactions may make themselves felt in more subtle ways.
%
%None the less, interactions are still expected to play an important 
%role \cite{Kotov2012}, and 
%
One area where they can have a profound effect is in the 
renormalisation of energy scales.
A prototype for this is provided by the strong peak in the optical response 
of bulk graphene in the UV spectrum, at \mbox{$\hbar \omega \sim 4.7$~eV} \cite{Eberlein2008, Mak2011}.  
Within a simple %(non--interacting) 
tight--binding model [cf. Eq.~(\ref{eq:H0})] 
this peak reflects transitions between single--electron states with 
energy $E \sim \pm t$, and with the usual parameterisation, 
$t = 2.8$~eV [\onlinecite{CastroNeto2009}], the peak would 
be expected to occur for \mbox{$\hbar \omega = 5.6$~eV}.   
However in bulk graphene the particle--hole pairs associated with 
this peak can be viewed as excitons, and interactions lead to a finite 
binding--energy, shifting the peak to lower energies~ \cite{Yang2009, Kravets2010,
Mak2011, Chae2011, Matkovic2012}.
The same should be true for the equivalent, ``$2t$'' peak 
in GQD's of intermediate to large size [cf. Fig.~\ref{fig:DosCondALL}], 
with the added feature that the one--dimensional character of the 
associated wave functions will enhance correlation effects 
[cf. Fig.~\ref{fig:1DWF}].
We also anticipate that interactions will lead to a shift 
in the peak at \mbox{$\hbar \omega \sim t$}, observed for 
GQD's with zigzag edges and/or vacancies [cf. Fig.~\ref{fig:SizeEvo}, 
Fig.~\ref{fig:Defects}].
This expectation remains to be verified, but we hope that the results
in this Article can provide a useful starting point for future studies.
}

%%%%%%%%%%%%%%%%%%%%%%%%%%%%%%%%%%%%%%%%%%%%%%%%%%

And for the time being, perhaps the most exciting prospect is the measurement of 
the optical response of GQD's in experiment.
Given that GQD's with regular and irregular shapes are now available 
[cf. e.g. Ref.~\onlinecite{Yan2012, Olle2012}], this seems a very real possibility.

%%%%%%%%%%%%%%%%%%%%%%%%%%%%%%%%%%%%%%%%%%%%%%%%%%
\begin{acknowledgments}
%%%%%%%%%%%%%%%%%%%%%%%%%%%%%%%%%%%%%%%%%%%%%%%%%%

\noindent
The authors are indebted to Judit Romhanyi for a critical reading of the manuscript, 
and helpful suggestions about symmetry analysis.
This work was supported by the Theory of Quantum Matter Unit of the Okinawa Institute 
of Science and Technology Graduate University. 
E.K. was partially supported by the European Union, Seventh Framework Programme 
(FP7-REGPOT-2012-2013-1) under grant agreement 316165.

%%%%%%%%%%%%%%%%%%%%%%%%%%%%%%%%%%%%%%%%%%%%%%%%%%
\end{acknowledgments}
%%%%%%%%%%%%%%%%%%%%%%%%%%%%%%%%%%%%%%%%%%%%%%%%%%

%%%%%%%%%%%%%%%%%%%%%%%%%%%%%%%%%%%%%%%%%%%%%%%%%%
% APPENDICES
%%%%%%%%%%%%%%%%%%%%%%%%%%%%%%%%%%%%%%%%%%%%%%%%%%

\appendix        
\label{sec:appendix}

%%%%%%%%%%%%%%%%%%%%%%%%%%%%%%%%%%%%%%%%%%%%%%%%%%
\section{Optical selection rules for hexagonal dots}				 
%%%%%%%%%%%%%%%%%%%%%%%%%%%%%%%%%%%%%%%%%%%%%%%%%%
\label{appendix:SymHex}

%%%%%%%%%%%%%%%%%%%%%%%%%%%%%%%%%%%%%%%%%%%%%%%%
\subsection{Group--theory Analysis}  
%%%%%%%%%%%%%%%%%%%%%%%%%%%%%%%%%%%%%%%%%%%%%%%%%%
\label{eq:group.theory.C6v}

%%%%%%%%%%%%%%%%%%%%%%%%%%%%%%%%%%%%%%%%%%%%%%%%%%
% Table VI
%%%%%%%%%%%%%%%%%%%%%%%%%%%%%%%%%%%%%%%%%%%%%%%%%%

\begin{table} [h]
\centering
	\begin{tabular}{ | c | c | c | c | c | c | c | c | p{1cm} | }     
   	\hline
  		$C_{6v}$ 	& $E$ & $2C_6(z)$ & $2C_3(z)$ & $C_2(z)$ & $3\sigma_v$ & $3\sigma_d$ & polar \\ 
				& 	& 	& 	& 	&	&	& vectors	\\ \hline
  		$A_1$ 	& 1 	& 1 	& 1 	&  1 	& 1 	& 1 	& z		\\ \hline
  		$A_2$ 	& 1 	& 1 	& 1 	&  1 	& -1 	& -1	&   		\\ \hline
   		$B_1$ 	& 1 	& -1 	& 1 	& -1 	& 1 	& -1	& 		\\ \hline
		$B_2$ 	& 1 	& -1 	& 1 	& -1 	& -1 	& 1 	&		\\ \hline
		$E_1$ 	& 2 	& 1  	& -1 	& -2 	& 0 	& 0 	& (x, y) 	\\ \hline
		$E_2$ 	& 2 	& -1 	& -1 	& 2  	& 0 	& 0 	&		\\ \hline
  	\end{tabular}
	\caption{
   	Character Table of the point--group $C_{6v}$, describing the 
    	symmetry of hexagonal graphene quantum dots (GQD's) of the type shown in 
    	Fig.~\ref{fig:lattice}(c) and Fig.~\ref{fig:lattice}(d). 
    	Eigenstates of a hexagonal GQD transform with irreducible representations 
    	(irreps) $A_1$, $A_2$, $B_1$, $B_2$, $E_1$ and $E_2$, 
    	while the $x$-- and $y$--components of the current operator $\hat{\mathpzc{J}}$ 
    	(a polar vector) transform with $E_1$ 
    	[cf. \onlinecite{Weyl,Heine,LandauLifshitz,Tinkham,Wagner,Jones}].
	The corresponding symmetry operations are the identity ($E$), 
	$2 \times \frac{2 \pi}{6}$ ($2C_6$), $2 \times \frac{2 \pi}{3}$ ($2C_3$), and one 
	$\pi$--rotation ($C_2$) about the principal axes,
	and 3 reflections on symmetry axes ($\sigma_v$ and $\sigma_{d}$), as 
   	shown in Fig.~\ref{fig:lattice}(c) and (d).
	}
\label{tab:CharacterTableHexagon}
\end{table}

%%%%%%%%%%%%%%%%%%%%%%%%%%%%%%%%%%%%%%%%%%%%%%%%%%
% Fig. 12
%%%%%%%%%%%%%%%%%%%%%%%%%%%%%%%%%%%%%%%%%%%%%%%%%%

\begin{figure*}  [t]
  	\centering
  		\includegraphics[width=0.6\textwidth]{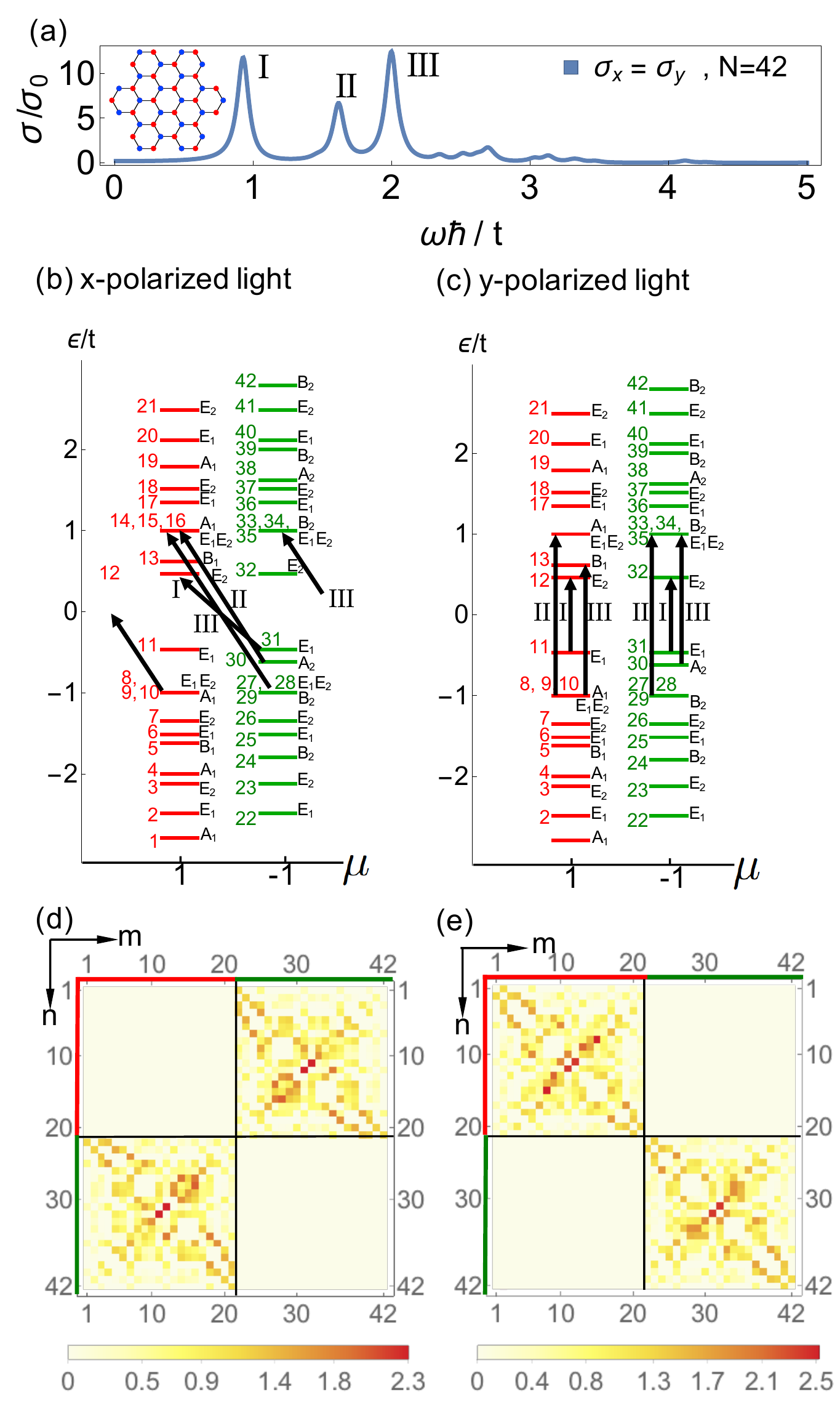}    
	\caption{
	Optical selection rules for the smallest possible hexagonal graphene quantum dot (GQD) 
	with armchair edges ($N=42$), in linearly--polarized light.
      	(a) Optical conductivity $\sigma_{\alpha}(\omega)$, showing the equivalence of results for 
       	$x$-- and $y$--polarized light.    
	(b) and (c) Spectrum of the corresponding tight--binding model [Eq.~(\ref{eq:H0})], 
	in the mirror basis Eq.~(\ref{eq:mirror}), showing the different 
	allowed transitions for $x$-- and $y$--polarized light.  
        (d) and (e) Matrix elements of the corresponding current operators 
      	$|\hat{\mathpzc{J}}_{nm}^{x}|^2$ and  $|\hat{\mathpzc{J}}_{nm}^{y}|^2$ 
	[cf.~Eq.~(\ref{eq:single.electron.matrix.element})], 
        	in units of $(e t / \hbar)^2$.  
        	Results for $\sigma_{\alpha}(\omega)$ were calculated from Eq.~(\ref{eq:sigma.tight.binding}), 
        	with a Lorentzian of FWHM $2\gamma = 0.1\ \text{t}$.
        Eigenstates are labelled according to their quantum number $n=1\ldots 42$, 
        	eigenvalue $\mu_n = \pm 1$ [Eq.~(\ref{eq:mirror})] and corresponding irrep 
        [cf. Table~\ref{tab:CharacterTableHexagon}]
	}
	\label{fig:SymHexMir}
\end{figure*}

%%%%%%%%%%%%%%%%%%%%%%%%%%%%%%%%%%%%%%%%%%%%%%%%%%
% Fig. 13
%%%%%%%%%%%%%%%%%%%%%%%%%%%%%%%%%%%%%%%%%%%%%%%%%%

\begin{figure*} [t]
 	\centering
  	\includegraphics[width=0.6\textwidth]{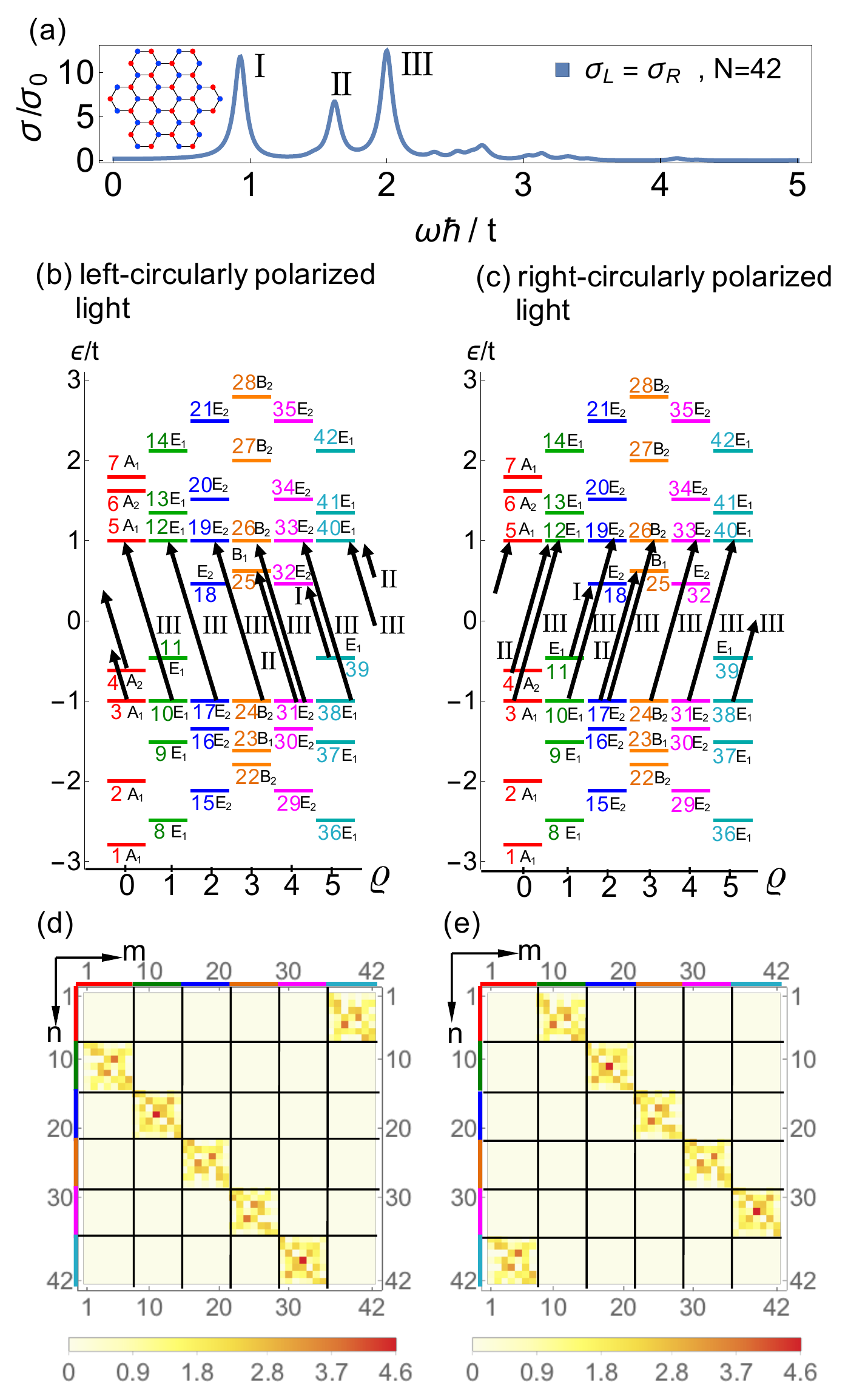}    
	\caption{
	Optical selection rules for the smallest possible hexagonal graphene quantum dot (GQD) 
	with armchair edges ($N=42$), in circularly--polarized light.
	(a) Optical conductivity $\sigma_{\alpha}(\omega)$, showing equivalence of results for 
	left-- and right-- circularly polarized light.
	(b) and (c) Spectrum of the corresponding tight--binding model Eq.~(\ref{eq:H0}),
	in the rotation basis Eq.~(\ref{eq:RotationOperator}), showing the different allowed transitions
	for left-- and right-- circularly polarized light. 
	(d) and (e) Matrix elements of the corresponding current operators $|\hat{\mathpzc{J}}_{nm}^{L}|^2$
	and $|\hat{\mathpzc{J}}_{nm}^{R}|^2$ [cf.~Eq.~(\ref{eq:single.electron.matrix.element})], 
	in units of $(e t / \hbar)^2$.
	Results for $\sigma_{\alpha}(\omega)$ were calculated from Eq.~(\ref{eq:sigma.tight.binding}), 
	with a Lorentzian of FWHM $2\gamma = 0.1\ \text{t}$.
	Eigenstates are labelled according to their quantum number $n=1\ldots 42$, 
    	eigenvalue $\mu_n  = e^{i\frac{2\pi}{6} \rho_n} \; , \; \rho_n = 0,1,2,3,4,5$ 
	[cf. Eq.~(\ref{eq:EV_RotationOperator})] and corresponding irrep 
	[Table~\ref{tab:CharacterTableHexagon}]. 
	}
\label{fig:SymHexRot}
\end{figure*}

%%%%%%%%%%%%%%%%%%%%%%%%%%%%%%%%%%%%%%%%%%%%%%%%%%

In Section~\ref{sec:SymTri}, we showed for a triangular GQD, that the optical conductivity 
is polarization--independent.
The given derivation is valid for any GQD where the in--plane ($x, y)$
components of the current operator transform under the same irrep.
And this is also the case for the point--group $C_{6v}$ describing an hexagonal GQD.

%
%It turns out, that this property is very generic for any type of dot with non--abelian point--group 
%symmetry, therefore also for hexagonal GQD's.

%%%%%%%%%%%%%%%%%%%%%%%%%%%%%%%%%%%%%%%%%%%%%%%%%%%

The symmetry analysis for a hexagonal GQD follows the same concept 
as shown in Section~\ref{sec:group.theory.triangular.GQD}, 
with the difference, that we must work with point--group symmetries 
of a hexagon, $C_{6v}$, as listed in Table~\ref{tab:CharacterTableHexagon}.
The group $C_{6v}$ comprises the identity ($E$), $2 \times \frac{2 \pi}{6}$ ($2C_6$), 
$2 \times \frac{2 \pi}{3}$ ($2C_3$), and one $\pi$--rotation ($C_2$) about the 
principal axes, and 3 reflections on symmetry axes ($\sigma_v$ and $\sigma_{d}$), 
as shown in Fig.~\ref{fig:lattice}(c) and (d).
Eigenstates of a hexagonal GQD transform with irreducible representations 
(irreps) $A_1$, $A_2$, $B_1$, $B_2$, $E_1$ and $E_2$, while the $x$-- and 
$y$--components of the current operator $\hat{\mathpzc{J}}$ (a polar vector) 
transform with $E_1$ 
[cf. \onlinecite{Weyl,Heine,LandauLifshitz,Tinkham,Wagner,Jones}].  
As explained for triangular GQD's, this is the reason for a polarisation--independent 
optical conductivity $\sigma_{\alpha}(\omega)$. 

%%%%%%%%%%%%%%%%%%%%%%%%%%%%%%%%%%%%%%%%%%%%%%%%%%%

Following the steps of Section~\ref{sec:group.theory.triangular.GQD}, by using the 
character Table~\ref{tab:CharacterTableHexagon}, one can resolve the optical selection 
rules for hexagonal GQD's by applying the general rules for the products of representations
\cite{Weyl,Heine,LandauLifshitz,Tinkham,Wagner,Jones}.  

\begin{center}
	\begin{tabular}{c c c l}
	$A_1$ 	& $\times$	& $E_1$	&	$\rightarrow E_1$	\; ,	\\
	$A_2$ 	& $\times$	& $E_1$	&	$\rightarrow E_1$	\; ,	\\
	$B_1$ 	& $\times$	& $E_1$	&	$\rightarrow E_2$	\; ,	\\
	$B_2$ 	& $\times$	& $E_1$	&	$\rightarrow E_2$	\; ,	\\
	$E_1$ 	& $\times$	& $E_1$	&	$\rightarrow A_1 + A_2 + E_2$	\; ,	\\
	$E_2$ 	& $\times$	& $E_1$	&	$\rightarrow B_1 + B_2 + E_1$	\; ,	\\
	\end{tabular}
\end{center}
This analysis leads to the optical selection rules
\begin{eqnarray}	
	A_1 \longleftrightarrow E_1 \ , \quad A_2 \longleftrightarrow E_1 \ , 
			\quad  E_2 \longleftrightarrow E_1 \ , \nonumber \\
	B_1 \longleftrightarrow E_2 \ , \quad B_2 \longleftrightarrow E_2 \ , 
			\quad  E_1 \longleftrightarrow E_2 \ ,
    	\label{eq:SelRulesHex} 
\end{eqnarray}
which are explicitly independent of polarisation.

%%%%%%%%%%%%%%%%%%%%%%%%%%%%%%%%%%%%%%%%%%%%%%%%
\subsection{Illustration of optical selection rules for linearly and circularly--polarised light}  
%%%%%%%%%%%%%%%%%%%%%%%%%%%%%%%%%%%%%%%%%%%%%%%%%%

The optical conductivity $\sigma_\alpha(\omega)$ of hexagonal GQD 
can be calculated explicitly using the non--interacting tight--binding model 
introduced in Section~\ref{sec:tight-binding.model}.
The result for linearly--polarised light incident on an hexagonal GQD 
of size $N = 42$ sites is shown in Fig.~\ref{fig:SymHexMir}(a).
As expected, the result is independent of polarisation, and is dominated
by peaks associated at three different values of $\omega$, labelled I---III.
The corresponding optical transitions, for x-- and y--polarised light, 
are identified in Fig.~\ref{fig:SymHexMir}(b) and Fig.~\ref{fig:SymHexMir}(c), 
where states have been labelled according to their irreps, 
and further classified according to their eigenvalues under the reflection 
operator $\hat{\mathpzc{M}_y}$ [Eq.~\ref{eq:mirror}].
All optical transitions satisfy the selection rules given in 
Eq.~(\ref{eq:SelRulesHex}).   
The corresponding matrix elements are illustrated in 
Fig.~\ref{fig:SymHexMir}(d) and (e).

%%%%%%%%%%%%%%%%%%%%%%%%%%%%%%%%%%%%%%%%%%%%%%%%%%

Equivalent results for circularly--polarised light are shown in Fig.~\ref{fig:SymHexRot}.
The result for $\sigma_\alpha(\omega)$, shown in Fig.~\ref{fig:SymHexRot}(a) 
is independent of polarisation, and identical to that found for linearly--polarised light 
[Fig.~\ref{fig:SymHexMir}(a)].
The corresponding optical transitions, for left-- and right--polarised light, 
are shown in Fig.~\ref{fig:SymHexRot}(d) and Fig.~\ref{fig:SymHexRot}(e), 
where states have been labelled according to their irreps, 
and further classified according to their eigenvalues under the rotation operator
$\hat{\mathpzc{R}}_{2\pi/6}$ [cf. Eq.~\ref{eq:RotationOperator}].
Once again, all transitions satisfy the selection rules given in 
Eq.~(\ref{eq:SelRulesHex}).  
The corresponding matrix elements are illustrated in Fig.~\ref{fig:SymHexRot}(d) 
and Fig.~\ref{fig:SymHexRot}(e).

%%%%%%%%%%%%%%%%%%%%%%%%%%%%%%%%%%%%%%%%%%%%%%%%%%

From the comparison of these two cases we see clearly that optical selection
rules are unaffected by the choice of linearly-- or circulary--polarised light, 
and are independent of polarisation in both cases.
This confirms the results of the group theory analysis given in 
Section~\ref{eq:group.theory.C6v}.

%%%%%%%%%%%%%%%%%%%%%%%%%%%%%%%%%%%%%%%%%%%%%%%%%%
\section{Kramers doublets in triangular graphene quantum dots}				 
%%%%%%%%%%%%%%%%%%%%%%%%%%%%%%%%%%%%%%%%%%%%%%%%%%
\label{appendix:KramersD}

%%%%%%%%%%%%%%%%%%%%%%%%%%%%%%%%%%%%%%%%%%%%%%%%%%
% Fig. 14
%%%%%%%%%%%%%%%%%%%%%%%%%%%%%%%%%%%%%%%%%%%%%%%%%%

\begin{figure} 
  \centering
  	\includegraphics[width=0.4\textwidth]{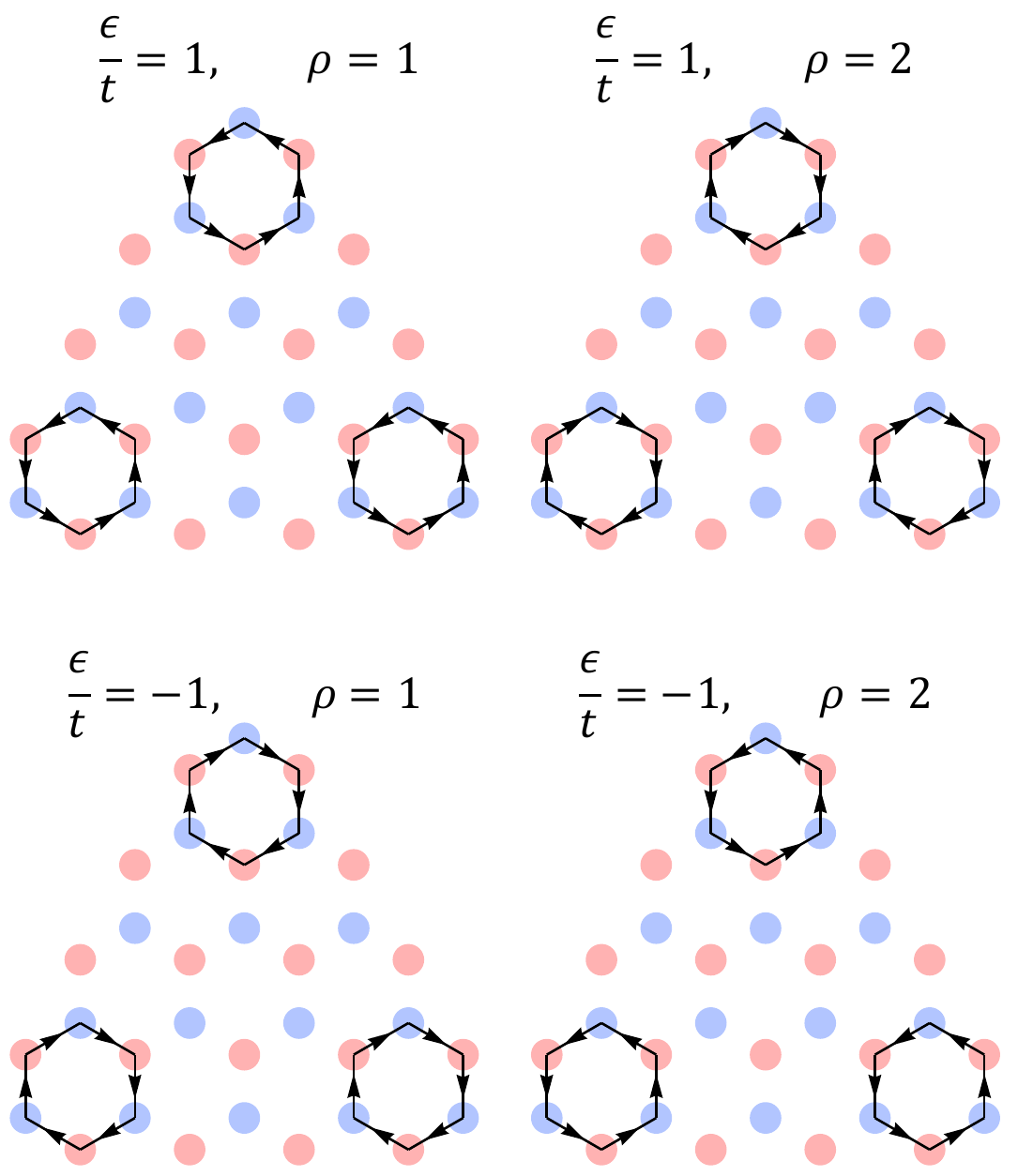}  
 	 \caption{
	 Distribution of the current on bonds within the Kramers--doublet states in triangular 
	 graphene quantum dots (GQD's). 
	 By choosing a vector basis of the Hamiltonian Eq.~(\ref{eq:H0}), 
	 which respects the rotational symmetry of the dot [Eq.~(\ref{eq:RotationOperator})],
	 we can distinguish between states of $\rho = 1,2$ [see Fig.~\ref{fig:SymTri_Rot}], 
	 showing a net circulation of current on their bonds.
	 Hereby, states at $\hbar \omega / t = \pm 1$ show non--zero currents at the edges 
	 of the GQD.
	 Results are shown for a GQD with zigzag edges of size $N = 33$.
	 }
\label{fig:KramersDoubletTri}
\end{figure}

%%%%%%%%%%%%%%%%%%%%%%%%%%%%%%%%%%%%%%%%%%%%%%%%%%
% Fig. 15
%%%%%%%%%%%%%%%%%%%%%%%%%%%%%%%%%%%%%%%%%%%%%%%%%%

\begin{figure} 
  \centering
  	\includegraphics[width=0.45\textwidth]{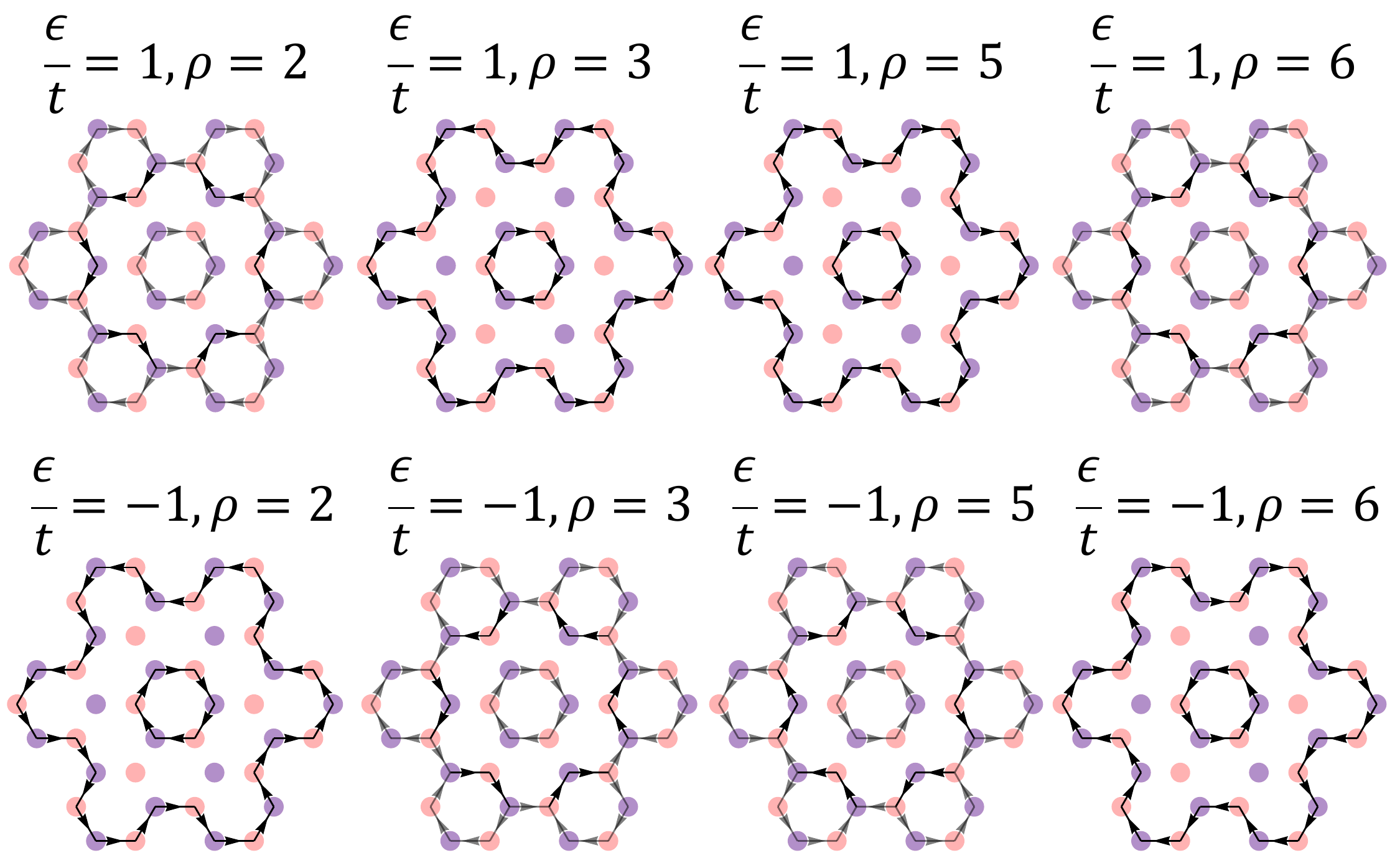}  
 	 \caption{
	 Distribution of the current on bonds within the Kramers--doublet states in hexagonal  
	 graphene quantum dots (GQD's). 
	 By choosing a vector basis of the Hamiltonian Eq.~(\ref{eq:H0}), 
	 which respects the rotational symmetry of the dot [see Eq.~(\ref{eq:RotationOperator})], 
	 we can distinguish between states of $\rho = 1,2,4,5$ [see Fig.~\ref{fig:SymHexRot}], 
	 showing a net circulation of current on their bonds.
	 Results are shown for a GQD with armchair edges of size $N = 42$.
	}
\label{fig:KramersDoubletHex}
\end{figure}

%%%%%%%%%%%%%%%%%%%%%%%%%%%%%%%%%%%%%%%%%%%%%%%%%%

We recognised in Sec.~\ref{sec:rot} the existence of doubly--degenerate states 
of Irrep $E$, forming Kramers doublets with time-reversal symmetry. 
The representation of the Hamiltonian in the basis of the rotational operator gives us 
access to these states.

%%%%%%%%%%%%%%%%%%%%%%%%%%%%%%%%%%%%%%%%%%%%%%%%%%

In Fig.~\ref{fig:KramersDoubletTri} and Fig.~\ref{fig:KramersDoubletHex} we plot the current (black 
arrows) within a triangular (\mbox{N = 33}) and hexagonal (\mbox{N = 42}) GQD for energies 
$\epsilon = \pm t$ for the Kramers doublets.
The currents flowing within each member of the Kramers doublet are oriented in opposite 
directions, such that the two states are connected by time--reversal.   
An external magnetic field would break this time--reversal symmetry 
and lift the degeneracy of the Kramers doublets.
We would then see changes in the optical absorption spectrum, which can lead to 
possible manipulations of magnetic moments in GQD's\cite{Kavousanaki2015}.

%%%%%%%%%%%%%%%%%%%%%%%%%%%%%%%%%%%%%%%%%%%%%%%%%%

In Fig.~\ref{fig:KramersDoubletTri} we find currents localised on the tips of the triangle, while 
in Fig.~\ref{fig:KramersDoubletHex} currents proceed in a circular fashion within the whole dot.
We find an absence of currents for states with $\epsilon = 0$.

%%%%%%%%%%%%%%%%%%%%%%%%%%%%%%%%%%%%%%%%%%%%%%%%%%
\section{Estimate of exciton binding energy within one--dimensional wave functions}			
%%%%%%%%%%%%%%%%%%%%%%%%%%%%%%%%%%%%%%%%%%%%%%%%%%
\label{appendix:HubbardU}

Calculations of $\sigma_{\alpha}(\omega)$ within a tight--binding model 
[Eq.~(\ref{eq:H0})] of an extended graphene sheet show a pronounced 
peak for $\hbar\omega \sim 2t$ [\onlinecite{Stauber2008, Buividovich2012}].
The same is true of a sufficiently large GQD [cf. Fig.~\ref{fig:DosCondALL}].
In experiments on graphene sheets, this peak is not observed at
$\hbar\omega = 2t \approx 5.6 \text{eV}$, but at the lower energy 
of $\hbar\omega = 4.6 \text{eV}$ [see e.g. Ref.~\onlinecite{Mak2011}]
--- a red--shift of \mbox{$\sim 1$~eV}.
This shift is usually ascribed to the binding--energy of an exciton
formed of particle--hole pairs, due to the interaction between electrons
 \cite{Herbut2008, Fritz2008, Yang2009, Yang2011a, Yang2011b, Yuan2011a}.

%%%%%%%%%%%%%%%%%%%%%%%%%%%%%%%%%%%%%%%%%%%%%%%%%%

Building on the insight that electronic states at energy $\epsilon = \pm t$ have a
one--dimensional character [cf. Fig.~\ref{fig:1DAll}, Section~\ref{sec:1-dimWaveFunction}], 
and are built of two electrons confined to two sites [cf. Fig.~\ref{fig:1DWF}], we can make 
a very simple estimate of the exciton binding energy by considering the two--site Hubbard model
\begin{multline}
	\hat{\mathpzc{H}} = -t ( 
		   \hat{\mathpzc{c}}^{\dagger}_{1,\uparrow} \hat{\mathpzc{c}}_{2,\uparrow}
		+ \hat{\mathpzc{c}}^{\dagger}_{2,\uparrow} \hat{\mathpzc{c}}_{1,\uparrow}
		+ \hat{\mathpzc{c}}^{\dagger}_{1,\downarrow} \hat{\mathpzc{c}}_{2,\downarrow}
		+ \hat{\mathpzc{c}}^{\dagger}_{2,\downarrow} \hat{\mathpzc{c}}_{1,\downarrow} )   \\
		+ U ( \hat{\mathpzc{n}}_{1,\uparrow} \hat{\mathpzc{n}}_{1,\downarrow} 
		+  \hat{\mathpzc{n}}_{2,\uparrow} \hat{\mathpzc{n}}_{2,\downarrow} )     \; .
\label{eq:H_ED} 	
\end{multline}
%
%%%%%%%%%%%%%%%%%%%%%%%%%%%%%%%%%%%%%%%%%%%%%%%%%%
%
Diagonalising this Hamiltonian, we find the eigenvalues
\begin{align}
	\epsilon_0 &= \frac{1}{2} \big(U - \sqrt{ 16 t^2 + U^2} \big)		\\
	\epsilon_1 &= 0										\\
	\epsilon_2 &= U									\\
	\epsilon_3 &= \frac{1}{2} \big(U + \sqrt{ 16 t^2 + U^2} \big)  	\; .
\label{eq:ED_Val} 	
\end{align}
Optical selection rules allow transitions from the lowest lying energy state with $\epsilon_0$ 
to the intermediate states $\epsilon_1, \epsilon_2$.
Since the interaction potential $U$ will lift their degeneracy, 
low--energy transitions will just occur between $\epsilon_0$ and 
$\epsilon_1$, where we find that
\begin{align}
	\Delta \epsilon = \epsilon_1 - \epsilon_0 = 4.6\ \text{eV}
\end{align}
for
\begin{align}
	U \approx 2.2\ \text{eV} 
\end{align} 
We note that this is substantially lower than estimates of the on--site
potential $U \sim 9.3 - 10.1 \text{eV}$ in the published 
literature \cite{Wehling2011,Schuler2013}.

%%%%%%%%%%%%%%%%%%%%%%%%%%%%%%%%%%%%%%%%%%%%%
\bibliography{article}
%%%%%%%%%%%%%%%%%%%%%%%%%%%%%%%%%%%%%%%%%%%%%%%%%%%%

\end{document}